\documentclass[twocolumn,tighten,twocolappendix]{aastex63}

\usepackage{mathtools} 
\usepackage{amsmath}
\usepackage{bbm,bm}
\usepackage{makecell}
\usepackage{textcomp, gensymb}
\usepackage{float}

\usepackage{multirow}
\usepackage{enumitem}

\graphicspath{{./}{figures/}}
\makeatletter
\g@addto@macro\@floatboxreset\centering
\makeatother

\let\oldAA\AA
\renewcommand{\AA}{\text{\normalfont\oldAA}}

\newcommand{\PI}{PI\ignorespaces}
\newcommand{\PE}{PE\ignorespaces}

\newcommand{\SB}{Starburst99\ignorespaces}
\newcommand{\Athena}{{\it Athena}\ignorespaces}

\newcommand{\p}{\partial}
\newcommand{\di}{\mathrm{d}}

\newcommand{\pc}{\,{\rm pc}}
\newcommand{\kpc}{\,{\rm kpc}}
\newcommand{\second}{\,{\rm s}}
\newcommand{\yr}{\,{\rm yr}}
\newcommand{\Myr}{\,{\rm Myr}}
\newcommand{\Gyr}{\,{\rm Gyr}}
\newcommand{\cm}{\,{\rm cm}}
\newcommand{\nm}{\,{\rm nm}}
\renewcommand{\micron}{\,\mu{\rm m}}

\newcommand{\pcc}{\,{\rm cm}^{-3}}

\newcommand{\kms}{\,{\rm km}\,{\rm s}^{-1}}
\newcommand{\Msun}{\,M_{\odot}}
\newcommand{\Lsun}{\,L_{\odot}}
\newcommand{\Kel}{\,{\rm K}}
\newcommand{\kB}{{k_{\rm B}}}
\newcommand{\erg}{{\,{\rm erg}}}
\newcommand{\eV}{{\,{\rm eV}}}

\newcommand{\MeV}{{\,{\rm MeV}}}
\newcommand{\Sunit}{\,M_{\odot}\,{\rm pc^{-2}}}

\newcommand{\mH}{m_{\rm H}}
\newcommand{\muH}{\mu_{\rm H}}

\newcommand{\NH}{N}
\newcommand{\NHH}{N_{\rm H_2}}

\newcommand{\nH}{n}

\newcommand{\nHH}{n({\rm H_2})}

\newcommand{\nHinit}{n_0}

\newcommand{\xHH}{x_{\rm H_2}}
\newcommand{\xHI}{x_{\rm H}}
\newcommand{\xHII}{x_{\rm H^+}}

\newcommand{\Ho}{${\rm H}$}
\newcommand{\Hplus}{${\rm H}^+$}
\newcommand{\HI}{\ion{H}{1}}
\newcommand{\HII}{\ion{H}{2}}
\newcommand{\HH}{${\rm H}_2$}

\newcommand{\xCtot}{x_{\rm C,tot}}
\newcommand{\xCI}{x_{\rm C}}
\newcommand{\xCII}{x_{\rm C^+}}
\newcommand{\CI}{${\rm C}$}
\newcommand{\CII}{${\rm C}^+$}

\newcommand{\xOtot}{x_{\rm O,tot}}

\newcommand{\xOII}{x_{\rm O^+}}
\newcommand{\OI}{${\rm O}$}
\newcommand{\OII}{${\rm O}^+$}

\newcommand{\xCO}{x_{\rm CO}}
\newcommand{\xEL}{x_{\rm e}}
\newcommand{\nEL}{n_{\rm e}}

\newcommand{\CO}{${\rm CO}$}

\newcommand{\chiHH}{\chi_{\rm H_2}}

\newcommand{\chiunatt}{\chi_{\rm 0}}

\newcommand{\xicr}{\xi_{\rm cr}}
\newcommand{\xicrunatt}{\xi_{\rm cr,0}}

\newcommand{\Lcool}{\mathcal{H}}

\newcommand{\kHHform}{k_{\rm gr,H_2}}

\newcommand{\Zd}{Z_{\rm d}^{\prime}}
\newcommand{\Zg}{Z_{\rm g}^{\prime}}
\newcommand{\Ztot}{Z^{\prime}}

\newcommand{\fshldH}{f_{\rm red,UVB}}

\newcommand{\REV}[1]{\textcolor{black}{#1}}

\begin{document}

\title{Photochemistry and Heating/Cooling of the Multiphase Interstellar Medium
  with UV Radiative Transfer for Magnetohydrodynamic Simulations}

\author[0000-0001-6228-8634]{Jeong-Gyu Kim}
\affiliation{Department of Astrophysical Sciences, Princeton University, Princeton, NJ 08544, USA}
\affiliation{Korea Astronomy and Space Science Institute, Daejeon 34055, Republic Of Korea}
\affiliation{Division of Science, National Astronomical Observatory of Japan, Mitaka, Tokyo 181-0015, Japan}

\author[0000-0003-1613-6263]{Munan Gong}
\affiliation{Max-Planck Institute for Extraterrestrial Physics, Garching by Munich, D-85748, Germany}

\author[0000-0003-2896-3725]{Chang-Goo Kim}
\affiliation{Department of Astrophysical Sciences, Princeton University, Princeton, NJ 08544, USA}

\author[0000-0002-0509-9113]{Eve C.~Ostriker}
\affiliation{Department of Astrophysical Sciences, Princeton University, Princeton, NJ 08544, USA}

\correspondingauthor{Jeong-Gyu Kim}
\email{jeonggyu.kim@nao.ac.jp}

\begin{abstract}
  We present an efficient heating/cooling method coupled with chemistry and
  ultraviolet (UV) radiative transfer, which can be applied to numerical
  simulations of the interstellar medium (ISM). We follow the time-dependent
  evolution of hydrogen species (H$_2$, H, H$^+$), assume carbon/oxygen species
  (C, C$^+$, CO, O, and O$^+$) are in formation-destruction balance given the
  non-steady hydrogen abundances, and include essential heating/cooling
  processes needed to capture thermodynamics of all ISM phases. UV radiation
  from discrete point sources and the diffuse background is followed through
  adaptive ray tracing and a six-ray approximation, respectively, allowing for
  H$_2$ self-shielding; cosmic ray (CR) heating and ionization are also
  included. To validate our methods and demonstrate their application for a
  range of density, metallicity, and radiation field, we conduct a series of
  tests, including the equilibrium curves of thermal pressure vs. density, the
  chemical and thermal structure in photo-dissociation regions, H I-to-H$_2$
  transitions, and the expansion of H II regions and radiative supernova
  remnants. Careful treatment of photochemistry and CR ionization is essential
  for many aspects of ISM physics, including identifying the thermal pressure at
  which cold and warm neutral phases co-exist. We caution that many current
  heating and cooling treatments used in galaxy formation simulations do not
  reproduce the correct thermal pressure and ionization fraction in the neutral
  ISM. Our new model is implemented in the MHD code {\it Athena} and
  incorporated in the TIGRESS simulation framework, for use in studying the
  star-forming ISM in a wide range of environments.
\end{abstract}


\section{Introduction}\label{s:intro}

The interstellar medium (ISM) consists of multiple phases spanning a wide range
of density, temperature, chemical, and ionization states. Understanding and
modeling the thermal and chemical properties of the ISM is a fascinating subject
in its own right, but also crucial to other fields in astronomy. Cosmic
evolution depends on how gas is converted into stars and flows in and out of
galaxies, and these processes in turn depend critically on cycling of ISM gas
through thermal phases. Our cosmological view of the Universe is also
foregrounded by the emission and absorption of the multiphase (dusty) ISM, and
propagation of radiation through this medium similarly affects observations of
stars and exoplanets.

The multiphase ISM is a result of complex interplay between radiative and other
microphysical processes and multiple dynamical stresses associated with gravity,
magnetic fields, turbulence, propagating shocks, and other large-scale flows.
Injection of energy and momentum as star formation ``feedback'' creates
localized \ion{H}{2} regions, wind bubbles, and supernova remnants (SNR) which
may merge as hot superbubbles. Young, hot stars also produce radiation that may
be absorbed either close to star-forming regions or at a large distance,
depending on ISM density structure and cross sections. Most of the energy
deposited in the ISM is lost relatively quickly via radiative cooling,
controlled by the microphysics of the ISM.

Traditionally, the ISM is divided into three main phases---warm neutral medium (WNM), cold neutral medium (CNM), and hot ionized medium (HIM)---that are in rough pressure balance and collectively fill most of the ISM volume \citep{MO1977}.
The modern paradigm additionally includes thermally-unstable neutral gas with temperatures intermediate between those of CNM and WNM \citep{Heiles03, Murray18} and a warm ionized medium (WIM) that fills much of the volume at high altitude \citep{Reynolds1989}. The other key components, although small in terms of volume, are self-gravitating, dense clouds and feedback-driven bubbles at early evolutionary stages that tend to be over-pressured relative to their environment \citep{Sun20, Barnes2021}.

To represent the complex ``ecosystem'' of the ISM theoretically, modeling heating/cooling processes accurately is important for several reasons.
(1) Radiative cooling enables the runaway gravitational collapse that initiates star formation. The thermal state of star-forming gas determines the fragmentation scale and the characteristic mass of the collapsing object, which is crucial to the stellar initial mass function \citep[e.g.,][]{Jappsen05, Gong15, Sharda22}.
(2) The balance between heating and cooling in warm gas sets its thermal pressure, contributing to vertical support against the ISM's gravitational weight and thereby modulating the large-scale star formation rate (SFR) required to maintain equilibrium \citep{Ostriker10, KimCG11, KimCG13, Ostriker2022}.
(3) Radiative cooling controls feedback-induced bubble expansion, which converts thermal to kinetic energy and thereby drives the turbulence that both supports the ISM against gravity and creates localized structure \citep[e.g.,][]{MO2007}. In particular, the efficiency of mechanical feedback depends primarily on cooling at leading shocks \citep[e.g.,][]{Cox72,Cioffi88} 
and in turbulent mixing layers that form at interfaces between warm and hot gas \citep[e.g.,][]{El-Badry2019, Fielding2020, Lancaster2021a}.
(4) Chemical composition and gas temperature directly affect the observables of the ISM including ionic, atomic, and molecular line emission. The chemical and thermal states in turn depend on gas density, metal and dust abundance, and strengths of heating/ionization sources such as the ultraviolet (UV) radiation field and cosmic rays (CRs).

Accurate and efficient modeling of the microphysical processes relevant for all ISM phases
still remains a major computational challenge. 
The full description of radiation field at a given time requires six variables (three spatial, two angular, and one frequency), and the radiation field at a given point is affected by the properties of matter intervening along sight lines toward all radiation sources. Because of this high dimensionality and non-locality, solving the radiation transfer equation 
can be computationally expensive even with several simplifying assumptions.
Calculating chemistry usually requires solving multiple coupled, nonlinear ordinary differential equations (ODEs), which has a high computational cost if a large number of chemical species 
is involved. Because both radiation transfer and chemistry calculations can easily become the computational bottleneck in numerical simulations, the ISM photochemistry is often modeled via simple local approximations and/or in post-processing decoupled from the gas dynamics \citep[e.g.,][]{Safranek-Shrader17, Gong18, Gong20, Armillotta20, Jeffreson20, Hu21, Hu22}.

To model thermodynamics, hydrodynamic (HD) and magneto-hydrodynamic (MHD) simulations of the ISM often employ simple prescriptions for heating and cooling without explicit chemistry and radiation transfer. Often, a spatially and/or temporally uniform heating rate is adopted and the cooling function is approximated only as a function of temperature \citep[e.g.,][]{Rosen95, Wada2001, Koyama02, Sanchez2002, Piontek2004, Audit2005, Joung2006, KimCG08, Hill12, Kobayashi20}. The TIGRESS simulation framework introduced by \citet{KimCG17} models the temporal variation in far-ultraviolet (FUV) radiation based on recent star formation, adopting a spatially-uniform intensity calculated using a single approximate attenuation factor. 
In other work, a more complete chemical and cooling model is implemented, 
combined with an approximate shielding treatment such as 
constant shielding length \citep[e.g.,][]{Dobbs2008}, ``two-ray'' \citep[e.g.,][]{Inoue12, Iwasaki19}, ``six-ray'' \citep[e.g.,][]{Glover07a, Glover07b, Glover10}, or a tree-based method \citep[e.g.,][]{Glover12a, Smith14, Walch15, Hu16, Simpson16, Gatto17, Rathjen21} applied to a temporally-fixed FUV radiation field;
the \citet{Rathjen21} SILCC simulations follow ionizing radiation (but not FUV for photoelectric (PE) heating or \HH\ dissociation) from stellar sources via a tree-based backward ray tracing method \citep{Wunsch21}.  See \autoref{s:comparison_methods} for more extensive comparison of photochemistry and radiation treatments.

To model chemistry and radiation transfer explicitly and self-consistently, several studies combined hydrogen (and helium) photochemistry with a (M1-closure) moment-based radiative transfer method \citep[e.g.,][]{Rosdahl13, Nickerson18, Kannan19, Kannan20a, Kannan20b, Hopkins20, Chan21}. Although the moment-based method has an advantage that the computational cost is independent of the number of sources, it suffers from artifacts when beams originating from multiple sources cross. This approach also relies on a local approximation for the self-shielding of \HH\ dissociating radiation since gas columns cannot be tracked directly.

There are a few codes that are capable of simulating detailed ISM thermodynamics and photochemistry coupled with an accurate radiative transfer method such as adaptive ray tracing \citep[e.g.,][]{Baczynski15} and Monte Carlo techniques \citep[e.g.,][]{Harries19}, but computational cost has been a major obstacle in their practical use for large-scale ISM simulations. Implementations of microphysics employed in the galaxy formation community sometimes include fairly detailed 
chemistry and cooling \citep[e.g.,][]{Grassi14, Smith17, Ploeckinger20}, although the radiation field treatments used for photoprocesses 
are sometimes not well suited for studying the star-forming ISM.

In this paper, we present a simple 
photochemistry model coupled with radiation transfer that can be used in simulations of the multiphase, star-forming ISM under a wide range of physical conditions. We select the most important microphysical processes that govern the chemical and thermal state of the ISM. The UV radiation field (both ionizing and FUV radiation) is obtained by ray tracing from radiation sources based on \citet{KimJG17}, and coupled self-consistently with the photochemistry and photoheating calculations.
Our model tracks time-dependent evolution of molecular, neutral atomic, and ionized hydrogen, while imposing
steady-state abundances for carbon- and oxygen-bearing species for a given (time-dependent) hydrogen abundances.
Our model is based on the chemical network of \citet{Gong17} that has been
tested against observations of chemical species in the diffuse ISM and those
from more complex photodissociation region (PDR) codes. The heating and cooling
rates in our model are calculated from chemical abundances in molecular,
neutral, and photoionized gas. For high-temperature, collisionally-ionized gas,
we apply cooling rates from interpolation tables provided by \citet{Gnat12}. We
perform a variety of tests including steady-state, one-zone models; steady-state PDR models; and the expansion of radiative SNRs and \HII\ regions coupled with full hydrodynamic simulations. We also make comparison with other heating/cooling models used in the ISM and galaxy formation communities and the resulting thermal equilibrium curves of the neutral ISM.

Our photochemistry module is implemented within the MHD code \Athena\ \citep{Stone08}, and has been deployed for the next generation TIGRESS simulations \citep[extending the methods of][]{KimCG17} that we will present in subsequent publications.
We emphasize that our chemistry and heating/cooling implementation is quite general, relatively simple, and computationally inexpensive.  It is therefore suitable for implementation in both numerical ISM and numerical galaxy formation models (at a range of redshifts).  While we employ adaptive ray tracing to obtain the radiation field with best accuracy, less costly treatments that apply approximate shielding to the time-dependent emission produced by nearby star formation may also yield good results.

The remainder of the paper is organized as follows. \autoref{s:overview} gives an overview of ISM phases and our model ingredients.
\autoref{s:chem} and \autoref{s:heating_cooling} describe chemical and heating/cooling processes, respectively. \autoref{s:radcr} describes methods of radiative transfer and CR attenuation. \autoref{s:updates} describes the method of numerical updates.
\autoref{s:tests} presents numerical tests for our photochemistry module. In \autoref{s:discussion}, we make comparison of our heating/cooling method and resulting thermal equilibrium curves with previous studies. \autoref{s:summary} gives a summary.

\section{Overview of model ingredients}\label{s:overview}

In this section, we present an overview of the physics required for modeling photon-driven chemistry and heating/cooling in the ISM. To make this section pedagogically useful, we begin by providing a high-level overview of ISM gas phases (\autoref{s:phases}). We then outline key physical ingredients, namely, gas dynamics, UV radiation, dust grains, and CRs (\autoref{s:mhd}--\ref{s:cr}). We briefly discuss parameter dependence of source terms in \autoref{s:param}. Detailed descriptions of chemical (including photochemical) processes, heating/cooling processes, methods of UV radiative transfer and CR attenuation, and the scheme for numerical updates will be given in Sections~\ref{s:chem}--\ref{s:updates}.

The microphysics elements included in our photochemistry and thermodynamics module are radiative heating and cooling of gas,
photochemistry of hydrogen (time-dependent)
and of key carbon- and oxygen-bearing species 
(steady-state abundances coupled to time-dependent hydrogen abundances), ray-tracing UV radiation transfer (both ionizing/non-ionizing continuum and \HH-dissociating Lyman-Werner bands), and heating and ionization by low-energy CRs. The primary application of our model will be to MHD simulations of star-forming molecular clouds ($\sim 10$--$10^2 \pc$ outer scale) and ISM patches or global galactic ISM disks ($\sim 1$--$10 \kpc$ outer scale) that include stellar feedback. These applications guide the choices in developing a simplified set of chemical reactions and radiative transitions that enables us to track time-dependent gas thermodynamics. However, we emphasize that our model is not limited to these applications, and we expect that it will be valuable for a very broad range of ISM and galaxy formation/evolution studies. In particular, our chemistry and heating/cooling allow the gas metallicity and dust abundance to be free parameters, enabling application in high-redshift as well as low-redshift modeling, provided that appropriate resolution is achieved.

Nevertheless, we caution that because photochemical and thermodynamic processes are tightly coupled, the validity of the treatments discussed here is only guaranteed when they are correctly and consistently used together. For example, our \PE\ heating rate formula should only be used if the FUV radiation field appropriate for the local population of recently-formed stars is known, and if the ionization fraction within neutral gas is also  known (which generally requires knowledge of the local CR ionization rate). As noted, use of our formalism is only appropriate at sufficiently high numerical resolution, and could give physically meaningless results if resolution is too low. For example, if the numerical resolution is low enough that a single computational element at the mean density would be completely self-shielded to Lyman-Werner dissociating radiation,
it does not imply that the ISM should be fully molecular. Rather, it implies that there is insufficient resolution to distinguish the molecular, atomic, and ionized portions of the ISM, which depend on the distribution of internal radiation sources and inhomogeneous gas structure.


Some limitations of this study are worth noting. Our model is not intended to capture detailed chemistry of gas in dense portions of star-forming clouds (e.g., cores, protostellar envelopes) and PDRs, and gas that cools behind shocks, where a variety of molecules form by rich gas-grain chemistry, and grain properties are different from those in the diffuse ISM. Our model does not include transport of CRs along magnetic fields, nor does it include the transfer of dust-emitted infrared radiation. Our model neglects the ionization and heating by X-ray photons \citep{Wolfire03}, although low-energy CRs play a similar role. Finally, our hot gas cooling function assumes collisional ionization equilibrium (CIE) conditions. By limiting the effects we follow, we are able to keep the computational expense of the photochemistry and heating-cooling modules sufficiently low that they do not significantly impact the computational cost of our MHD simulations, while still incorporating the key microphysical processes that affect dynamical evolution of the system.

\subsection{Phases of interstellar gas}\label{s:phases}

Here we give a high-level overview of ISM phases with an emphasis on dominant chemical and heating/cooling processes that are the focus of this paper. The topic of ISM phases and thermodynamics is vast, and the reader is referred to numerous reviews and books for more comprehensive coverage of the subject \citep[e.g.,][]{Ferriere01, Tielens05, Cox05, Osterbrock06, Draine11book, Girichidis20}. This very short primer is intended as an orientation for those who are new to studies of the ISM, and may be skipped by experts.

\begin{itemize}
\item {\bf Molecular hydrogen gas} (\HH) is the dominant component of star-forming clouds and is a prerequisite for the formation of many other molecules. In the higher-density, inner regions of galaxies, \HH\ is the main ISM constituent by mass.
Since \HH\ is not directly observable in low-temperature molecular gas (due to the lack of excited dipole transitions), the second-most abundant molecule, \CO, is the most commonly-used molecular gas tracer \citep[see, e.g.,][]{Bolatto13}.
Except in the early universe, the most efficient route for \HH-formation is catalytic reactions on surfaces of dust grains \citep{Gould63, Dalgarno76, Cazaux04}.\footnote{Even in the present-day Universe, gas-phase H$_2$ formation can become important if dust is depleted locally, e.g., as a result of settling in proto-planetary disks \citep{Glassgold04} or sublimation in protostellar jets \citep{Tabone20}.}
The main \HH\ destruction channels are dissociating UV radiation, CRs (in UV shielded regions), and collisional dissociation (for shock-heated gas). In diffuse ($\nH \lesssim 10^2 \pcc$) molecular gas with moderate shielding, both the PE effect and CRs are important heating sources, and the main coolants are fine-structure lines from \CII\ and \CI~ \citep[e.g.,][]{Nelson97, Glover07a}.
In dense ($\nH \gtrsim 10^3 \pcc$) molecular clouds where UV starlight is strongly shielded, the gas temperature is determined by heating by low-energy CRs, cooling by rotational lines of \CO\ and other molecules \citep[such as H$_2$O and OH;][]{Goldsmith78, Neufeld95}, and (at very high density) by dust-gas interaction \citep{Goldsmith01}. Cooling by ro-vibrational transitions of \HH\ can be significant in warmer (shock-heated) molecular gas \citep{Neufeld93}.

\item {\bf Atomic hydrogen gas} (\HI)\footnote{We use \Ho\ to denote neutral atomic hydrogen as a  chemical species, but an alternative notation \ion{H}{1} (as in \HI-to-\HH\ transition) is also used to avoid \Ho\ being misinterpreted as total hydrogen.},
observable via 21 cm emission and absorption, is the main constituent overall of the ISM in galactic disks by mass and volume. In the solar neighborhood and throughout the majority of galactic atomic regions, the cold neutral medium (CNM; $\nH \sim 10$--$10^2 \pcc$ and $T \sim 30$--$200 \Kel$ in the solar neighborhood) and warm neutral medium (WNM; $\nH \sim 0.01$--$1\pcc$ and $T \sim 6000$--$10^4 \Kel$ in the solar neighborhood) can co-exist in thermal pressure balance \citep{FGH1969, Wolfire03, Bialy19}, while gas at intermediate temperatures is subject to thermal instability \citep{Field65}.\footnote{Although the thermal equilibrium curve for atomic gas admits two separate stable solutions at intermediate pressure over a large range of conditions, this is not always true. In extremely low-metallicity environments, increased \HH\ cooling can smooth out multiphase structure \citep[e.g.,][]{Inoue2015, Bialy19}.}
The two-phase equilibrium regime shifts to higher density and pressure at higher heating rates (in inner galaxies and at higher SFRs), lower metallicity (in outer galaxies and at higher redshift), or enhanced dust-to-gas ratio \citep{Wolfire95, Wolfire03}. The primary heating mechanism in the CNM and WNM is by the \PE\ effect induced when FUV photons are absorbed by small dust grains (which is sensitive to grain charging) and by low-energy CR ionization \citep{Watson72, deJong77, Bakes94}. The main metal coolants of atomic gas are fine-structure lines from \CII\ and \OI\ with minor contributions from \CI, ${\rm Si^+}$, ${\rm Fe}$, ${\rm Fe^+}$, etc. \citep[e.g.,][]{Wolfire95}. In the WNM, cooling by resonance Ly$\alpha$ lines and recombination of electrons on polycyclic aromatic hydrocarbon (PAH) molecules become important. In the WNM, the primary sources of free electrons are CR-ionized hydrogen and helium, while in the CNM, singly ionized carbon is also a key contributor \citep{Draine11book}.

\item {\bf \HII\ regions} are regions of warm gas, photoionized
by the Lyman continuum (LyC) radiation produced by young massive stars. Although they occupy a small fraction of the ISM volume, \HII\ regions play a unique astronomical role as observational signposts of star formation and laboratories for discovering repercussions of massive star feedback \citep[e.g.,][]{Deharveng10, Barnes2021}. The heating is dominated by thermalization of the excess kinetic energy (a few eV) carried by electrons generated by PI of hydrogen and helium \citep{Spitzer50}, with \PE\ heating from FUV on dust grains also contributing \citep{Weingartner01b} unless small carbonaceous grains and PAHs are destroyed by the harsh UV radiation field \citep[e.g.,][]{Chastenet2019}. The cooling is dominated by collisionally excited lines of metal ions, free-free emission, and radiative recombination \citep[e.g.,][]{Osterbrock65, Osterbrock06, Draine11book}. The equilibrium temperature is $\sim 10^4 \Kel$ and tends to decrease at high metallicity because of higher cooling efficiency of metal lines. Dust in \HII\ regions also absorbs radiation, with the photon momentum producing a force that pushes the coupled gas-dust mixture away from stellar sources \citep{Draine11}. If the drift velocity of dust relative to gas is very large (in central regions), however, grains may be destroyed.

\item {\bf Warm ionized medium} (WIM, or diffuse ionized gas: DIG) is warm ($T \sim 10^4\Kel$), low-density ($\nH \sim 10^{-2}$--$10^{-1}\pcc$), ionized gas in the diffuse ISM, often extending up to kiloparsecs away from the disk midplane \citep{Haffner09}. DIG is responsible for a significant fraction of the total H$\alpha$ emission and accounts for most of the ionized gas mass in galaxies.
In normal star-forming galaxies, the most likely source of ionization and heating of the WIM/DIG is LyC photons that escape from star-forming clouds after emission by O-stars, with propagation over large distances in some directions enabled by low-density hot-gas channels \citep[e.g.,][and references therein]{Kado-Fong2020}. Observations of line intensity ratios suggest that DIG is generally warmer (from higher [\ion{S}{2}]/H$\alpha$ and [\ion{N}{2}]/H$\alpha$) and ionized by a softer spectrum (from lower [\ion{O}{3}]/H$\alpha$ and \ion{He}{1}/H$\alpha$) than gas in classical \HII\ regions \citep[e.g.,][]{Madsen06}.
Heating and cooling processes are similar to those in \HII\ regions, although the CR heating and grain \PE\ heating play a greater role in low density ionized gas \citep[e.g.,][]{Weingartner01b, Dong11}.

\item {\bf Hot ionized medium} (HIM) is created by shock waves from SNe and, to a lesser extent, stellar winds \citep[e.g.,][]{Cox74, MO1977}. Mainly traced by UV absorption lines of high-ionization metals and X-ray emission, the HIM is thought to occupy a significant fraction of the local ISM's volume ($\sim 20$--$50\%$; \citealt{Ferriere01, Cox05}), but direct observational constraints are lacking.
Superbubbles created by multiple SNe are crucial for driving of galactic winds and fountains \citep[e.g.,][]{Norman89, MacLow88, KimCG18}. Dominant coolants in the hot gas are line emission from collisionally excited metal ions (for $T \sim 10^{5}$--$10^{7} \Kel$) and bremsstrahlung emission (for $T \gtrsim 10^7\Kel$) \citep[e.g.,][]{Sutherland93}. Accounting for both thermal and dynamical processes involving the hot gas is crucial for modeling the ISM phase balance and kinematics, as well as larger-scale multiphase outflows from galaxies to the circumgalactic medium. Although turbulent mixing of hot gas with cool/warm gas at fractal interfaces can greatly enhance radiative energy loss \citep[e.g.,][]{El-Badry2019, Fielding2020, Lancaster2021a, Lancaster2021b}, acceleration driven by SN shocks is still believed to be the main source of turbulence in the multiphase ISM.
\end{itemize}

In addition to the above, the term photodissociation regions (PDRs) refers classically to surfaces of interstellar clouds marking the boundary between an \HII\ region and molecular clouds, where the gas converts
from atomic to molecular form due to the attenuation of photodissociating radiation \citep{TH1985, Wolfire22}. In a broader context, PDRs include all neutral atomic/molecular regions in which FUV radiation regulates the gas chemistry and temperature. Much of the dust and fine-structure FIR emission (from \CII, \OI, \CI\, etc.) in galaxies is thought to originate from PDRs \citep{Hollenbach99}. While the term ``PDR'' in the latter sense encompasses the majority of atomic and molecular regions in star-forming galaxies (and hence the \HH\ and \HI\ components described above), there remains a conceptual emphasis on a one-dimensional succession of species abundances and corresponding heating/cooling processes associated with a monotonically varying FUV intensity \citep[e.g.,][]{Sternberg95}.

The above pr\'ecis indicates that modeling the multiphase ISM realistically involves not only gas dynamics but also a range of microphysical processes and UV radiative transfer.  In the next several sections, we provide additional details regarding key physics elements.

\subsection{Basic Equations}\label{s:mhd}

\begin{deluxetable*}{lll} 
\tabletypesize{\scriptsize}
\tablecaption{Parameters for Heating/Cooling and Photochemistry Source Terms \label{t:param}}
\tablewidth{0pt}
\tablehead{
\colhead{Symbol} & 
\colhead{Meaning} &
\colhead{Notes}
}
\startdata
\multicolumn{3}{l}{\bf Gas property} \\
$\nH$ & number density of hydrogen nuclei & $\nH = \rho/(\muH m_{\rm H})$ with $\muH \approx 1.40 - 0.02 (\Zg - 1)$ \\
$x_{\rm s}\equiv n_{\rm s}/\nH$ & species abundance & \HH, H, H$^+$, C$^+$, \CO, \CI, e \\
$T$ & gas temperature &  $T = P/[(1.1 + \xEL - \xHH)\kB \nH]$ \\
$\langle |\di v/\di r| \rangle$ & mean of local velocity gradient & used for LVG approximation in \CO\ cooling \\
\tableline
\multicolumn{3}{l}{\bf Radiation field} \\
$\mathcal{E}_{\rm LyC}$ & energy density of LyC photons ($\lambda < 91.2\,{\rm nm}$) & H/H$_2$ photoionization rate: $\zeta_{\rm pi,H/H_2} = c\sigma_{\rm pi,H/H_2} \mathcal{E}_{\rm LyC}/(h\nu_{\rm LyC})$ 
\\
$\chi_{\rm PE}$ & normalized intensity for PE band & relative to Draine (1978) ISRF for $110.1\,{\rm nm}$--$206.6\,{\rm nm}$ \\%
$\chi_{\rm LW}$ & normalized intensity for LW band & relative to Draine (1978) ISRF for $91.2\,{\rm nm}$--$110.8\,{\rm nm}$ \\
$\chiHH$ & normalized intensity for \HH-dissociating radiation & $\chiHH = \chi_{\rm LW} f_{\rm shld,H_2,eff}$; dissociation rate $\zeta_{\rm pd,H_2}=5.7\times 10^{-11}\second^{-1} \chiHH$ \\
$\chi_{\rm C}$ & normalized intensity for C-ionizing radiation & $\chi_{\rm C} = \chi_{\rm LW}f_{\rm shld,C,eff}$ ; ionization rate: $\zeta_{\rm pi,C} = 3.5\times 10^{-10}\second^{-1}\chi_{\rm C}$\\
\tableline
\multicolumn{3}{l}{\bf Other parameters:} \\
$\xicr$ & primary CR ionization rate per hydrogen nucleus & 
\makecell[l]{
Solar neighborhood value (unattenuated): $\xicrunatt = 2\times 10^{-16}\second^{-1}$ \\ 
attenuation based on effective column (Section~\ref{s:cr_atten})} \\
$\Zg$ & scaled gas metallicity & $\xCtot = 1.6 \times 10^{-4} \Zg$ and $\xOtot = 3.2 \times 10^{-4} \Zg$ \\
$\Zd$ & scaled dust abundance & relative to the \citet{Weingartner01a} model \\
\enddata
\tablecomments{$\zeta$ always denotes a rate per particle of a given species.}
\end{deluxetable*}

The system of fluid equations we solve is\footnote{Although we present MHD equations here for generality, for the tests presented in this paper we consider hydrodynamic flows only.}
\begin{equation}\label{e:cont}
  \dfrac{\p \rho}{\p t} + \nabla \cdot (\rho \bm{v}) = 0\,
\end{equation}
\begin{equation}\label{e:momentum}
  \dfrac{\p (\rho\bm{v})}{\p t} + \nabla \cdot \left[ \rho\bm{v}\bm{v} + P^* \bm{I} - \frac{\bm{B}\bm{B}}{4\pi} \right] = \bm{f}_{\rm rad}\,, 
\end{equation}
\begin{equation}\label{e:energy}
  \dfrac{\p E}{\p t} + \nabla \cdot \left[ \left( E + P^* \right) \bm{v} - \frac{\bm{B}(\bm{B}\cdot \bm{v})}{4\pi} \right] = \mathcal{G} - \mathcal{L} + \bm{v}\cdot \bm{f}_{\rm rad}\,,
\end{equation}
\begin{equation}\label{e:induction}
\frac{\partial \bm{B}}{\partial t} - \nabla \times \left( \bm{v} \times \bm{B} \right) = 0\,,
\end{equation}
\begin{equation}\label{e:scalar}
  \dfrac{\p n_s}{\p t} + \nabla \cdot (n_s \bm{v}) = \nH\mathcal{C}_s\,.
\end{equation}
Here $\rho = \muH m_{\rm H}\nH$ is the gas density, $\nH$ is the number density of hydrogen nuclei, $\muH$ is the mean molecular weight per H nucleon, $m_{\rm H}$ is the mass of a hydrogen atom,
$\bm{v}$ is the velocity, $\bm{B}$ is the magnetic field, $P^* = P + B^2/(8\pi)$ is the sum of gas pressure and magnetic pressure, $E = e + \rho v^2/2 + B^2/(8\pi)$ 
is the total energy density, and $e$ is the internal energy density.
We implement our photochemical and heating/cooling modules in the \Athena\ MHD code, and the numerical methods for solving the above set of partial differential equations are described in the code method papers of \citet{Stone08, Stone2009}. Temporal updates of the operator-split source terms on the right hand side of the equations are discussed in \autoref{s:updates}.

The source term $\bm{f}_{\rm rad}$ in the momentum and energy equations represents the radiative force per unit volume acting on the gas-dust mixture. This is obtained by summing $\rho \kappa_j{\bf F}_j/c$, where ${\bf F}_j$ is the flux in a given radiation band $j$ and $\kappa_j$ is the corresponding opacity (see \autoref{s:radcr}). We treat gas and dust as one fluid assuming that dust is tightly coupled to gas. 
We note that more generally, other forces such as gravity are included in the momentum source terms, but are omitted here as they are not essential for the tests presented.

In the energy equation, the volumetric heating and cooling rates are denoted by $\mathcal{G}$ and $\mathcal{L}$, respectively. These are detailed in \autoref{s:heating_cooling}.
We note in passing that because \autoref{e:energy} is a total energy equation, dissipation of turbulence (e.g. by advection of oppositely-signed momenta into a given computational cell) automatically creates an increase in the thermal energy density, without a need for explicit viscosity.
Equation~\eqref{e:scalar} represents the continuity equation for a species $s$, with the number density $n_s = x_s\nH$, where $x_s$ is the fractional abundance relative to hydrogen nuclei (note that for fully molecular gas, $x_{\rm H_2}=0.5$).
The source term $\nH\mathcal{C}_s$ is the net creation rate by chemical reactions.  The chemical network is described in \autoref{s:chem}.

We adopt the ideal gas law $e = P/(\gamma-1)$, where the adiabatic index $\gamma=5/3$.\footnote{Here we ignore the rotational and vibrational degrees of freedom in \HH. In warm 
molecular gas, the internal energy of molecular gas depends on ro-vibrational states of \HH, which makes $\gamma$ a function of temperature and ortho-to-para ratio \citep[e.g.,][]{Boley07}.} 
The gas pressure and temperature are related by $P = \rho \kB T/(\mu \mH) = (1.1 + \xEL - \xHH) n\kB T$, where 
$\xEL$ is the free electron abundance and 
we use the closure relation $\xHI + \xHII + 2\xHH = 1$, assume the abundance of helium $x_{\rm He,tot}=0.1$, and neglect the abundance of other trace species. The mean mass per particle is $\mu = \muH/(1.1 + \xEL - \xHH)$, and we adopt $\muH \approx 1.40 - 0.02 (\Zg - 1)$, where $\Zg = Z_{\rm g}/Z_{\rm g,\odot}$ is the gas metallicity normalized to the solar neighborhood value.
Adopting the protosolar abundances of the elements with atomic number less than 32 \citep{Asplund09}, the fractional mass of metals is $Z_{\rm g,\odot} = 0.014$.

\subsection{UV Radiation}\label{s:rt}

To model the propagation of continuum UV radiation through dusty gas, we solve in each frequency band the time-independent radiative transfer equation
\begin{equation}
  \hat{\bm{k}} \cdot \nabla I = -\alpha I \label{e:RT}
\end{equation}
where $I$ is the intensity, $\hat{\bm{k}}$ is the unit vector specifying the direction of radiation propagation, and $\alpha$ is the absorption cross section per unit volume. Our numerical ray-tracing method is summarized in \autoref{s:radcr} and detailed in \citet{KimJG17}. We consider absorption by dust grains (for both ionizing and non-ionizing radiation) and neutral hydrogen (for ionizing radiation) as the main sources of opacity. The ray tracing method does not allow for 
dust scattering.\footnote{The \citet{Weingartner01a} dust model suggests a dust albedo $\sim 0.2$--$0.4$ and scattering asymmetry parameter $\langle \cos \theta \rangle \gtrsim 0.6$ at UV wavelengths. To zeroth order, the relatively low albedo and preferential forward scattering justify the neglect of scattering. \citet{Parravano03} estimated that large-angle scattering is expected to produce diffuse radiation that is about 10\% of the total FUV radiation field.}

We divide the UV radiation field into three bands for the purpose of radiation transfer:
\begin{enumerate}
\item Photoelectric (PE; $110.8\nm < \lambda \le 206.6\nm$): Absorption of these FUV photons ($6.0 \eV < h\nu < 11.2 \eV$) by small dust grains leads to photoelecton emission, representing an important source of heating for neutral gas.

\item Lyman-Werner (LW; $91.2\nm < \lambda \le 110.8\nm$): Photons in this energy range ($11.2 \eV < h\nu < 13.6\eV$)\footnote{Most of the ${\rm H}_2$ line absorption at low-lying rotational levels of the ground vibrational state occur in this band \citep{Sternberg14}. The threshold wavelengths for C ionization and CO dissociation are about $\sim 110 \nm$ \citep{Heays17}.} are responsible for dissociating ${\rm H}_2$ and ${\rm CO}$ molecules and for ionizing C. They also contribute to the \PE\ heating.

\item Lyman Continuum (LyC; $\lambda \le 91.2\nm$): Photons in this energy band ionize \Ho\ ($h\nu > 13.6 \eV$) and \HH\ ($h\nu > 15.4 \eV$). Even though LyC photons with $13.6 < h\nu < 15.4 \eV$ can photodissociate \HH, they are mostly absorbed by atomic hydrogen and dust in ionized regions.
\end{enumerate}
Under this division, the LW and PE bands together constitute FUV radiation. For our purposes, the LyC band is synonymous with EUV radiation.

In the implementation described in this paper, we do not solve the transfer of optical photons (OPT; $206.6 \nm \le \lambda \lesssim 10^3\nm$; strictly speaking this also includes near-UV). Although optical photons do not have enough energy to contribute to gas heating by inducing the grain \PE\ effect, old stars produce significant radiation in the optical band, which is an important source of dust heating. Optical photons also exert a non-negligible radiation pressure force on dust grains. We treat the latter effect approximately (see \autoref{s:radforce}).

In a given band, the radiation energy density and mean intensity are related to the intensity by $\mathcal{E} = (1/c) \int I d\Omega$ and $J = 1/(4\pi) \int I d\Omega$, respectively. In our numerical calculation, given the contribution of photons from all rays passing through a given grid cell, we compute the energy density $\cal E$ (as well as the flux $\bf F$) in each band, and set $J=c{\cal E}/(4\pi)$. For notational convenience, we introduce a dimensionless variable for the mean intensity in the PE and LW bands, normalized by the \citet{Draine78} interstellar radiation field (ISRF) at the solar circle as 
\begin{eqnarray}\label{eq:chi_defn_PE}
\chi_{\rm PE} & = J_{\rm PE}/J^{\rm Draine}_{\rm PE}\,, \\
\chi_{\rm LW} & = J_{\rm LW}/J^{\rm Draine}_{\rm LW}\,,
\label{eq:chi_defn_LW}
\end{eqnarray}
where 
\begin{eqnarray}
J^{\rm Draine}_{\rm PE} 
&= 1.8 \times 10^{-4} \erg \cm^{-2}\second^{-1}\,{\rm sr}^{-1}\,, \\
J_{\rm LW}^{\rm Draine} & = 3.0 \times 10^{-5}\,\erg \cm^{-2}\second^{-1}\,{\rm sr}^{-1} \,.
\end{eqnarray}
The normalized FUV intensity is 
\begin{equation}\label{e:chi_FUV}
\chi_{\rm FUV} = (J_{\rm PE} + J_{\rm LW})/(2.1\times 10^{-4} \erg \cm^{-2}\second^{-1}\,{\rm sr}^{-1}) \,.
\end{equation}

In Appendix~\ref{s:sb99}, we present the evolution of UV spectra and luminosity per unit stellar mass emitted by a coeval stellar population sampling the Kroupa initial mass function (IMF; \citealt{Kroupa01}), computed from \SB\ \citep{Leitherer99, Leitherer14}.  \autoref{t:sb99} in Appendix~\ref{s:sb99}  summarizes the mean  timescales and energies for each radiation band. In Appendix~\ref{s:ISRF}, we also compare the spectrum weighted by the star formation history in the solar neighborhood to the \citet{Draine78} and other models of the local ISRF.

For all three UV bands, we use ray-tracing to compute a value of $\cal E$ in each cell, which for PE and LW is then normalized to obtain $\chi_{\rm PE}$ and $\chi_{\rm LW}$. The \PI\ of H/\HH\ and the corresponding heating rates as well as the \PE\ heating rate are then obtained by multiplying by appropriate rate coefficients (see 
\autoref{t:param}
and \autoref{s:heating}).

Photodissociation of \HH\ is more complicated because it is driven by a set of discrete lines (LW absorption band) which readily become optically thick. The full radiative transfer allowing for the effects of line overlapping and dust absorption/scattering is computationally infeasible in MHD simulations. Instead, as explained in \citet{Gong17}, at each point along every ray we apply a simple self-shielding factor from \citet[][]{Draine96} that depends on the total \HH\ column from the source \citep[see also][]{Sternberg14}. Thus, in addition to computing dust attenuation of LW photons, in our ray-tracing we compute the cumulative \HH\ column to be used in the self-shielding factor $f_{\rm shld,H_2,eff}$. Both dust- and self- shielding effects are taken into account in computing $\chi_{\rm H_2}$ and the dissociation rate (see \autoref{t:param} and  \autoref{e:Erad_H2},\autoref{e:fshld_H2_DB}).
C-ionizing radiation is followed in a similar way, accounting for the self-shielding and cross-shielding by \HH.

Detailed radiative transfer for the CO-dissociating band in principle could be computed similarly, with both self-shielding and cross-shielding. But because CO chemistry is complex and computationally expensive, here we adopt a simpler prescription for the CO abundance (\autoref{e:CO_fit}) based on a fit to the full photochemical models in \citet{Gong17}.

\subsection{Dust Physics and Cross Sections}\label{s:dust}

Dust grains play several important roles in the thermodynamics and photochemistry of the ISM. Grains are a major source of opacity at UV wavelengths. Most of the absorbed UV energy heats up the grains and is re-radiated in the infrared, but a small fraction goes into gas heating via the PE effect. The \PE\ effect is the most important heating source for the diffuse ISM in the Milky Way \citep[e.g.,][]{Bakes94, Wolfire95, Wolfire03} and more generally down to quite low dust abundances (at which point \PE\ heating drops below CR heating; see, e.g., \citealt{Bialy19}). Grains also act as coolants in diffuse warm gas (via recombination of ions on grains) and in dense molecular gas (via collisional interaction with the gas). Grain surfaces are key catalysts for molecule formation and ion recombination in the dense ISM. Finally, grains also transfer momentum gained from photons to gas via collisional and Coulomb interactions, with the resulting UV and IR radiation pressure force potentially driving gas flows over a range of scales.

In this work, we adopt the \citet{Weingartner01a} grain model. It consists of a separate population of carbonaceous grains, silicate grains, and very small grains (including PAHs), and reproduces the observed starlight extinction in the local ISM. 
We use $\Zd = Z_{\rm d}/Z_{\rm d,\odot}$ to denote the dust abundance normalized to the solar neighborhood value. For $\Zd = 1$, the mass of grain material relative to gas mass is $0.0081$ \citep{Weingartner01a}. The wavelength-dependent extinction and scattering cross sections for $\Zd=1$ are shown in Appendix~\ref{s:sigma_ave}. When $\Zd \ne 1$, we simply scale the cross sections by a factor $\Zd$.

We assume that dust grains are the only source of continuum opacity for the PE and LW bands (for \HH\ dissociation and C ionization, self-shielding and cross-shielding factors from lines are also applied), while for LyC we include absorption by \Ho\ and \HH. Thus, the continuum cross section per unit volume is $\alpha_{\rm LW}=\nH \sigma_{\rm d,LW}$
and $\alpha_{\rm PE}=\nH \sigma_{\rm d,PE}$
in the LW and PE bands, and $\alpha_{\rm LyC} = \xHI \nH \sigma_{\rm pi,H} + \xHH \nH \sigma_{\rm pi,H_2} + \nH \sigma_{\rm d,LyC}$ 
in the LyC band. Here, $\sigma_{\rm pi}$ denotes a \PI\ cross section (see \autoref{s:HII} for H and \autoref{s:HH} for \HH) and $\sigma_{\rm d}$ denotes a dust absorption cross section per hydrogen nuclei. In \autoref{s:sigma_ave} we calculate the mean dust cross sections, where we average in each band over the time-varying spectral energy distribution (SED) from \SB, assuming a coeval fully-sampled IMF that ages $50 \Myr$.
We find $\sigma_{\rm d,ext}/(10^{-21}\Zd \cm^2 {\rm H}^{-1}) = 2.5$, $2.1$, $1.4$, and $0.8$ and $\sigma_{\rm d,abs}/(10^{-21}\Zd \cm^2 {\rm H}^{-1}) = 1.9$, $1.5$, $0.9$, and $0.3$ for LyC, LW, PE, and OPT, respectively.
We generally adopt absorption cross sections, which is not a bad approximation since dust grains are strongly forward-scattering at UV wavelengths \citep{Weingartner01a}. \autoref{t:sb99} in \autoref{s:sigma_ave} also summarizes the mean cross sections relevant for computing radiation pressure.
The dependence of cross sections on the cluster age is weak (less than $\sim 5\%$ ($10\%$) variation for the FUV (LyC) band) over a timescale of $\sim 20\Myr$.

We assume that dust grains are tightly coupled to gas with a constant dust-to-gas mass ratio, although we have a simple model for dust destruction by thermal and non-thermal sputtering that suppresses radiation interactions under 
extreme conditions (see \autoref{s:updates} and \autoref{s:dust_dest}). 

\subsection{Cosmic rays}\label{s:cr}

CRs are highly energetic charged particles with an extended non-thermal spectrum that constitute a key component of the ISM. In the Milky Way, the total CR energy density 
is comparable to that of thermal gas,
with the majority of the energy carried by protons with kinetic energy of order a GeV \citep{Grenier15}.
CRs are mainly produced at SNe shocks via diffusive shock acceleration \citep[e.g.,][]{Baade34, Blandford78} with $\sim 10\%$ energy conversion efficiency \citep{Caprioli14}.
Since the majority of SNe in star-forming galaxies are associated with core collapse of short-lived, massive stars, the production rate of CRs is approximately proportional to the SFR. The energy density in CRs depends on both this input rate and on CR transport.  

The propagation of CRs is nearly collisionless, but they scatter off small-scale magnetic fluctuations (created by resonant streaming instabilities or cascades of ISM turbulence). CRs are tied to the ISM magnetic fields and therefore are rapidly advected by high-velocity gas, while also diffusing down the CR pressure gradient. The magnetic fluctuations that scatter CRs are subject to strong damping by ion-neutral collisions in the atomic and molecular gas that fills much of the ISM volume near the galactic midplane, while high-ionization, low-density gas is better able to support waves that confine CRs \citep[][and references therein]{Plotnikov2021,Armillotta2021}.

Although not dominant in the total CR energy budget, low-energy ($\lesssim 10^2 \MeV$) CR protons are an important source of ionization and heating in the atomic and molecular portion of the ISM \citep[e.g.,][]{Draine11book, Padovani2020}. Low-energy CRs suffer energy loss via ionization and Coulomb scattering by thermal gas. Ionization by CRs creates a secondary electron with mean kinetic energy $\sim 35\eV$. Some of this energy is used to excite bound states of \Ho\ and \HH, while another portion is converted to thermal energy of the gas via Coulomb scattering \citep{Dalgarno99}. Observations and theoretical models suggest that the CR ionization rate decreases with increasing column in the denser portion of the ISM \citep{Neufeld17, Silsbee2019}. 

In this work, we consider only low-energy CRs and do not include self-consistent CR transport. We adopt either a constant CR ionization rate or account for CR attenuation utilizing the effective column density obtained from radiative transfer (see Section~\ref{s:cr_atten}).

For solar neighborhood ISM conditions, we choose the standard value of the primary CR ionization rate per H nucleon as $\xicrunatt = 2.0\times 10^{-16}\second^{-1}$. This value is consistent with observational constraints inferred from absorption lines of molecular ions at the solar circle (\citealt{Indriolo07, Indriolo15, Bacalla19}; see also \citealt{Neufeld17}). As noted above, the CR energy density and therefore $\xicr$ is expected to increase with SFR, although this may be sublinear because CR collisional losses are fractionally higher in the high surface density regions where SFRs are high and advection velocities transporting CRs out of galaxies may also increase \citep{Armillotta2022}.

\subsection{Parametric Dependence of Source Terms}\label{s:param}

In general, the source terms in Equations~\eqref{e:energy} and \eqref{e:scalar} depend on several physical processes which in turn depend on both fluid and radiation variables as well as microphysical parameters (see \autoref{t:param} for a summary). For example, processes contributing to net species creation rates $\mathcal{C}_s$ include two-body reactions such as collisional ionization/dissociation and recombinations (radiative and dielectronic); the formation of \HH\ on grain-surfaces; recombination of ions on small grains and PAHs; \PI\ of H, \HH\, and \CI; photodissociation of \HH\ and \CO; ionization/dissociation of H and \HH\ by cosmic rays, and so on. Evaluating reaction rates thus requires knowledge of the local gas density, temperature, chemical composition, dust abundance, mean intensity (or energy density) of radiation in different frequency bins, and the CR ionization rate. Similarly, the heating and cooling rates depend on density, temperature, abundances of coolants and dust grains, and the radiation field. 

To make clear the dependence on quantities that vary in space and time within a simulation, we write
\begin{align}
  \mathcal{C}_s & = \mathcal{C}_s (\nH, T, x_s, \Zd, \Zg, \mathcal{E}, \xicr) \,, \\
  \mathcal{G} & \equiv \nH\Gamma(\nH, T, x_s, \Zd, \Zg,  \mathcal{E}, \xicr) \,, \label{eq:Gamma_def}
  \\
  \mathcal{L} & \equiv \nH^2\Lambda(\nH,T, x_s, \Zd, \Zg, \mathcal{E}, 
  |dv/dr|) \,.
  \label{eq:Lambda_def}
\end{align}
The specific heating rate $\Gamma$ and cooling function\footnote{As \citet{Gnat12} noted, several non-standard terms are used to denote $\Lambda$ in the literature: cooling coefficient, energy loss function, cooling function, and emissivity coefficient, to name a few. In this paper, we use the term ``cooling function'' following \citet{Dalgarno72}. 
We also note that, although less common, 
some authors normalize the cooling rate $\mathcal{L}$ to $\nEL\nH$ rather than $\nH^2$.} $\Lambda$ are in units of $\erg\second^{-1}$ and $\erg\cm^3\second^{-1}$, respectively. Note that 
$\Lambda$ has 
dependence on the velocity gradient $|dv/dr|$ to evaluate the CO cooling with radiative trapping effect (see \autoref{s:C-O_cooling}), but depends also on $\xicr$ implicitly through $\xEL$. In addition to abundances of carbon and oxygen-bearing species, accurate and consistent evaluation of the electron abundance $x_e$ is important in cold and warm gas. This is because not only do several line cooling emission terms depend on electron-impact collisional excitation, but also the efficiency of \PE\ heating depends sensitively on $\xEL$ through the grain charging parameter.
Failure to correctly estimate $x_e$ can lead to large 
over- or under-estimates of the role of heating and cooling (see \autoref{s:discussion}).
We also emphasize that the CR as well as the EUV and FUV radiation energy densities that enter the source terms depend on {\it non-local} conditions (i.e. outside a given simulation cell). The radiation field in the ISM and CGM generally depends on recent star formation within the galaxy (unlike the situation in the intergalactic medium, which depends on metagalactic radiation).

\section{Chemical Processes}\label{s:chem}

This section describes the chemical processes we include.
\autoref{t:chem_cooling} summarizes (1) the chemical (including photon- and CR-driven) reactions and associated heating/cooling processes we follow explicitly for selected species, (2) treatment of abundances and associated cooling for other species, and (3) treatment of metals, dust, and CRs.


\begin{deluxetable*}{lllllll}
\tabletypesize{\footnotesize}
\tablewidth{0pt}
\tablecaption{Summary of Chemistry, Radiation, Heating, and Cooling processes}
\tablehead{
\colhead{Species} &
\colhead{Abundance Calc.} &
\colhead{Reactions/Interactions\tablenotemark{$\dagger$}} &
\colhead{Radiation\tablenotemark{$\ddagger$}} &
\colhead{Heating} &
\colhead{Cooling} &
\colhead{Ref.\tablenotemark{$\mathsection$}}
}
\startdata
\multicolumn{6}{l}{Hydrogen:} \\
$\mathrm{H_2}$ & time-dependent & & & & & \\
& & $\mathrm{H + H + gr \rightarrow H_2 + gr}$ & & $\mathrm{H_2}$ formation & & Eq.\ref{e:H2form},\ref{e:Gamma_H2form} \\
& & $\mathrm{H_2 + cr \rightarrow H_2^+ + e}$ & & CR &  & Eq.\ref{eq:CR_heat} \\
& & $\mathrm{H_2 + \gamma_{pd} \rightarrow H + H}$ & LW(d,\HH) & photodiss. & & Eqs.\ref{eq:H2_pd},\ref{e:Gamma_H2diss} \\
& & $\mathrm{H_2 + \gamma_{pi} \rightarrow H_2^+ + e}$ & LyC(d,\Ho,\HH) & photoion. & & Eq.\ref{eq:H2_pi} \\
& & $\mathrm{H_2}$--$\mathrm{\gamma_{pd}}$ & LW(d,\HH) & UV pumping & & V18, Eq.\ref{e:Gamma_H2pump} \\
& & $\mathrm{H_2 + H \rightarrow 3H}$ & & & coll. dissoc. & G1-22, Eq.\ref{e:Lambda_cd_H2} \\
& & $\mathrm{H_2 + H_2 \rightarrow H_2 + 2H}$ & & & coll. dissoc. & G1-23,Eq.\ref{e:Lambda_cd_H2} \\
& & $\mathrm{H_2}$--$\mathrm{H_2/H}$ & & & ro-vib. lines & M21 \\
$\mathrm{H^+}$ & time-dependent & & & & & \\
& & $\mathrm{H + \gamma_{pi} \rightarrow H^+ + e}$ & LyC(d,\Ho,\HH) & photoion. & & Eqs.\ref{e:H_pi}, \ref{eq:photoion_heat} \\
& & $\mathrm{H + cr\rightarrow H^+ + e}$ & & CR & & G2-6,Eq.\ref{eq:CR_heat} \\ 
& & $\mathrm{H + e \rightarrow H^+ + 2e}$ & & & coll. ion. & G1-24, Eq.\ref{e:Lambda_ci_H} \\
& & $\mathrm{H^+ + e \rightarrow H}$ & & & radiative recomb. & G1-21,D11 \\
& & $\mathrm{H^+ + e + gr \rightarrow H + gr}$ & & & grain-assisted recomb. & G2-2,W01b \\
& & $\mathrm{H^+}$--$\mathrm{e}$ & & & free-free & D11 \\
$\mathrm{H}$ & conservation law & & & & & \\
& & $\mathrm{H}$--$\mathrm{e}$ & & & Ly$\alpha$ & Eq.\ref{e:k_lya} \\
\tableline
\multicolumn{6}{l}{Carbon and oxygen:} \\
\CII & equilibrium & & & & & \\ 
& & $\mathrm{C + \gamma_{pi}\rightarrow C^+ + e}$ & LW(d,\HH,\CI) & & & G2-14 \\
& & $\mathrm{C + cr\rightarrow C^+ + e}$ & & & & G2-9 \\
& & $\mathrm{C + \gamma_{cr} \rightarrow C^+ + e}$ & & & & G2-11 \\
& & $\mathrm{C^+ + e + gr\rightarrow C + gr}$ & & & & G2-3 \\
& & $\mathrm{C^+ + e \rightarrow C}$ & & & & G1-17 \\
& & $\mathrm{C^+ + H_2 \rightarrow CH_2^+}$ & & & & G1-9, G1-10 \\
& & $\mathrm{C^+}$--$\mathrm{H_2/H/e}$ & & & fine structure lines & G3 \\
\CO\ & fit to equilibrium & & LW(d)\tablenotemark{$\P$} & & & Eq.\ref{e:CO} \\
& & \CO--\HH/H/e & & & rotational lines\tablenotemark{$\ast$} & G3 \\
$\mathrm{C}$ & conservation law & & & & & \\
& & C--\HH/H/e & & & fine structure lines & G3 \\
$\mathrm{O^+}$ & $x_{\rm O^+} = \xHII$ &  & & &  & D11\\
$\mathrm{O}$ & conservation law & & & & & \\
& & O--\HH/H/e & & & fine structure lines & G3 \\
\tableline
\multicolumn{6}{l}{Others:} \\
$\mathrm{e}$ & charge conservation & & & & & Eqs.\ref{e:xEL}, \ref{e:xELhot} \\
$\mathrm{He}$ & constant $x_{\rm He,tot}$ & & & & CIE cooling in hot gas & GF12 \\
metals & constant $\Zg$ & & & & CIE cooling in hot gas & GF12 \\
 &  & & & & nebular metal lines & Eq.\ref{e:Lambda_neb}\\ 
dust & constant $\Zd$ & & & & & \\
& & gr--$\gamma_{\rm pe}$ & PE(d)+LW(d) & photoelectric & & Eq.\ref{eq:photoel_heat} \\
& & gr--e & & & grain-assisted recomb. & W01c \\
& & gr--\HH & & & dust-gas interaction & Eq.\ref{eq:gas-dust} \\
CR & attenuation & & & & & Eq.\ref{e:cratt} \\
\enddata
\tablenotetext{\dagger}{Chemical reactions are indicated with reactants and products. Other interactions leading to heating or cooling are indicated with key participants; e.g.~among the \HH\ interactions, \HH--\HH/\Ho\ indicates collisions between \HH\ and either \HH\ or \Ho\ that leads to ro-vibrational cooling.}
\tablenotetext{\ddagger}{Relevant radiation bands are indicated for photon-driven reactions, followed by source of opacity in parentheses.}
\tablenotetext{\mathsection}{References for reaction/interaction rates are to Table 1, 2, or 3 in \citet{Gong17}, equations in this paper, or an external reference. For example, G2-1 means the reaction and its rate are taken from Table 2 reaction 1 in \citet{Gong17}, G3 means the interaction and cooling/heating rate are given in Table 3 of \citet{Gong17}, and Eq.16 means that the reaction rate is specified in Equation (16) in this paper. Other references are as follows: V18 is \citet{Visser2018}, M21 is \citet{Moseley21}, D11 is \citet{Draine11book},  W01b is \citet{Weingartner01a}, W01c is \citet{Weingartner01c}, GF12 is \citet{Gnat12}. Where a specific equation is not listed in the table, heating/cooling rate information is given within the text of \autoref{s:heating}/\autoref{s:cooling}, respectively.}
\tablenotetext{\P}{Fitting formula uses dust-shielded LW radiation.}
\tablenotetext{\ast}{Radiative trapping effect taken into account with the LVG approximation.}
\label{t:chem_cooling}
\end{deluxetable*}

\subsection{Hydrogen}\label{s:hydrogen}

\subsubsection{Molecular hydrogen}\label{s:HH}

As the most abundant molecule in the universe, \HH\ is crucial to chemistry of interstellar clouds and facilitates the formation of other molecules. The main pathway for forming \HH\ is via catalytic reactions on surfaces of dust grains \citep{Gould63, Hollenbach71}.\footnote{The gas-phase formation of \HH\ becomes important only for very low $\Zd \lesssim 10^{-3}$
  \citep[e.g.,][]{Glover03, Cazaux04}. However, some observations suggest that the dust-to-gas ratio may drop more rapidly than gas metallicity in low-metallicity environments ($\Zd < \Zg$; e.g., \citealt{Remy-Ruyer14}). Considering this, 
  the transition to gas-phase \HH\ formation may occur at  
  $\Zg$ higher than $10^{-3}$ \citep[e.g.,][]{Bialy15, Bialy19}.}
The main \HH\ destruction channels are dissociative ionization by CRs and dissociation by FUV photons. We also include \HH\ destruction by \PI\ and collisional ionization.

The net creation rate for \HH\ can be written as
\begin{equation}\label{e:HH}
\begin{split}
\mathcal{C}_{\rm H_2} = & \kHHform \nH \xHI - (1.65\zeta_{\rm
cr,H_2} + \zeta_{\rm pd,H_2} \\ & + \zeta_{\rm pi,H_2} + \zeta_{\rm cd,H_2})\xHH\,. \\
\end{split}
\end{equation}
The first term on the right hand side represents the formation on grain surfaces. The rate coefficient $\kHHform$ (also commonly denoted by $R_{\rm H_2}$) depends on gas temperature and grain properties such as size distribution,
surface characteristics,
and temperature.
The functional form of $\kHHform$ we adopt is
 \begin{equation}\label{e:H2form}
 \begin{split}
     \kHHform = & 3.0\times 10^{-17} \Zd\\
     &\times \frac{2T_2^{1/2} }{1 + 0.4 T_2^{1/2} + 0.2 T_2 + 0.08 T_2^2}\cm^3\second^{-1},
 \end{split}
 \end{equation}
where 
$T_2 = T/(10^2\Kel)$. This functional form is taken from Eq. (3.8) in \citet{Hollenbach79}, and 
neglects the dependence on dust temperature.
The overall coefficient is adjusted such that $\kHHform$ equals to $3.0\times 10^{-17}\cm^3 \second^{-1}$ at $T_2 \approx 0.5$, and is proportional to $T_2^{1/2}$ for $T_2 \lesssim 1$. $\kHHform$ has a peak value of 
$3.7 \times 10^{-17} \cm^3 \second^{-1}$,
and decreases rapidly at higher temperatures ($T_2 \gtrsim 3$) due to the reduced sticking probability of hydrogen atoms on dust grains.
We note that there exist more up-to-date models of \HH\ formation on grain surfaces accounting for detailed surface chemistry \citep[e.g.,][]{LeBourlot12}. However, these models depend on parameters such as the grain size distribution and surface properties of dust grains that are uncertain and difficult to model in MHD simulations.
Our adopted coefficient is consistent with the observational estimates of $\kHHform$ in the diffuse ISM in the solar neighborhood
$3$--$4 \times 10^{-17}\,{\rm cm}^{-3}\,{\rm s}^{-1}$ \cite[e.g.,][]{Jura75, Gry02}, and captures the temperature-dependence of the formation rate reasonably well.
We also note that, although some evidence suggests that $\kHHform$ may be higher in dense PDRs \citep[e.g.,][]{Habart04, Wakelam17}, our model does not account for this.

For the \HH\ destruction by CR ionization, we use
\begin{equation}\label{eq:H2_cr}
    \zeta_{\rm cr,H_2} = 2 \xicr (2.3 \xHH + 1.5 \xHI),
\end{equation}
which includes ionization by secondary electrons \citep{Glassgold74}.\footnote{The number of secondary ionizations is $\sim 0.67$ in predominantly neutral atomic gas and decreases with increasing fractional ionization (Eq. 13.12 in \citealt{Draine11book}). \citet{Glassgold74} takes into account the difference in \HH\ and HI primary ionization, as well as the effect of secondary ionization. More recent calculations from \citet{Ivlev2021} found a slightly higher secondary ionization rate for \HH. Given the large uncertainty in the  low-energy CR spectrum which affects the total ionization rate, we choose to adopt the simple prescription in \citet{Glassgold74} which is the standard in the literature \citep[see e.g.,][]{Neufeld17}.} Here $\xicr$ is the primary CR ionization rate per H nucleon allowing for CR attenuation as explained in \autoref{s:cr_atten}. The constant factor 1.65 in \autoref{e:HH} comes from the destruction of \HH\ by ${\rm H}_2^+$ and the formation of \HH\ by ${\rm H_3^+}$ (see Eq. (18) in \citealt{Gong18}).


The dominant pathway for the photodissociation of \HH\ 
is through excitation from the ground electronic state (X${}^1\Sigma_g^+$) into the first or second electronic excited states (B${}^1\Sigma_g^+$ and C${}^1\Pi_u$) by absorption of resonance line photons (LW absorption band), followed by spontaneous decay into the vibrational continuum of the ground electronic state \citep[e.g.,][]{Stecher67, Sternberg14, Heays17}.
Because the cross sections have sharp peaks at discrete wavelengths, it is important to account for the effects of self-shielding. The photodissociation rate can be written as
\begin{equation}\label{eq:H2_pd}
\zeta_{\rm pd,H_2} = \zeta_{\rm pd,H_2}^{{\rm ISRF}} \chi_{\rm H_2} \,,
\end{equation}
where $\zeta_{\rm pd,H_2}^{{\rm ISRF}} = 5.7 \times 10^{-11} \second^{-1}$ is the  
free-space (or unshielded) \HH-dissociation rate in the standard \citet{Draine78} ISRF \citep{Heays17}, and $\chi_{\rm H_2}$ is the normalized intensity for \HH\ dissociation accounting for dust- and self-shielding effects. Section~\ref{s:ART} describes how $\chi_{\rm H_2}$ is computed from adaptive ray tracing with both shielding effects (see also \autoref{s:sixray} for discussion of the six-ray approximation).


\HH\ can be photoionized by LyC photons with energy $h\nu > 15.4 \eV$ to become an ${\rm H}_2^+$ ion.\footnote{The fraction of LyC photons absorbed by \HH\ is very small in \HII\ regions (see, e.g., \autoref{s:HIIreg}), but we have included this process for completeness.} Since we do not directly track molecular ions ${\rm H}_2^+$, we assume that they instantly become two hydrogen atoms by dissociative recombination. This is a good approximation given the large recombination coefficient and large electron density at \HH\ ionization fronts \citep{Baczynski15}. The \PI\ rate of \HH\ is 
\begin{equation}\label{eq:H2_pi}
    \zeta_{\rm pi,H_2} =  \int \sigma_{\rm pi,H_2}(\nu) \frac{c \mathcal{E}_{\nu}}{h\nu} d\nu = \sigma_{\rm pi,H_2} \frac{c \mathcal{E}_{\rm LyC}}{h\nu_{\rm LyC}},
\end{equation}
where $\mathcal{E}_{\rm LyC}$ is the radiation energy density of the LyC radiation ($\lambda < 91.2\nm$, see Section \ref{s:radcr} for the calculation of radiation transfer).
The integral is over the LyC portion of the spectrum, with mean photon energy $h\nu_{\rm LyC} = 19.4\eV$ and mean \PI\ cross section $\sigma_{\rm pi,H_2} = 4.6\times 10^{-18} \cm^2$ computed based on a temporal average of the spectrum from \SB, adopting a Kroupa IMF (see \autoref{s:sb99} and \ref{s:sigma_ave}).


Finally, molecular hydrogen can be collisionally dissociated when the passage of shock waves propagating at speeds $\gtrsim 25$--$40 \kms$ heat the gas \citep{Hollenbach80}. The collisional dissociation rate is computed as $\zeta_{\rm cd,H_2} = \sum_{\rm c} k_{\rm cd,H_2\mbox{-}c}n_{\rm c} $, where the subscript c denotes the collisional partner (\Ho\ and \HH). We take density- and temperature-dependent rate coefficients given by \citet[][see also \citealt{Gong17} Table 1]{Glover07a}.


\subsubsection{Ionized hydrogen}\label{s:HII}

The net creation rate for ionized hydrogen (\Hplus) is
\begin{equation}\label{e:HII}
\begin{split}
\mathcal{C}_{\rm H^+} = &(\zeta_{\rm pi,H} + \zeta_{\rm cr,H} + \zeta_{\rm ci,H})\xHI \\ & - (\alpha_{\rm rr,H^+}\nEL + \alpha_{\rm gr,H^+}\nH) \xHII
\end{split}
\end{equation}
The creation terms represent PI, CR ionization, and collisional ionization of H. The \PI\ rate by LyC radiation is calculated as
\begin{equation}\label{e:H_pi}
\zeta_{\rm pi,H}  = \int \sigma_{\rm pi,H}(\nu) \frac{c \mathcal{E}_{\nu}}{h\nu} d\nu  
 = \sigma_{\rm pi,H}\frac{c\mathcal{E}_{\rm LyC}}{h\nu_{\rm LyC}}.
\end{equation}
The integral is over the LyC portion of the spectrum, with mean photon energy $h\nu_{\rm LyC} = 19.4\eV$ and mean \PI\ cross section $\sigma_{\rm pi,H} = 3.1\times 10^{-18} \cm^2$ computed based on a temporal average of the spectrum from \SB, adopting a Kroupa IMF (see \autoref{s:sb99}).
For the CR ionization rate, we adopt $\zeta_{\rm cr,H} = \zeta_{\rm cr,H_2}/2$ (see \autoref{eq:H2_cr}). We consider the collisional ionization by electrons, $\zeta_{\rm ci,H}=k_{\rm ci, H\mbox{-}e}n_{\rm e}$, where the collisional ionization rate $k_{\rm ci, H\mbox{-}e}$ is taken from \citet{Janev87}.


The main destruction channels for \Hplus\ are  radiative recombination and grain-assisted recombination.
For radiative recombination, we use the  on-the-spot approximation and adopt the case B recombination rate coefficient $\alpha_{\rm rr,H^+}$ from \citet{Glover10}.
Grain-assisted recombination can be important in lowering the electron abundance in the CNM \citep{Draine11book}. We adopt the rate coefficient $\alpha_{\rm gr,H^+}$ given by \citet{Weingartner01c}.

\subsubsection{Atomic hydrogen}\label{s:HI}

The abundance of neutral hydrogen H is determined from the conservation of hydrogen nuclei $\xHI = 1 - \xHII - 2\xHH$, assuming all hydrogen is in the forms of H, \Hplus, or \HH. This is a good assumption: the abundances of other species that contain hydrogen, such as ${\rm H_3^+}$, ${\rm H_2^+}$, ${\rm CH_x}$, ${\rm OH_x}$ and ${\rm HCO^+}$, are comparatively very low \citep{Gong17}.

\subsection{Carbon and Oxygen}\label{s:CandO}

Carbon and oxygen are the most abundant elements after hydrogen and helium. The carbon- and oxygen-containing species we consider here (C, C$^+$, O, O$^+$, CO) are (together with H) the principal cooling agents of the warm and cold ISM. We derive the abundances of these species assuming steady-state solutions under the given (time-dependent) hydrogen abundances.
Following \citet{Gong17}, we adopt the the standard total gas phase abundances of carbon and oxygen atoms to be $\xCtot = 1.6\times 10^{-4} \Zg$ \citep{Sofia04} and $\xOtot = 3.2\times 10^{-4} \Zg$ \citep{Savage96}.



\subsubsection{Carbon}\label{s:carbon}

For carbon, we consider \CI, \CII, and \CO, neglecting the abundances of other trace carbon bearing species. Since the ionization potential of C is $11.3 \eV$, most of carbon in the diffuse ISM exposed to FUV radiation is in the form of \CII\ \citep{Werner70}. In these regions, the equilibrium approximation is fairly good for \CI\ and \CII\ because the ionization timescale is short (see below). In molecular clouds where most carbon is in the form of \CO, the chemical timescale for \CO\ formation is short compared to the dynamical time, and the steady-state assumption is also reasonably good \citep{Gong18,Hu21}.

The net rate for \CII\ creation is
\begin{equation}
\begin{split}
\mathcal{C}_{\rm C^+} = & (\zeta_{\rm pi,C} + \zeta_{\rm cr,C})\xCI - [ \alpha_{\rm gr,C^+}\nH \\
& \quad + \alpha_{\rm rr+dr,C^+}\nEL
+ k_{\rm C^+\mbox{-}H_2} n_{\rm H_2} ] \xCII,
\end{split}
\end{equation}
where $\zeta_{\rm pi,C}$ is the \PI\ rate, $\zeta_{\rm cr,C} = 3.85 \xicr$ the CR ionization rate,  $\alpha_{\rm gr,C^+}$ the grain-assisted recombination rate coefficient, $\alpha_{\rm rr+dr,C^+}$ is the sum of radiative and dielectronic recombination rate coefficients, and $k_{\rm C^+\mbox{-}H_2}$ the reaction rate for ${\rm C^+ + H_2 \rightarrow CH_2^+}$ \citep{Gong17}.


We take the \PI\ rate of C as
\begin{equation}
  \zeta_{\rm pi,C} = \zeta_{\rm pi,C}^{\rm ISRF}\chi_{\rm C} + 520 \times 2\xHH \xicr\,.
\end{equation}
For the first term, the (unshielded) \PI\ rate in the standard \citet{Draine78} ISRF is  $\zeta_{\rm pi,C}^{\rm ISRF} = 3.43\times 10^{-10} \second^{-1}$, giving a very short timescale of $1/\zeta_{\rm pi,C}^{\rm ISRF}\approx 10^2 \yr$. This indicates that the C/\CII\ transition is likely to be close to
the steady-state solution,
as also found in the numerical simulations in \citet{Hu21}. The normalized radiation field for PI of C
$\chi_{\rm C}$ accounts for dust-, self-, and cross shielding by \HH\ (see \autoref{s:ART} and Section 2.1.2 of \citealt{Gong17}).
The second term represents the ionization by CR-generated LW-band photons in molecular gas (``CR-induced photoionization'', e.g., \citealt{Gredel87, Draine11book}).
We assume that all CR-generated photons are absorbed locally by gas (not by dust) and take the factor 520 from \citet{Heays17}.

Setting $\mathcal{C}_{\rm C^+} = 0$ and using $\xCI = \xCtot - \xCII$ (assuming negligible CO in C/\CII\ transition regions), the (approximate) equilibrium abundance of \CII\ can be obtained,
\begin{equation}\label{eq:C+_frac}
  \dfrac{\xCII}{\xCtot} = \dfrac{\zeta_{\rm pi,C} + \zeta_{\rm cr,C}}{\zeta_{\rm pi,C} + \zeta_{\rm cr,C} +  \alpha_{\rm gr,C^+}\nH + \alpha_{\rm rr+dr,C^+}\nEL + k_{\rm C^+\mbox{-}H_2} n_{\rm H_2}}\,.
\end{equation}

Once the \CII\ (and \OII) abundance is determined, we compute the \CO\ abundance from
\begin{equation}\label{e:CO}
\dfrac{\xCO}{\xCtot} = \dfrac{2\xHH \left( 1 - {\rm max}\left(\tfrac{x_{\rm C^+}}{\xCtot}, \tfrac{x_{\rm O^+}}{\xOtot}\right) \right)}{1 + (n_{\rm CO,crit}/\nH)^2}.
\end{equation}
Here the numerator limits the maximum abundance of CO. The adopted form limits $\xCO/\xCtot$ to be below $2\xHH$, motivated by the fact that \HH\ is a prerequisite for \CO\ formation
and that \CO\ dissociating radiation in general penetrates deeper into molecular clouds than the H$_2$-dissociating radiation, which results in ``CO-dark'' molecular gas in the outer regions of molecular clouds \citep[see][and references therein]{Gong17}.
Considering the conservation of carbon and oxygen atoms, the \CO\ abundance is also limited by the abundances of \CII\ and \OII.
The denominator enables a smooth transition from CO-poor to CO-rich gas as the density increases, and this density dependence is chosen to reasonably match the CO abundance in \citet{Gong17}. The critical density $n_{\rm CO,crit}$, above which $\xCO/\xCtot > 0.5$ in molecular gas, is obtained from a fit to equilibrium chemistry calculations (Eq. 25 in \citealt{Gong17}),
\begin{equation}\label{e:CO_fit}
n_{\rm CO,crit} = \left(4 \times 10^3 \Zd \xi_{\rm cr,16}^{-2} \right)^{\chi_{\rm CO}^{1/3}} \left( \frac{50 \xi_{{\rm cr},16}}{Z_{\rm d}^{\prime 1.4}} \right) \pcc,
\end{equation}
where $\xi_{{\rm cr},16} = \xicr/(10^{-16}\second^{-1})$, and $\chi_{\rm CO}$ is the normalized radiation intensity for CO dissociating radiation in the LW band accounting only for dust-shielding.
Since the threshold wavelength for CO dissociation ($110\nm$) is the nearly same as the maximum wavelength of the LW band, we adopt $\chi_{\rm CO} = \chi_{\rm LW}$ in this work.
\footnote{In \citet{Gong17}, the dust-, self-, and cross-shielding were all taken into account in computing the CO abundance in PDR models. The dust-shielded normalized radiation intensity $\chi_{\rm CO}$ was used only as a fitting parameter to obtain \autoref{e:CO_fit}. The fit was obtained under the assumption that $\Ztot = \Zg = \Zd$. The applicable parameter range for the fit is $\nH \approx 10$--$10^4\pcc$, $\xi_{\rm cr,16} \approx 0.1$--$10$, and $\Ztot = 0.1$--$1$.}

We note that the functional form adopted in \autoref{e:CO}
  is intended as a convenient approximation rather than a detailed fit. By performing a uniform-density PDR test and comparing the result with the \citet{Gong17} model, we find that our model reproduces the gas column at which the C/CO transition occurs within 30\% (see \autoref{s:plane_par}).

Finally, the abundance of \CI\ is calculated from the conservation of carbon atoms, $\xCI = \xCtot - \xCII - \xCO$.

\subsubsection{Oxygen}\label{s:oxygen}

We assume that all ionized oxygen is in the singly ionized state of \OII. Because the ionization potential of O ($13.62 \eV$) is almost identical to that of H ($13.60 \eV$), we adopt a simple approximation $\xOII/\xOtot = \xHII$. Even in partially ionized gas, the rapid charge exchange ($\mathrm{O^+ + H \leftrightarrow O + H^+}$) keeps the oxygen ionization fraction close to that of hydrogen for $T \gtrsim 10^3 \Kel$ (see e.g., Fig 14.5 in \citealt{Draine11book}). In lower temperature gas, our approximation overestimates the amount of \OII, but it makes no practical difference for cooling as other coolants are more important at these temperatures.


After computing the abundance of O$^+$ and CO (from \autoref{e:CO}), the abundance of neutral oxygen is obtained from $x_{\rm O} = \xOtot - x_{\rm CO} - x_{\rm O^+}$.

\subsection{Free electrons}\label{s:electrons}

For gas temperatures below $T < 2 \times 10^4 \Kel$, we assume that free electrons come from \Hplus, \CII, \OII, and molecular ions, represented by $M{\rm H}^+$.  Thus, $\nEL=\xEL \nH$ with 
\begin{equation}\label{e:xEL}
\xEL = \xHII + \xCII + \xOII + x_{M{\rm H}^+}.
\end{equation}
In neutral atomic gas, the main source of free electrons is from H$^+$ and C$^+$. 
We ignore the fact that the CR- and X-ray-ionized Helium can contribute up to $20\%$--$30\%$ of electron abundance \citep{Wolfire03}, but in dense gas it is even less \citep{Gong17}.
The contribution to $\xEL$ by \OII\ is relatively small in both neutral and ionized gas, but is included for consistency.

In dense molecular gas with $\xEL \lesssim 10^{-5}$, electrons resulting from the CR ionization of molecular hydrogen and other molecular ions such as HCO$^+$ are important \citep{McKee89}.
The abundance of molecular ion $M{\rm H}^+$ is obtained from Equation (16.25) in \citet{Draine11book},\footnote{We used Eq. (13.12) of \citet{Draine11book} to calculate the parameter for secondary ionization $\phi_s$.}
and is multiplied by an additional factor of $2\xHH$ to account for the molecular gas fraction. 

At higher temperatures ($T \gtrsim 2 \times 10^4\Kel$), the main source of electrons comes from collisionally ionized hydrogen and helium.
Collisionally ionized metals contribute a small fraction (up to $\sim \Zg \%$).
The electron abundance in hot gas is therefore calculated as
\begin{equation}\label{e:xELhot}
\xEL = \xHII + x_{\rm e,He,CIE} + \Zg x_{\rm e,metal,CIE}\,.
\end{equation}
We calculate the time-dependent $\xHII$ self-consistently across the full temperature range, and use the tabulated CIE values for helium and metals as a function of temperature from \citet{Gnat12}.

\section{Heating and Cooling}\label{s:heating_cooling}


\subsection{Heating}\label{s:heating}

We model heating associated with absorption of FUV and EUV radiation, impacts of cosmic rays, and  exothermic reactions involving \HH.

With \autoref{eq:Gamma_def}, the heating rate per H nucleon is calculated from
\begin{equation}\label{e:heating}
\Gamma = \Gamma_{\rm pe} + \Gamma_{\rm cr} + \Gamma_{\rm pi} + \Gamma_{\rm H_2},
\end{equation}
where the individual terms denote heating from  the \PE\ effect on dust grains, CR ionization, PI, and \HH\ (by formation, dissociation, and UV pumping).

\subsection{Grain Photoelectric Heating}\label{s:heating_pe}

The dominant gas heating mechanism in the diffuse neutral atomic medium (and extending into low-extinction molecular gas) is the absorption of FUV photons ($6.0 \eV < h\nu < 13.6 \eV$) by small dust grains followed by \PE\ emission.
The \PE\ heating rate is proportional to the local FUV intensity, with the rate given by \citet{Weingartner01b}
\begin{equation}\label{eq:photoel_heat}
\Gamma_{\rm pe} = 10^{-26} G_0 \epsilon_{\rm pe} \Zd \erg\second^{-1},
\end{equation}
where $G_0 = 1.7\chi_{\rm FUV}$ is the FUV radiation strength in \citet{Habing68} units. The heating efficiency function $\epsilon_{\rm pe}$ depends on the grain charging parameter $\psi = G_0 \sqrt{T}/\nEL$, which is proportional to the ratio between the ionization and recombination rate on grains. Grains are more positively charged at high $\psi$, which makes it harder for a photoejected electron to escape the grain potential well. 
We adopt the functional form of $\epsilon_{\rm pe}$ from \citet{Weingartner01b} (for the grain size distribution A, $R_V=3.1$ and $b_{\rm C} = 4.0 \times 10^{-5}$, see their Eq. (44) and \autoref{t:chem_cooling}); i.e. 
\begin{equation}\label{e:PE_efficiency}
 \epsilon_{\rm pe}=\frac{C_0 + C_1 T^{C_4}}{1+ C_2 \psi^{C_5}\left[1+C_3 \psi^{C_6} \right]}  
\end{equation}
with $C_0=5.22$, $C_1=2.25$, $C_2=0.04996$, $C_3=0.00430$, $C_4=0.147$,$C_5=0.431$, $C_6=0.692$.
The heating efficiency $\epsilon_{\rm pe}$ 
is roughly inversely proportional to $\psi$ for $\psi\gtrsim 10^4 \Kel^{1/2} \cm^3$
and flattens to 
$\sim 5$--$10$ 
at low $\psi \lesssim 10^2 \Kel^{1/2} \cm^3$.\footnote{The absorption rate of FUV photons by dust grains per H nucleon for the \citeauthor{Draine78} ISRF is $c \int \mathcal{E}_{\nu} \sigma_{\rm d,abs,\nu} d\nu = 2.74 \times 10^{-24} \erg \second^{-1}$. Therefore, $\sim 0.6 \epsilon_{\rm pe}$ is the fraction of FUV energy (in percent) absorbed by dust grains going into \PE\ heating.}
Because $\psi$ can vary considerably, $\Gamma_{\rm pe}$ can vary significantly even for a fixed radiation field.

A few cautionary remarks on \PE\ heating efficiency are worthwhile. First, relatively large errors ($\gtrsim 20\%$) may arise when using \autoref{e:PE_efficiency} outside the ranges recommended by \citet{Weingartner01b}: $10 \Kel \le T \le 10^4 \Kel$ and $10^2 \Kel^{1/2} \cm^3 \le \psi \le 10^6 \Kel^{1/2} \cm^3$. In practice, we add $50 \Kel^{1/2} \cm^3$ to $\psi$ in order to prevent it from becoming too small in strongly shielded cold gas. Second, although we adopt the fixed functional form for the \PE\ heating efficiency (and the cooling function for grain-assisted recombination $\Lambda_{\rm gr,rec}$ in \autoref{s:gr_rec}), it is sensitive to the adopted spectrum of radiation field, abundances, and size distributions of very small grains and PAHs, and their (poorly constrained) shapes, which can introduce a factor of few uncertainty in \PE\ heating efficiency (B.T. Draine, private communication).

\subsection{Cosmic Ray Heating}\label{s:heating_cr}

The CR heating rate is written as
\begin{equation}\label{eq:CR_heat}
\Gamma_{\rm cr} = q_{\rm cr} \xicr
\end{equation}
where $q_{\rm cr}$ is the energy added to the gas per primary ionization.\footnote{\citet{Gong17} adopted the total ionization rate rather than the primary ionization rate in \autoref{eq:CR_heat}, which resulted in $\sim 50\%$ overestimate of the CR heating rate in atomic and molecular gas.}

In diffuse atomic gas, CR heating occurs because a fraction of the secondary electron's excess kinetic energy ($\sim 35\eV$) is thermalized via Coulomb interactions with free electrons in thermal gas. The heating efficiency increases from $\sim 20\%$ in neutral gas to $\sim$ 100\% in fully ionized gas.
The CR heating in molecular gas is more complicated as multiple heating mechanisms exist: Coulomb losses, vibrational/rotational excitation of \HH\ followed by collisional de-excitation, dissociation of \HH, and exothermic chemical reactions of CR-produced ions with heavy neutral species and electrons. The heating efficiency is then a complex function of density, abundances of electrons, \HH\ and heavy molecules, and temperature \citep{Dalgarno99, Glassgold12}.

Given the complexity of heating mechanisms, we adopt the simple approach introduced by \citet{Krumholz14} (see also Eqs.(30)--(32) in \citet{Gong17}),
\begin{equation}
    q_{\rm cr} = \xHI q_{\rm cr,H} + 2\xHH q_{\rm cr,H_2}
\end{equation}
where $q_{\rm cr,H}$ (in the range $7.5$--$32\eV$) is from the fitting formula recommended by \citet[][Ch. 30]{Draine11book} and $q_{\rm cr,H_2}$ (in the range $10$--$18\eV$) is a density-dependent fit to the heating efficiency in molecular gas presented in \citet[][Table 6]{Glassgold12}.

\subsection{Photoionization Heating}\label{s:heating_pi}

\PI\ of neutral hydrogen produces a free electron with excess kinetic energy which heats up the gas, with a rate per H nucleon
\begin{equation}\label{eq:photoion_heat}
\Gamma_{\rm pi,H} = \xHI \int_{h\nu_0}^{\infty} \frac{c\mathcal{E}_{\nu}}{h\nu} (h\nu - h\nu_0) \sigma_{\rm pi,H}(\nu) d\nu = \xHI  q_{\rm pi,H} \zeta_{\rm pi,H},
\end{equation}
where $h\nu_0 = 13.6 \eV$ is the ionization threshold, $q_{\rm pi}$ is the mean excess kinetic energy which depends on the local spectrum of the ionizing radiation, and $\zeta_{\rm pi,H}$ is the \PI\ rate given by \autoref{e:H_pi}.
Heating from \HH\ \PI\ is similar to that of \Ho, with $\xHH$ replacing $\xHI$ and $\sigma_{\rm pi,H_2}$ replacing $\sigma_{\rm pi,H}$ in the above formula:
\begin{equation}\label{eq:photoion_heat_H2}
    \Gamma_{\rm pi,H_2} = \xHH q_{\rm pi,H_2}\zeta_{\rm pi,H_2}\,,
\end{equation}
for $\zeta_{\rm pi,H_2}$ given in \autoref{eq:H2_pi}.
In \autoref{s:sigma_ave}, we present the mean photoelectron energies $q_{\rm pi,H}$ and $q_{\rm pi,H_2}$ averaged over the SED of a simple stellar population predicted by \SB.
The photon rate weighted, time-average values are $q_{\rm pi,H}=3.4\eV$ and $q_{\rm pi,H_2}=4.4\eV$).

%

\subsection{\texorpdfstring{${\rm H}_2$}{H2} Heating}\label{s:heating-h2}

For the \HH\ heating ($\Gamma_{\rm H_2}$), we include heating by \HH\ formation, UV pumping, and photodissociation. The formation of \HH\ is exothermic, releasing $4.5\eV$ per \HH. Some fraction of the released energy is converted to heat by the ejection of \HH\ from dust grains or by the excitation of \HH\ to higher ro-vibrational levels followed by collisional de-excitation.
UV pumping is initiated by the absorption of LW band photons, which excites \HH\ to higher electronic states. About $10\%$--$15\%$ of the photoexcited \HH\ decays into the vibrational continuum of the ground electronic state, leading to dissociation of \HH\ and deposition of thermal energy to the gas ($0.4\eV$ per dissociation). Most of the excited \HH\ decays into vibrationally excited levels,
and then to the ground vibrational state. When the density is high enough, however, some fraction of the energy can be converted into heat by collisional de-excitation \citep{Sternberg89}. 

Similarly to \citet{Hollenbach79} (see also \citealt{Baczynski15, Bialy19}), we calculate $\Gamma_{\rm H_2}$ as
\begin{align}
    \Gamma_{\rm H_2,form} & = \kHHform \nH \xHI \left( 0.2 \eV + \frac{4.2\eV}{1 + n_{\rm crit}/\nH} \right) \label{e:Gamma_H2form} \,, \\
    \Gamma_{\rm H_2,pump} & = \zeta_{\rm pd,H_2} \xHH f_{\rm pump} \times \frac{2\eV}{1 + n_{\rm crit}/\nH} \label{e:Gamma_H2pump} \,, \\
    \Gamma_{\rm H_2,pd} & = \zeta_{\rm pd,H_2}\xHH  \times 0.4\eV \label{e:Gamma_H2diss} \,,
\end{align}
where $\left[1 + (n_{\rm crit}/\nH)\right]^{-1}$ is the efficiency reduction factor for heating by collisional de-excitation and $f_{\rm pump}=8$ is the ratio between the rate at which the vibrationally excited \HH\ (${\rm H_2}^{\ast}$) is produced and the photodissociation rate \citep{Sternberg14}. The critical density is calculated as 
\begin{equation}
n_{\rm crit} = \frac{A_{\rm eff} + \zeta_{\rm pd,H_2}}{\xHI k_{\rm H_2\textnormal{-}H} + \xHH k_{\rm H_2\textnormal{-}H_2}}\,,
\end{equation}
where $A_{\rm eff} = 2\times 10^{-7}\second^{-1}$ is the effective radiative decay rate of ${\rm H_2}^{\ast}$ \citep{Burton90}. We take $\zeta_{\rm pd,H_2}$ as the effective dissociation rate of ${\rm H_2}^{\ast}$ and adopt the updated collisional rates $k_{\rm H_2\textnormal{-}H/H_2}$ as described in Appendix A of \citet{Visser2018}.

\subsection{Cooling}\label{s:cooling}

We model the cooling processes in gas at $T<2\times 10^4~\Kel$
explicitly, allowing for dependencies on density, temperature, chemical abundances, local radiation field (via grain-assisted recombination), and so on (see \autoref{t:param} and \autoref{t:chem_cooling}). This includes cooling by free-free emission, recombination, 
and forbidden lines of ionized metals (photoionized gas); 
the Ly$\alpha$ line and 
grain-assisted recombination (warm neutral gas); fine structure lines of \CII, \CI, and \OI\  (warm and cold neutral gas), \CO\ rotational lines (molecular gas); \HH\ ro-vibrational lines (shock-heated molecular gas); and gas--dust coupling (dense molecular gas). For gas with $T > 3.5 \times 10^4~\Kel$, we use the tabulated CIE cooling function from \citet{Gnat12} for metals and helium, while calculating hydrogen cooling explicitly.
A similar approach has been adopted in, e.g., \citet{Hu17} and the SILCC simulations \citep{Walch15}, with the transition temperature of $3\times 10^4\Kel$ and $10^4\Kel$, respectively.
For temperatures $T_{\rm t,1} < T < T_{\rm t,2}$,
where $T_{\rm t,1} = 2\times 10^4~\Kel$ and $T_{\rm t,2} = 3.5 \times 10^4~\Kel$, 
we transition between our explicit cooling calculation and the CIE cooling function. All of our cooling calculations except for cooling by \CO\ assume optically-thin conditions.

The cooling function is expressed as
\begin{align}\label{eq:cooling_fnct}
  \Lambda = &  \Lambda_{\rm hyd} + \nonumber \\
  & (1-S)\times ( \Lambda_{\rm C^+} + \Lambda_{\rm C} + \Lambda_{\rm O} + \Lambda_{\rm CO} + \nonumber \\
  & \Lambda_{\rm gr,rec} + \Lambda_{\rm neb} + \Lambda_{\rm g\textnormal{-}d} ) + S \times \Lambda_{\rm CIE},
\end{align}
where $\Lambda_{\rm hyd}$ is the total cooling rate from hydrogen in atomic neutral, ionized, and molecular forms,
the term proportional to $(1-S)$ represents other 
coolants in cold and warm photo-ionized gas, and 
the term proportional to $S$ represents coolants in warm and hot collisionally ionized gas.  
The factor $S=S(T)$ appearing in \autoref{eq:cooling_fnct} is a sigmoid function that smoothly increases from 0 to 1 between $T_{\rm t,1}$  and $T_{\rm t,2}$: $S = 1/[1 + \exp (-a(T - 0.5 (T_{\rm t,1} + T_{\rm t,2}))/(T_{\rm t,2} - T_{\rm t,1}))]$, 
where we take $a = 10$. 
The hydrogen cooling,
\begin{align}
\Lambda_{\rm hyd} & = \Lambda_{\rm Ly\alpha} 
+ \Lambda_{\rm ci,H}  
+ \Lambda_{\rm ro\textnormal{-}vib,H_2} 
+ \Lambda_{\rm cd,H_2} 
+ \Lambda_{\rm ff} 
+ \Lambda_{\rm rr},
\end{align}
is calculated using non-equilibrium abundances across the full temperature range.  Below we describe each cooling term in detail.


\begin{figure}
  \epsscale{1.1}\plotone{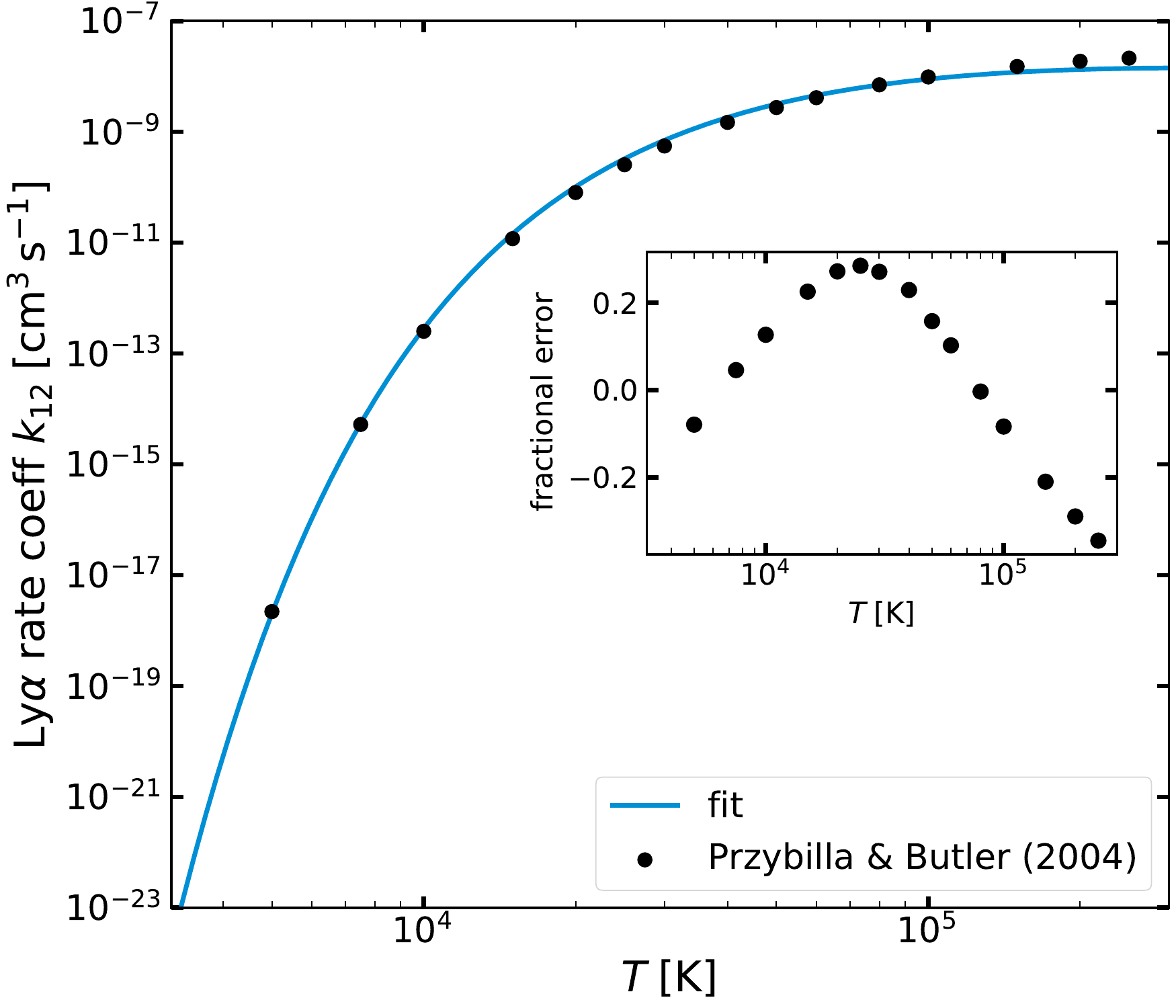}
  \caption{Electron-impact excitation rate for Ly$\alpha$ transition ($1 \rightarrow 2$) of atomic hydrogen as a function of temperature. The black circles indicate the quantum mechanical calculations from \citet{Przybilla04}, and the blue line is our fit to the data (see \autoref{e:k_lya}). The inset shows the fractional error of the fit.}\label{f:Lyalpha}
\end{figure}

\subsubsection{Hydrogen}


%
The hydrogen line emission resulting from collisional excitation by free electrons is the main cooling channel for warm neutral gas with $8000 \Kel \lesssim T \lesssim$ a few $10^4 \Kel$. In particular, the Ly$\alpha$ resonance line (2p $\rightarrow$ 1s)
dominates the overall cooling. If we regard a neutral hydrogen atom as a simple two-level system, the specific cooling rate by Ly$\alpha$ can be written as $\nH\Lambda_{\rm Ly\alpha} = \xHI A_{21} E_{21} [r_{12}/(r_{21} + r_{12} + A_{21})]$, where $r_{21} = \nEL k_{21}$ and $r_{12} = \nEL k_{12} = \nEL (g_1/g_2)k_{21} \exp( -E_{21}/(k_{\rm B}T) )$. Here, $g_1=2$ and $g_2=6$ are the statistical weights of 1s and 2p, $A_{21} = 6.265 \times 10^8 \second^{-1}$ is the Einstein coefficient, $E_{21}/\kB = 1.184\times 10^5\Kel$, and $k_{21} = 8.63 \times 10^{-6} \Upsilon_{21} / (g_2 T^{1/2}) \cm^3\second^{-1}$ is the collisional de-excitation rate coefficient with the velocity-averaged collision strength $\Upsilon_{21}=\Upsilon_{12}$ \citep[e.g.,][]{Osterbrock06}. Because the critical density at which collisional de-excitation is equal to the radiative decay $A_{21}/k_{21} \sim 10^{16}\cm^{-3}$ is so large, the specific cooling rate can be simplified as $\nH\Lambda_{\rm Ly\alpha} = \xHI E_{21} \nEL k_{12}$.

\citet{Przybilla04} performed ab initio, quantum mechanical calculations of 
collisional strengths for selected electron temperatures from $5\times 10^3 \Kel$ to $2.5\times 10^5\Kel$. Using their data for the ground-to-first excited level transition ($\Upsilon_{12}$), we show the collisional rate coefficient $k_{12}$ as circles in \autoref{f:Lyalpha}. Our fit to the data is
\begin{equation}\label{e:k_lya}
  k_{\rm 12} = 5.31\times 10^{-8} \frac{T_4^{0.15}}{1 + (T_4/5)^{0.65}} e^{-11.84/T_4} \,{\rm cm}^3\,{\rm s}^{-1}\,,
\end{equation}
shown as a blue line.\footnote{The fit given by \citet{Gong17} is,
\begin{equation*}
  k_{\rm 12,G17} = 6.38\times 10^{-9} T_4^{1.17} \exp (-11.84/T_4),
\end{equation*}
which underestimates the Ly$\alpha$ cooling rate by a factor of $2$--$5$ for $T \lesssim 2\times 10^4 \Kel$.} The inset shows that the fractional error from our fit compared to the data is less than 25\% 
for the range of temperature where the Ly$\alpha$ cooling is most significant ($0.8 \lesssim T_4 \lesssim 5$).

For $0.8 \lesssim T_4 \lesssim 5$, our adopted Ly$\alpha$ cooling rate is slightly ($\lesssim 10\%$) higher than the fit of  \citet{Smith22} to the \HI\ cooling rate by collisional excitation, which allows for excitation into not just 1p (which results in Ly$\alpha$ emission) but also 2s (which results in two photon emission) and higher excited states.
This level of discrepancy is acceptable because (1) the quantum mechanical calculations of $\Upsilon$ themselves are uncertain at the $10$--$20\%$ level \citep{Dijkstra19} and (2) the fact that the Ly$\alpha$ emissivity is sensitive to gas temperature and species abundance ($\Lambda_{\rm Ly\alpha} \propto \xEL \xHI \exp(-11.84/T_4)$) suggests that the accuracy of hydrodynamic and photochemical evolution in simulations would be more important sources of uncertainty \citep{Smith22}.

We also include the cooling by collisional ionization/dissociation of H/\HH\ as 
\begin{align}
    \Lambda_{\rm ci,H} & = 13.6\eV \times \xHI \xEL k_{\rm ci,H\textnormal{-}e} \,, \label{e:Lambda_ci_H}\\
    \Lambda_{\rm cd,H_2} & = 4.48\eV \times \xHH \sum_{\rm c} x_{\rm c} k_{\rm cd,H_2\textnormal{-}c} \label{e:Lambda_cd_H2} \,,
\end{align}
assuming that each event takes away $13.6 \eV$/$4.48 \eV$  of energy \citep[e.g.,][]{Grassi14}.  Collision rates for these reactions are given in \autoref{s:HII}/\autoref{s:HH}.

In warm molecular gas, the cooling by rotation-vibrational lines of \HH\ ($\Lambda_{\rm ro\textnormal{-}vib,H_2}$) can be important and dominate over other atomic coolants such as \CII\ and O. We adopt the new analytic function for $\Lambda_{\rm ro\textnormal{-}vib,H_2}$ provided by \citet{Moseley21}, which is valid for a wide range of density, temperature, and molecular fractions.\footnote{\citet{Moseley21} assumed that level populations are excited by collisional processes only and the effect of UV pumping (photo-excitation) is ignored. This assumption would fail in regions where UV pumping plays an important role and should be taken as a caveat.}

In ionized gas, electrons deflecting off ions result in the emission of continuum photons; free electrons recombining with ions leads to the loss of gas kinetic energy. 
We include hydrogen free-free (bremsstrahlung, $\Lambda_{\rm ff}$) and (case B) radiative recombination ($\Lambda_{\rm rr}$) cooling following Eqs. 10.12, 27.14, and 27.23 in \citet{Draine11book}.

\subsubsection{Carbon and oxygen}\label{s:C-O_cooling}

For cooling in cold and warm neutral gas, the terms 
$\Lambda_{\rm C^+} + \Lambda_{\rm C} + \Lambda_{\rm O}$ 
represent fine structure lines of \CII\ (2-level system; \ion{C}{2} $158\micron$), \CI\ (3-level system; \ion{C}{1}$^{*}$ $610\micron$ and \ion{C}{1}$^{**}$ $370\micron$), and \OI\ (3-level system; \ion{O}{1}$^{*}$ $63\micron$ and \ion{O}{1}$^{**}$ $146\micron$). We adopt the rate coefficients given by \citet{Gong17}. For important transitions, the critical density at which radiative and collisional de-excitation rates are equal is as follows (see 
\citealt{Draine11book}):
for \ion{C}{2} $158\micron$, $n_{\rm e,crit} \approx 53 T_4^{1/2} \pcc$ and $n_{\rm H,crit} \approx 3\times 10^3 T_2^{-0.1281-0.0087\ln T_2}\pcc$. For \ion{O}{1}$^{*}$ $63\micron$, $n_{\rm H,crit} = 2.5\times 10^5\pcc$ and $4.9\times 10^4 \pcc$ at $T=10\Kel$ and $5000\Kel$, respectively; for \ion{C}{1}$^{*}$ $610\micron$, $n_{\rm H,crit} = 620\pcc$ and $150\pcc$ at $T=10\Kel$ and $5000\Kel$, respectively. Therefore, the cooling is proportional to $\nH^2$ at typical CNM and WNM densities.

The rotational lines of CO are the main coolant in cold molecular gas.
\footnote{As noted in \autoref{s:phases}, line emission from other molecules (such as H$_2$O and OH) also become important in dense ($\nH \gtrsim 10^4\pcc$) molecular gas \citep[e.g.,][]{Hollenbach79, Goldsmith78, Neufeld95, Goldsmith01}, and our model can underestimate cooling in dense molecular gas to some extent. Because of thermalization of level populations and depletion of molecules onto grain surfaces, however, the dominant cooling at higher densities switches to gas-grain collisional coupling, which is included in our model.} Because CO lines are optically thick in molecular gas, the escape probability method with large velocity gradient approximation (LVG) is often used to determine the cooling rate \citep{Neufeld93}. Following the approach in \citet{Gong17, Gong18}, for $\Lambda_{\rm CO}$ we use the tabulated data in \citet{Omukai10}, where the CO cooling rate is given as a function of the effective column $\widetilde{N}= n_{\rm CO} / |\nabla \bm{v}|$ and $T$. 
The magnitude of the local velocity gradient $|\nabla \bm{v}|$ is calculated as the mean of the absolute velocity gradient (based on centered difference) along the three Cartesian axes.
(based on centered difference) along the three Cartesian axes.

\subsubsection{Grain-assisted recombination}\label{s:gr_rec}

Small dust grains and PAHs can transfer electrons to colliding ions if the ionization potential exceeds the energy required to ionize the grain.
The associated recombination cooling ($\Lambda_{\rm gr,rec}$) can be significant in warm gas.
For $\Lambda_{\rm gr,rec}$, we include cooling by grain-assisted recombination, adopting the parameters presented in Table 3 of \citet{Weingartner01b} ($R_V = 3.1$, $b_{\rm C} = 4.0$, and ISRF).

\subsubsection{Gas-dust interaction}

The cooling by gas-dust interaction can be important in dense gas ($\nH \gtrsim 10^5 Z_{\rm d}^{\prime -1} \pcc$).
Here, gas and dust grains exchange energy through inelastic collisions, leading to gas cooling with a rate coefficient,
\begin{equation}\label{eq:gas-dust}
\Lambda_{\rm g\textnormal{-}d} = \alpha_{\rm g\textnormal{-}d} T^{1/2} (T - T_{\rm d}) \Zd \,,
\end{equation}
where $T_{\rm d}$ is the equilibrium dust temperature. The coupling coefficient $\alpha_{\rm g\textnormal{-}d}$ depends on the degree of inelasticity of collisions and chemical composition \citep[e.g.,][]{Burke83}. Following \citet{Krumholz14}, we take the value appropriate for molecular gas recommended by \citet{Goldsmith01}: $\alpha_{\rm g\textnormal{-}d} = 3.2\times 10^{-34} \erg \cm^3 \second^{-1} \Kel^{-3/2}$.

To calculate the equilibrium dust temperature, we consider the internal energy equation for dust grains (see also \citet{Krumholz14}, although the notation is slightly different from ours)
\begin{equation}\label{eq:dust_en}
\frac{d e_{\rm d}}{dt} = \nH\Gamma_{\rm d,UV} + \nH\Gamma_{\rm d,other} 
+ \nH^2 \Lambda_{\rm g\textnormal{-}d}
- \nH\Psi_{\rm d,IR} \,.
\end{equation}
where $e_{\rm d}$ is 
thermal energy
of dust per unit volume, and the first three terms on the right hand side represent specific heating per H nucleon by absorption of UV photons, absorption of photons of all other wavelengths (e.g., optical, dust-reprocessed IR, and CMB), and heating via gas-dust interaction (for $T > T_{\rm d}$), respectively.
The UV heating is the main heating source except for gas in strongly shielded regions, and we calculate this term directly as $\Gamma_{\rm d,UV} = \sum_k \sigma_{\rm d,k} c \mathcal{E}_k$, where the index $k$ runs over LyC, LW, and PE frequency bins. 

The cooling by (optically-thin) thermal IR radiation is $\Psi_{\rm d,IR} = \sigma_{\rm d,IR} ca T_{\rm d}^4$, where $a=7.57 \times 10^{-15}\erg \cm^{-3} \Kel^{-4}$ is the radiation density constant, and we take $\sigma_{\rm d,IR}(T_{\rm d}) = 2.0\times 10^{-25}\Zd \cm^{2}\,{\rm H}^{-1} (T_{\rm d}/10\Kel)^2$ for the Planck-averaged dust-cross section per H nucleon, which is valid up to $T_{\rm d} \lesssim 150 \Kel$ \citep[e.g.,][]{Semenov03}.
Although we do not model the radiation field in the optical and IR wavelengths self-consistently, these can be specified appropriately depending on the problem. For simulations of the local diffuse ISM, for example, the $\Gamma_{\rm d,OPT} = \sigma_{\rm d,OPT} c \mathcal{E}_{\rm OPT}$ with $\mathcal{E}_{\rm OPT} = \mathcal{E}_{\rm OPT,0} \exp( - \tau_{\rm eff,OPT} )$ and $\Gamma_{\rm d,CMB} = \sigma_{\rm d,IR} ca T_{\rm CMB}^4$, with $\mathcal{E}_{\rm OPT,0} \sim \mathcal{E}_{\rm FUV,0}$\footnote{For the local ISRF, $\mathcal{E}_{\rm OPT,0}$ is several times higher than $\mathcal{E}_{\rm FUV,0}$ (see Table 12.1 in \citealt{Draine11book} and \autoref{s:ISRF}).} and $T_{\rm CMB}=2.73\Kel$ and we may take 
$\tau_{\rm eff, OPT} \sim \tau_{\rm eff,PE}/3$.

Using a root finding algorithm, we use \autoref{eq:dust_en} to calculate the equilibrium dust temperature assuming that dust grains are in instantaneous thermal equilibrium ($d e_{\rm d,sp}/dt = 0$). This is a good approximation given the relatively small heat capacity of dust. The equilibrium  $T_{\rm d}$ is then used in \autoref{eq:gas-dust}.

\begin{figure*}[t!]
  \includegraphics[width=\linewidth]{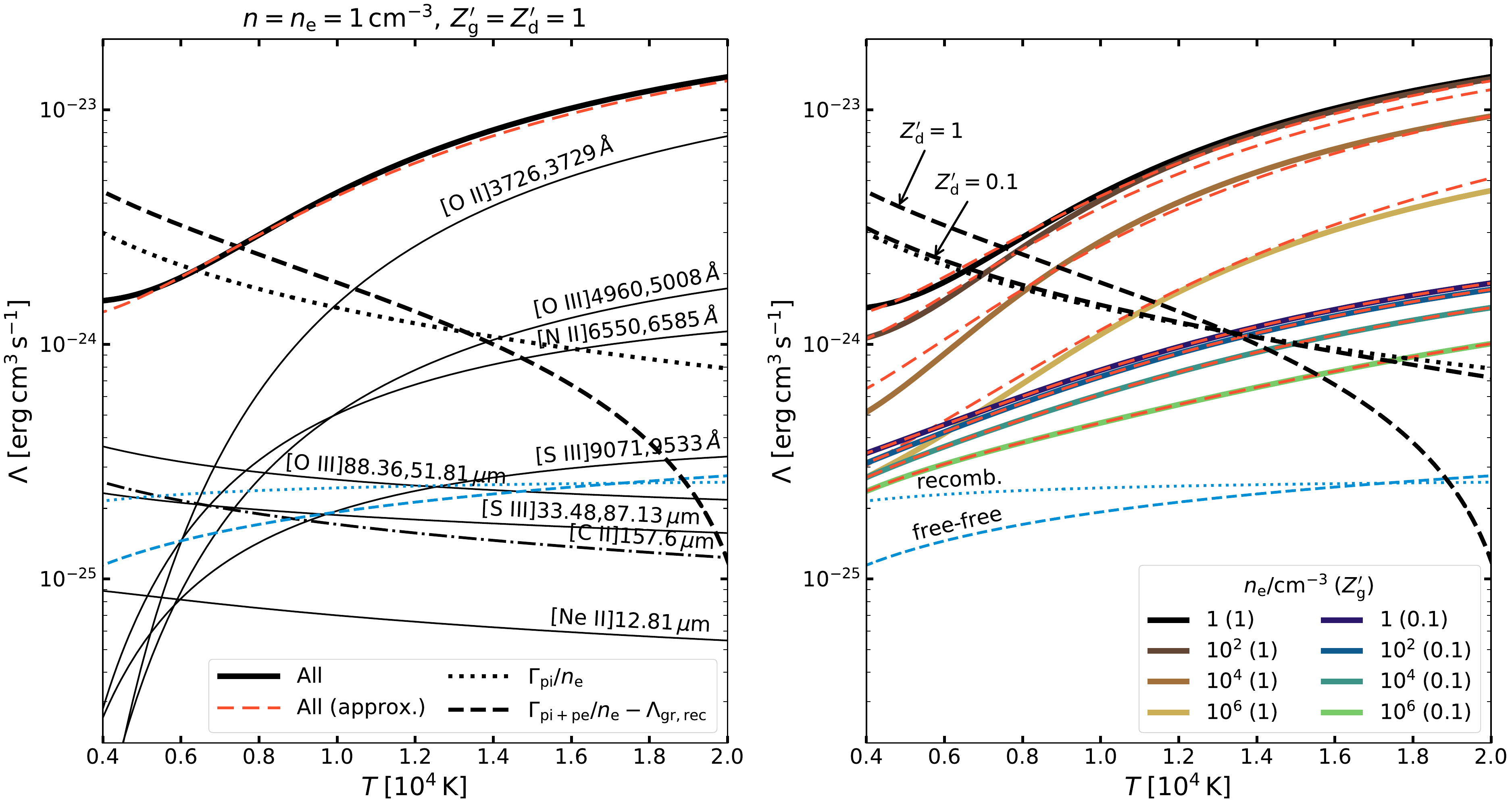}
  \caption{(Left) Cooling function in photoionized gas with $\nH = \nEL = 1\pcc$ for the solar neighborhood gas metallicity and dust abundance $\Zg = \Zd = 1$. The thick solid black line shows the total cooling function, which is the sum of cooling by collisionally excited lines of ionized metals
  (C, N, O, Ne, S),
  free-free emission (blue dashed), and hydrogen radiative recombination (blue dotted). We assume that 80\% (20\%) of O, N, and Ne is singly (doubly) ionized and 50\% (50\%) of S is singly (doubly) ionized. Thin black solid lines show notable nebular cooling lines from metal ions that contribute to $\Lambda_{\rm neb}$. (Right) The thick solid lines show the total cooling function at different gas densities ($\nH=\nEL=1,10^2,\cdots,10^6\pcc$) for $\Zg=1$ and $0.1$. Also shown in both panels are the \PI\ heating function $\Gamma_{\rm pi}/\nEL = 3.5\eV \times \alpha_{\rm rr,H^+}$ (assuming PI-recombination equilibrium; black dotted) and the total effective heating function, which is the sum of \PI\ and grain \PE\ heating, minus the cooling by grain-assisted recombination of H$^+$, assuming normalized FUV radiation intensity of $\chi_{\rm FUV} = 10^2$ (black dashed). The equilibrium temperature of photoionized gas is where the total cooling and the total effective heating functions meet. The red dashed lines show our adopted approximation for the total cooling, using $\Lambda_{\rm neb}$ (\autoref{e:Lambda_neb}), $\Lambda_{\rm rr}$, $\Lambda_{\rm ff}$, and $\Lambda_{\rm C^+}$.}\label{f:photoionized}
\end{figure*}

\subsubsection{Photoionized gas}\label{s:cooling_phot}

The main cooling channel in photoionized gas consists of collisionally excited forbidden ``nebular'' lines of metal ions such as O$^+$, O$^{++}$, N$^{+}$, S$^+$, S$^{++}$, and Ne$^+$. 
In low-metallicity gas ($\Zg \lesssim 0.1$), cooling from ${\rm H}^+$ via radiative recombination and bremsstrahlung (both included in $\Lambda_{\rm hyd}$) becomes more important. Important metal lines are from electronic excitations in the optical (e.g, $[{\rm O\,II}]3726, 3729\AA$, $[{\rm O\,III}]4960, 5008\AA$, $[{\rm N\,II}]6550, 6585\AA$) and fine-structure transitions in the mid- and far-infrared (e.g., $[{\rm O\,III}]51.81, 88.36\micron$, $[{\rm Ne\,II}]12.81\micron$).
The resulting cooling function depends on
detailed ionization balance of metal ions, which is determined by the local intensity and hardness of radiation. Although dedicated \PI\ codes \citep[e.g,][]{Ferland17} and recent radiation hydrodynamics codes \citep[e.g,][]{Bisbas15, Vandenbroucke18, Sarkar21a, Sarkar21b, Katz22} directly track photochemistry of multiple ion species, here we take a simplified approach.

We calculate the cooling function from metal lines using the atomic data available in the CMacIonize code by \citet{Vandenbroucke18} as functions of gas density ($\nH = \nEL = 1$,$10^2$,$\cdots$,$10^6\pcc$) and temperature ($2\times 10^3 \Kel \le T \le 5\times 10^4$). We assume a fixed ionization state such that 80\% (20\%) of O, N, and Ne are singly (doubly) ionized and 50\% (50\%) of S is singly (doubly) ionized.
These adopted ratios are broadly consistent with observational estimates in Galactic \HII\ regions \citep[e.g.,][]{Esteban05, Esteban18}. The cooling from carbon is assumed to come from ${\rm C}^+$ only, which we model explicitly. We then find an analytical fitting function $\Lambda_{\rm neb}(\nH,T)$ that approximates cooling from nebular lines
(except for those from ${\rm C}^+$):
\begin{equation}\label{e:Lambda_neb}
    \Lambda_{\rm neb} = \Zg \xEL \xHII \frac{3.68 \times 10^{-23}e^{-\frac{3.86}{T_4}} f_{\rm neb}(T)}{T_4^{1/2}(1 + 0.12 n_{\rm e,2}^{0.38 - 0.12 \ln T_4})}
\end{equation}
where $\Lambda_{\rm neb}$ is in $\cm^3\second^{-1}$, $T_4 = T/10^4 \Kel$, $n_{\rm e,2} = \nEL/(10^2 \pcc)$, and $\log_{10}\,f_{\rm neb}(T) = \sum_{i=0}^{6} a_i (\ln T_4)^i$ with $(a_0,a_1,\cdots,a_6) = (6.92 \cdot 10^{-1}, -5.86 \cdot 10^{-1}, 8.16 \cdot 10^{-1}, -5.05 \cdot 10^{-1}, 1.18 \cdot 10^{-1}, 7.66 \cdot 10^{-3}, -5.08 \cdot 10^{-3})$. The factor $\exp(-3.86/T_4)/T_4^{1/2}$ captures the temperature dependence of the collisional excitation rate coefficient, assuming the energy difference of the line $[{\rm O\,II}]3729\AA$; this is the dominant coolant in warm ionized gas except at very low temperature and metallicity (see \autoref{f:photoionized}). The decreasing cooling efficiency in dense gas due to collisional de-excitation is captured by the term in the parentheses in the denominator. We find that a simple functional form like $(1 + \nEL/n_{\rm e,crit})^{-1}$ does not work well since the nebular cooling consists of multiple lines with widely varying critical density $n_{\rm e,crit} \sim 10$--$10^6 \pcc$ (e.g., Table 3.15 in \citealt{Osterbrock06}).

The left panel of \autoref{f:photoionized} shows the cooling functions from important metal lines (thin solid black), hydrogen recombination lines (dotted blue), and free-free emission (dashed blue) for $\Zg = 1$. We assume $\nH = \nEL = 1\pcc$ so that the effect of collisional de-excitation is negligible. Also shown are the hydrogen \PI\ heating function under photoionization-recombination equilibrium ($\Gamma_{\rm pi}/\nEL = q_{\rm pi,H} \alpha_{\rm rr,H^+} $ with $q_{\rm pi,H} = 3.45 \eV$; dotted black), the net heating function accounting for \PE\ effect and grain recombination cooling ($(\Gamma_{\rm pi} + \Gamma_{\rm pe})/\nEL - \Lambda_{\rm gr,rec}$ with $\chi_{\rm FUV} = 10^2$ and $\Zd = 1$; dashed black), showing that small dust grains (and PAHs) can be a significant heating source in regions where both the ionizing and non-ionizing radiation is strong \citep{Weingartner01b} unless they are destroyed by harsh ionizing radiation \citep[e.g.,][]{Chastenet2019}.
The dominant coolant is $[{\rm O\,II}]3726, 3729\AA$, but several other coolants contribute to the total at equilibrium temperature where the total cooling equals heating, at $T \sim 7500\Kel$.

The equilibrium temperature of photoionized gas becomes higher with decreasing gas metallicity and with increasing gas density. This is demonstrated in the right panel of \autoref{f:photoionized}, which shows the total cooling for various $\nEL$ at two different metallicities $\Zg=1$ and $0.1$, as well as net heating function at $\Zd=1$ and $0.1$. In both panels, our approximation ($\Lambda_{\rm neb} + \Lambda_{\rm ff} + \Lambda_{\rm rr} + \Lambda_{\rm C^+}$; dashed red) is in very good agreement with the total cooling (solid thick black).

We caution that, although our fitting function for $\Lambda_{\rm neb}$ shows good agreement with the detailed calculation, it assumes a fixed ionization state, which while suitable for photoionized ISM gas is not intended to be applicable in all generality.


\subsubsection{Transition to CIE cooling in hot gas}\label{s:CIE}


In hot gas, energy loss occurs mainly via emission from collisionally excited lines 
of various species in a range of ionization states. Although we calculate the cooling from hydrogen (Ly$\alpha$, free-free, recombination) self-consistently across the full temperature range, we do not explicitly follow the ionization state of helium and metals. 
Instead, we use a tabulated CIE cooling function as described below.

Using Cloudy (Ver 10.00), \citet{Gnat12} calculated the ionization state of 30 elements (from H to Zn) as a function of temperature assuming CIE. The ion-by-ion cooling is tabulated as a function of temperature\footnote{\url{http://wise-obs.tau.ac.il/~orlyg/ion_by_ion}}, which we use to calculate the CIE cooling from helium and metals (assuming the \citet{Asplund09} element abundance ratios),
\begin{equation}\label{e:Lambda_CIE}
    \Lambda_{\rm CIE} = 
    \Lambda_{\rm He}(T) + \Zg \Lambda_{\rm metal, CIE}(T) \,.
\end{equation}
Note that in \autoref{eq:cooling_fnct}, $\Lambda_{\rm CIE}$ is multiplied by the sigmoid function $S(T)$.
The choice of the transition temperature $T_{\rm t,2}=3.5\times 10^4\Kel$ is where collisionally ionized metal ions that we do not directly model (such as silicon ions) begin to contribute to the total cooling non-negligibly (e.g., Figs. 2 and 3 in \citealt{Gnat12}).
We multiply the cooling rates from 
\OI, \CI, \CII\ and \CO\  
(see \autoref{s:C-O_cooling}) as well as nebular lines (see \autoref{s:cooling_phot}) by  $(1-S)$ in order to smoothly transition from self-consistent cooling based on chemistry to the tabulated CIE cooling.

\begin{figure*}[t!]
  \includegraphics[width=0.9\linewidth]{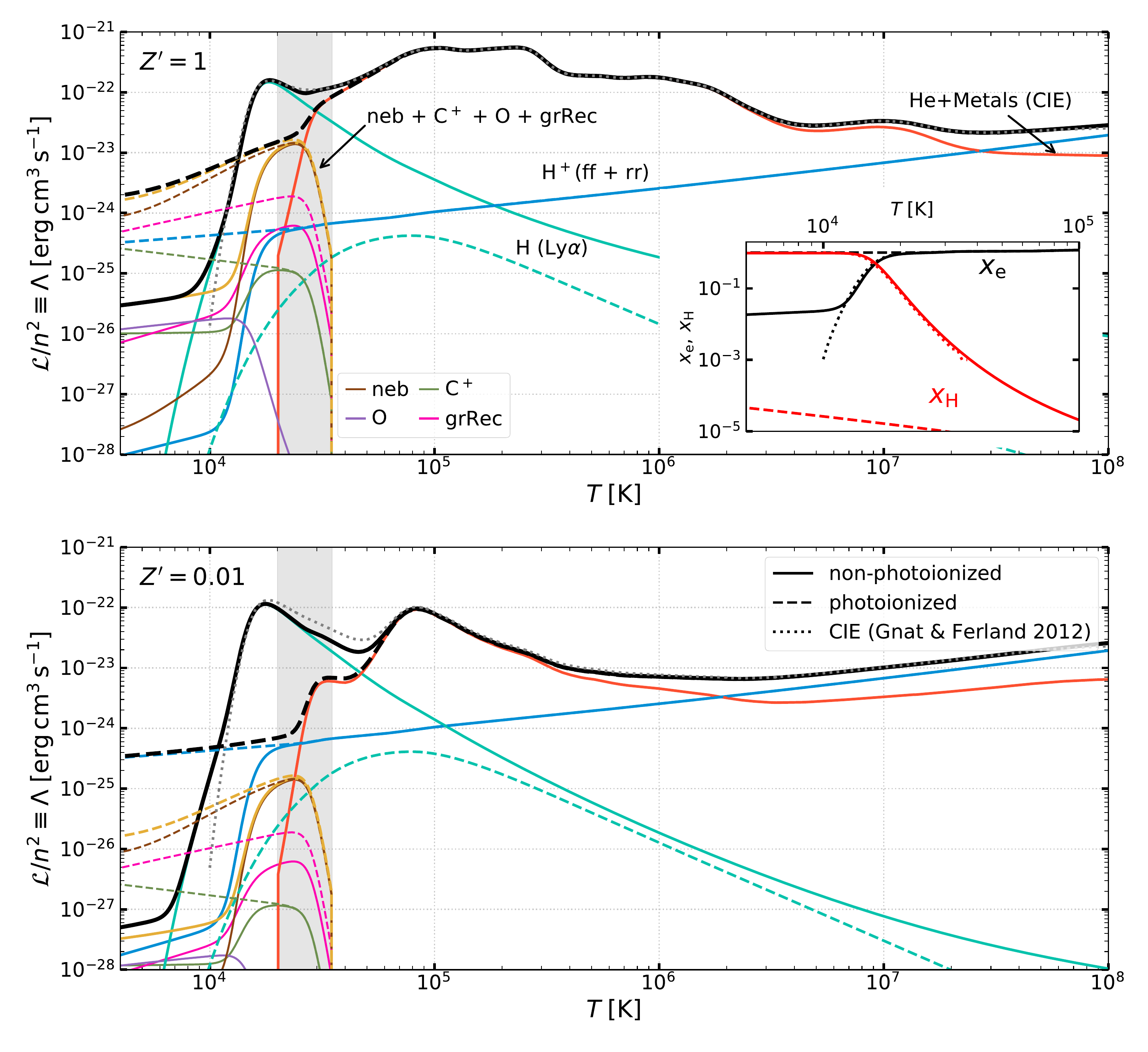}
  \caption{Cooling function in warm and hot gas with $\nH=1\pcc$ under ionization equilibrium for metal and dust scaled abundance $\Ztot = 1$ (top) and $\Ztot = 0.01$ (bottom). We consider two different physical conditions for warm gas: cooling functions for non-photoionized gas ($\xicr = 2\times 10^{-16}$, $\chi_{\rm FUV} = 1$, and $\zeta_{\rm pi} = 0$) are shown as solid lines, while those for photoionized gas ($\xicr = 2\times 10^{-16}$, $\chi_{\rm FUV} = 10^2$, and $\zeta_{\rm pi} = 10^{-8} \second^{-1}$) are shown as dashed lines. In both cases, we adopt the CIE cooling function of \citet{Gnat12} for helium and metals at $T > 3.5 \times 10^4\Kel$ (see \autoref{e:Lambda_CIE}). The black lines show the total cooling. The cooling function contributions for hydrogen (Ly$\alpha$, radiative recombination, and free-free) and other important coolants are shown separately as colored lines. The grey shaded area indicates the temperature range $2 \times 10^4 \Kel < T < 3.5\times 10^{4} \Kel$, in which we make a smooth transition from self-consistent, chemistry-based cooling to the CIE cooling. The dotted lines show the total (H+He+metals) CIE cooling function of \citet{Gnat12} assuming the \citet{Asplund09} element abundance ratios. The inset in the bottom panel shows the equilibrium electron fraction (black) and neutral hydrogen fraction (red) for $Z=1$.}\label{f:transition}
\end{figure*}

\autoref{f:transition} shows the cooling function in warm and hot gas ($4\times 10^3\Kel < T < 10^8\Kel$) for $\Ztot = 1$ (top) and $\Ztot = 0.01$ (bottom). For warm gas with $T < T_{\rm t,2}$, we consider two different physical conditions for calculating the ionization equilibrium. First, we use the condition for the diffuse, non-photoionized ISM in the solar neighborhood with $\chi_\mathrm{FUV} = 1$, $\xicr = 2\times 10^{-16}\second^{-1}$, and $\zeta_{\rm pi,H} = 0$ (solid lines). Second, we use the standard CR ionization rate, but strong FUV and ionizing radiation with $\chi_\mathrm{FUV} = 10^2$ and $\zeta_{\rm pi,H} = 10^{-8}\second^{-1}$, corresponding to gas in the vicinity of a massive star\footnote{An optically-thin radiation field $d=5\pc$ away from a massive star with ionizing photon rate $Q=10^{49}\second^{-1}$ would correspond to $\zeta_{\rm pi,H} = \sigma_{\rm pi,H} \times Q/(4\pi d^2) \approx 10^{-8}\second^{-1}$.} (dashed lines). This would be relevant for cooling in \HII\ regions and diffuse ionized gas. Additionally, we also show results for
cooling under CIE conditions as calculated by \citet{Gnat12}, without any CR ionization, PI, and grain-assisted recombination (dotted lines). In the first two cases, we calculate the cooling function at a given temperature using equilibrium abundances of hydrogen, carbon, and free electrons assuming $\xHH = 0$. We first calculate the equilibrium solution of \autoref{e:HII} with $\xHI = 1- \xHII$ (quadratic equation for $\xHII$), 
accounting for
the dependence of the grain-assisted recombination rate $\alpha_{\rm gr}$ on $\xEL$.
We iteratively solve for $\xHII$, $\xEL$, and $\xCII$ until a self-consistent converged solution is obtained (see Eq. 16.4 in \citealt{Draine11book} without $\zeta_{\rm pi,H}$). In all cases, we use a low gas density $\nH = 1\pcc$ so that the effect of collisional de-excitation is negligible. 

At $\Ztot = 1$, cooling in hot gas is dominated by CIE cooling of He and metals at 
$T \lesssim 3\times 10^7\Kel$, 
and by free-free and recombination at higher temperatures. The peak cooling rate occurs at $T \sim $ a few $10^5\Kel$ with $\Lambda_{\rm max} = 5.5\times 10^{-22} \erg \cm^3 \second^{-1}$, 
where highly ionized oxygen species (\ion{O}{4}, $
\cdots$, \ion{O}{7}) are the dominant coolants. In the absence of photoionizing radiation, the gas becomes neutral ($\xHII < 0.5$) below $T \approx 15300 \Kel$ (see top panel inset).
In warm, non-photoionized gas, the grain-assisted recombination rate is comparable to the radiative recombination rate ($\alpha_{\rm gr,H^+}\nH \sim \alpha_{\rm rr,H^+}\nEL$).
The cooling of warm neutral gas between $9000 \Kel \lesssim T \lesssim 10^4 \Kel$ is dominated by Ly$\alpha$ with minor contributions  from \CII\ and grain-assisted recombination. Since $x_H$ is much reduced below $0.1$ at $T \gtrsim 2\times 10^4\Kel$ by collisional ionization (solid red line in inset), the Ly$\alpha$ cooling drops. We note that the Ly$\alpha$ cooling is greatly reduced in warm photoionized gas (black dashed), where the abundance of neutral hydrogen is very low (red dashed line in the inset). In this case, cooling is dominated by metal lines (labeled by ``neb'' in the figure). In non-photoionized gas, the CIE cooling (gray dotted) tracks our cooling function from $T\sim 1-3.5 \times 10^4\Kel$, However, CR ionization enhances the electron abundance in warm gas with $T \lesssim 10^4\Kel$ relative to CIE (black solid vs. dotted in inset), which makes the overall cooling rate (black solid) much larger than that from CIE cooling (gray dotted) in this regime. 

For $\Ztot = 0.01$, cooling is dominated by hydrogen free-free and recombination above $2\times 10^6\Kel$. The cooling in hot gas with $T \sim 10^5$--$10^6 \Kel$ is significantly reduced due to the lower metallicity. Similarly, cooling by carbon and oxygen is much reduced, but Ly$\alpha$ cooling is still dominant between $T\sim 1$--$5 \times 10^4 \Kel$ in non-photoionized gas. The effect of grain-assisted recombination is negligible and abundances of \Ho\ and free electrons are quite similar to those in the $\Ztot=1$ (shown in the upper panel).


\begin{figure}[t!]
  \epsscale{1.1}\plotone{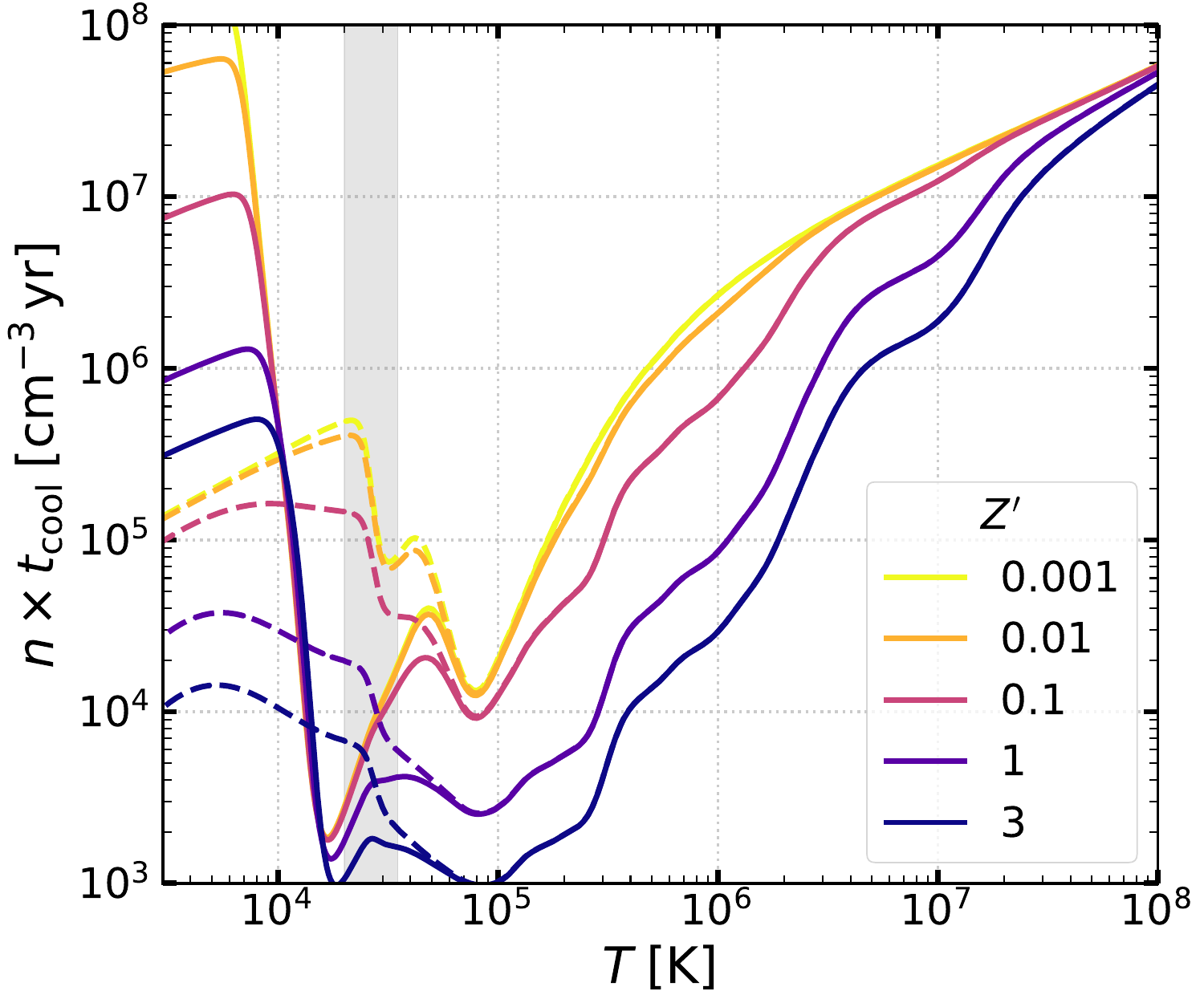}
  \caption{Characteristic cooling time $t_{\rm cool} \equiv P/(\nH^2 \Lambda)$ multiplied by the hydrogen number density $\nH$ as a function of temperature for different values of $\Ztot$. The dashed and solid lines show the cases with and without ionizing radiation, respectively (see \autoref{f:transition}). As in \autoref{f:transition}, the grey shaded area shows the temperature range in which we make transition to the CIE cooling.}\label{f:tcool}
\end{figure}


\autoref{f:tcool} shows the product of number density and cooling time $\nH t_\mathrm{cool}=P/(\nH\Lambda)$ for warm and hot gas at different metallicity $\Ztot$. We consider both the non-photoionized (solid) and photoionized (dashed) environments, similar to Figure \ref{f:transition}. The cooling time is as short as $(\nH/1\pcc)^{-1} (2$--$3)\times 10^3 \yr$ at $T \sim (2$--$10) \times 10^4\Kel$ for solar metallicity. In low-metallicity environments, the cooling time in warm photoionized gas (dashed) can be much longer than that in warm neutral gas (solid) due to the reduced cooling from Ly$\alpha$.

\section{Radiation and Cosmic Rays}\label{s:radcr}

In this section, we describe UV radiation transfer using the adaptive ray tracing and six-ray methods. We use the former for the transfer of radiation produced by point sources in the computational domain and the latter to treat shielding of diffuse background radiation originating from outside the computational domain. An example of an application in which this distinction is useful is the simulations of isolated star-forming clouds \citep[e.g.,][]{KimJG21}.
We also compute the attenuation of low-energy CRs based on effective gas column density calculated from ray tracing.

Given the radiation in the various bands as computed following the methods below, energy densities in each cell are used for computing ionization rates of H and C as respectively described in \autoref{s:HII} and \autoref{s:carbon}, and dissociation and ionization of \HH\ as described in \autoref{s:HH}; as well as photoheating rates as described in \autoref{s:heating}.

\subsection{Adaptive Ray Tracing}\label{s:ART}

The radiation field resulting from discrete point sources is calculated using the adaptive ray tracing method of \citet{Abel02}, as implemented in \Athena\ by \citet{KimJG17}. The reader is referred to \citet{KimJG17} for detailed description of the algorithm and test results. Here we give a brief summary of the method, with new additions made for calculating the \HH\ dissociating radiation field.

At the beginning of the ray tracing step, we inject $12 \times 4^4$ photon packets at the location of each source, corresponding to HEALPix level 4 \citep{Gorski05}. Walking along a ray, for each cell we calculate the length of the line segment $\Delta s$ between entering and exiting cell faces 
and the corresponding optical depth $\Delta \tau_j = \alpha_j \Delta s$ for each frequency bin $j$, where $\alpha_j$ is the product of cross section and density in the cell. The volume-averaged radiation energy density and flux are calculated as
\begin{equation}\label{e:Erad}
  \mathcal{E}_j = \dfrac{1}{\alpha_j \Delta V} \sum_{\rm rays}\dfrac{\Delta L_{{\rm ray},j}}{c}\,,
\end{equation}
\begin{equation}\label{e:Frad}
  \bm{F}_j = \dfrac{1}{\alpha_j \Delta V} \sum_{\rm rays} \Delta L_{{\rm ray},j}\bm{\hat{n}}\,,
\end{equation}
where $\Delta V$ is the cell volume, $\bm{\hat{n}}$ is the directional unit vector of the ray determined by the HEALPix algorithm,
and $\Delta L_{{\rm ray},j} = L_{{\rm ray,in},j} (1 - e^{-\Delta \tau_j})$ is the absorption rate of radiation energy. Here, $L_{{\rm ray,in},j} = L_{{\rm src},j} e^{-\tau_j}/N_{\rm ray}(\ell)$ is the luminosity of the incoming photon packet with the source luminosity $L_{\rm src,j}$, $N_{\rm ray}(\ell) = 12 \times 4^{\ell}$ is the number of rays in the HealPix level $\ell$, and $\tau_{j}$ is the optical depth measured from the source to the incoming face of the current cell. We compute $J_j=c\mathcal{E}_j/(4\pi)$ and normalized radiation field $\chi_j $ from \autoref{eq:chi_defn_PE}, \autoref{eq:chi_defn_LW} for PE and LW bands. 
Note that with this definition, $\chi_{\rm LW}$ is the scaled LW radiation field accounting for dust attenuation but not self-shielding by \HH.

For evaluation of the \HH\ photodissociation rate, we additionally compute the pseudo-radiation energy density in the LW band as
\begin{equation}\label{e:Erad_H2}
  \tilde{\mathcal{E}}_{\rm LW,H_2} = \dfrac{1}{\alpha_{\rm LW} \Delta V} \sum_{\rm rays} \dfrac{\Delta L_{{\rm ray,LW}}}{c} f_{\rm shld,H_2}(\NHH)\,,
\end{equation}%
where the computation of $L_{\rm ray,in,LW}$ includes attenuation by dust along the ray.
We adopt the \citet{Draine96} \HH\ shielding function, computed on each ray as:
\begin{equation}
\begin{split}
\label{e:fshld_H2_DB}
f_{\rm shld,H_2}(\NHH; b_5) = & \frac{0.965}{(1 + x/b_5)^2} \\ + 
& \frac{0.035}{(1 + x)^{1/2}}e^{-8.5\times 10^{-4}(1 + x)^{1/2}} \,,
\end{split}
\end{equation}
for $x = \NHH/(5\times 10^{14}\cm^{-2})$ with the \HH\ column density $\NHH$ integrated from the source to the center of the ray-cell intersection\footnote{There is an ambiguity about what fraction of the current cell's \HH\ column should be included in calculating $f_{\rm shld,H_2}$; the integration up to the cell face where a ray enters/exits the current cell underestimates/overestimates the shielding column. We find through tests that including more than 50\% of the current cell's column gives results in good agreement with the analytic solution.} and the Doppler broadening parameter $b = \sqrt{2} \sigma_{v} = b_5 \kms$ parametrizes the one-dimensional gas velocity dispersion $\sigma_v$. \footnote{\citet{Draine96} proposed Equation~\ref{e:fshld_H2_DB} as an approximate self-shielding function for cold ($T \lesssim 300 \Kel$) gas in local thermodynamic equilibrium \citep[see also][]{Sternberg14}. \citet{Wolcott-Green11} proposed a slightly modified form of \autoref{e:fshld_H2_DB} that is applicable to dust-free, warm ($300 \Kel \lesssim T \lesssim 10^4 \Kel$) gas in the early universe.}
In this work, we adopt a constant Doppler broadening parameter $b_5 = 3$; 
the shielding function is insensitive to the choice of $b_5$ for $\NHH \gtrsim 10^{17}\cm^{-2}$.
The first term in \autoref{e:fshld_H2_DB} captures Doppler line cores becoming optically thick for $\NHH \gtrsim 10^{14} \cm^{-2}$.  For intermediate columns $10^{17} \cm^{-2} \lesssim \NHH \lesssim 5\times 10^{20} \cm^{-2}$, $f_{\rm shld,H_2,eff} \propto x^{-1/2}$ because of absorption in the Lorentzian wings; line overlapping becomes important for $N_{\rm H_2} \gtrsim 5\times 10^{20} \cm^2$ \citep[see][]{Sternberg14}.

We define the dimensionless radiation intensity for \HH-dissociation as
\begin{equation}\label{e:chi_H2}
\chi_{\rm H_2} = \frac{\tilde{J}_{\rm LW,H_2}}{J_{\rm LW}^{\rm Draine}} \equiv \chi_{\rm LW} f_{\rm shld,H_2,eff}
\end{equation}
where $\tilde{J}_{\rm LW,H_2} = \tfrac{c}{4\pi}\tilde{\mathcal{E}}_{\rm LW,H_2}$ and the effective self-shielding factor $f_{\rm shld,H_2,eff} = \tilde{\mathcal{E}}_{\rm LW,H_2}/\mathcal{E}_{\rm LW}$ is 
the self-shielding factor averaged over (dust-attenuated) LW photons from individual sources.
An analogous calculation is done for self-shielding of C with cross-shielding by \HH, using the shielding factor $f_{\rm shld,C,eff}(N_{\rm C},\NHH)$ of \citet{TH1985} (also described in \citealt{Gong17}). The scaled radiation field for C is then $\chi_{\rm C} = \chi_{\rm LW} f_{\rm shld,C,eff}$.
A similar calculation could be done for the radiation that dissociates CO, but here we instead apply a simple fit for the CO abundance based on \citet{Gong17} (see \autoref{s:carbon}).

To make the angular resolution adaptive to the local grid resolution, a ray is split into 4 sub-rays if the number of rays traversing a cell per source falls below a certain number (default value is 4). We stop following a ray and destroy the photon packet if it exits the computational boundary or the optical depth measured from the source exceeds a critical value (default value is 30) in all frequency bins.


\subsection{Six-ray Approximation}\label{s:sixray}

The six-ray approximation calculates radiation intensity directed along the principal axes of the Cartesian grid. Despite its low angular resolution, this method yields a reasonable estimate of non-local shielding effects and has been used to treat the shielding of diffuse background UV radiation originating from outside the computational domain \citep[e.g.,][]{Nelson97, Yoshida03, Glover07a, Glover07b, Safranek-Shrader17, Gong20}.

In a spirit similar to the adaptive ray tracing method, we follow rays launched from six boundary faces of the computational domain and calculate optical depths and \HH\ columns.
For a ray directed along the +$x$ axis in uniform cubic grids for example, the injected photon packet luminosity is $L_{\rm ray,j,0} = c\mathcal{E}_{j,0} (\Delta x)^2$, where $\mathcal{E}_{j,0}$ is the ``unattenuated'' radiation energy density at the domain boundary and $(\Delta x)^2$ is the cell area. For the isotropic Draine ISRF, $\mathcal{E}_{j,0} = \mathcal{E}_j^{\rm Draine}/6$. The volume-averaged radiation energy density of a cell is obtained from \autoref{e:Erad} with $\Delta s = \Delta x$, $\tau_j$ the optical depth integrated from the domain boundary, and the summation taken over the six rays intersecting the cell. Similarly, the \HH-dissociating radiation field is obtained from \autoref{e:Erad_H2} using the \HH\ column density. 
As mentioned above, the six-ray method is used in applications where ``external'' radiation is incident on a cloud, and in this case we consider only LW and PE frequency bins and ignore LyC radiation.


\subsection{CR attenuation}\label{s:cr_atten}

Low-energy CRs lose their energy mainly by ionization of neutral gas. Due to this  interaction with intervening gas, the CR rate is reduced in regions of high column density.  Theoretically, a power-law decrease of ionization rate with increasing column is expected if the cosmic ray spectrum is itself a power law, with a steeper index in the diffusive limit than in the free-streaming limit \citep{Padovani2018,Silsbee2019}. Unfortunately, however, the low-energy CR spectral index is not well constrained, and it is unclear the extent to which a diffusive vs. free-streaming limit is appropriate.
Empirically, the observed column density ratios of molecular ions in diffuse molecular clouds suggest that the CR ionization rate decreases with increasing total cloud extinction as roughly  $\xicr \propto A_V^{-1}$ for $A_V \ge 0.5$ \citep{Neufeld17}.  Since our model does not include CR transport, we mimic the empirical CR attenuation in dense clouds in an approximate way using the effective column density determined by the ray tracing methods. 
We first assume an ``unattenuated'' CR ionization rate $\xicrunatt$, 
typically scaling the solar-neighborhood empirical CR rate ($\xicrunatt = 2.0\times 10^{-16}\second^{-1}$) by the ratio of the local SFR over the past 40 Myr to the solar neighborhood SFR, since the production rate of CRs is determined by the core-collapse SN rate  (assuming that the contribution from type Ia SNe is small).

To obtain an effective column density $N_{\rm eff}$ to each cell, the ray tracing (either adaptive ray tracing or six-ray approximation) 
makes use of the radiation energy density in the ``unattenuated'' PE band $\mathcal{E}_{\rm PE,unatt}$, as calculated in the ray-tracing step. In the unattenuated PE band, the injected photon packets are the same as in the PE band but the opacity is negligibly small.
Given the attenuated and unattenuated PE radiation energy densities in each cell,
we compute the effective dust optical depth for PE radiation,
\begin{equation}\label{e:tau_eff}
\tau_{\rm PE,eff} \equiv -\ln (\mathcal{E}_{\rm PE}/\mathcal{E}_{\rm PE,unatt}). 
\end{equation}
The effective column density to the cell is then $N_{\rm H,eff} = \tau_{\rm PE,eff}/\sigma_{\rm d,PE}$ where $\sigma_{\rm d,PE} = 10^{-21} \Zd \cm^{-2}\,{\rm H}^{-1}$.

The local primary CR ionization rate per H nucleon is then calculated as
\begin{equation}\label{e:cratt}
  \xicr =
  \begin{cases}
    \xicrunatt & \mbox{if } N_{\rm eff} \le N_{0} \,; \\
\xicrunatt \left(\frac{N_{\rm eff}}{N_{\rm eff,0}}\right)^{-1} & 
\mbox{if } N_{\rm eff} > N_{0} \,,
\end{cases}
\end{equation}
where $N_{0} = 9.35\times 10^{20}\,\cm^{-2}$. 

We emphasize that given the uncertainties in CR transport and the underlying spectrum of low energy CRs, \autoref{e:cratt} should only be considered provisional. The attenuation relation might, for example, have an index closer to $-0.3$ if the low-energy CR spectral index is $-0.8$ and CRs  free-stream along non-tangled magnetic field lines \citep{Padovani2018}.

\section{Method of Numerical Updates}\label{s:updates}

We update the source terms in Equations~\eqref{e:momentum}--\eqref{e:scalar} explicitly using the first-order operator splitting method. We first integrate Equations~\eqref{e:cont}--\eqref{e:scalar} 
without the source terms (the hyperbolic part)
over a time interval $\Delta t_{\rm mhd}$ using a higher-order Godunov's method solution, where $\Delta t_{\rm mhd}$ is determined by the usual Courant-Friedrichs-Lewy (CFL) condition \citep{Stone08}. Among many possible combinations of integrators, Riemann solvers, and reconstruction orders available in \Athena, our standard choice for the tests shown in \autoref{s:tests} and other applications combines the van Leer second-order integrator \citep{Stone2009}, Roe's Riemann solver, and piecewise linear reconstruction. 
For robustness, we turn on first-order flux correction \citep{Lemaster09} and H-correction \citep{Sanders98} during the integration step, adopting the CFL number $0.3$.

With the updated density and species abundances, the adaptive ray tracing module (and/or six-ray approximation) calculates radiation fields and the effective column density $N_{\rm eff}$. The heating/cooling module then takes this information and evolves the internal energy and species abundances over $\Delta t_{\rm mhd}$. 
At the end of each time step, we check all gas cells and correct any cells with negative density and pressure using $3^3$ neighbor cell averaging.

\subsection{Heating/Cooling and Chemistry}

Because the heating/cooling functions and reaction rates may have associated time scales shorter than $\Delta t_{\rm mhd}$ for gas temperature and chemical abundance evolution, we subcycle the cooling and chemistry updates relative to the MHD and radiation updates, with density and radiation variables held constant. The coupled ODEs we solve to update the internal energy density $e$ and species abundances $x_s$ due to heating/cooling and chemistry are
\begin{equation}\label{e:ODE1}
  \frac{de}{dt} = K \dfrac{d T_{\mu}}{dt} = \mathcal{G} - \mathcal{L} \equiv \Lcool \,,
\end{equation}
\begin{equation}\label{e:ODE2}
  \frac{d x_s}{dt} = \mathcal{C}_s \,, \;\;\;\; \text{(s: ${\rm H}_2$ and ${\rm H}^+$)} \,,
\end{equation}
where $T_{\mu} \equiv T/\mu$, $\Lcool$ is the net heating rate defined from \autoref{eq:Gamma_def} and \autoref{eq:Lambda_def} (with individual terms discussed in \autoref{s:heating_cooling}), $\mathcal{C}_s$ are the net creation rate coefficients for hydrogen species (defined in \autoref{e:scalar} and explained in \autoref{s:HH} and \autoref{s:HII}) and $K = \nH \muH \kB/(\gamma-1)$ is taken as a constant. 

Let the superscript $n$ denote the quantities evaluated at the $n$-th substep. At the beginning of each substep of a cell, we evaluate the source terms (from quantities at the current time) and determine the time step size as 
\begin{equation}
  \Delta t_{\rm sub}^n = \min \left( \Delta t_{\rm mhd}, \frac{0.1e^n}{|\Lcool^n|}, \frac{0.1}{|\mathcal{C}_{\rm H_2}^n|}, \frac{0.1}{|\mathcal{C}_{\rm H^+}^n|} \right)
\end{equation}
where the factor 0.1 ensures that the maximum relative change in internal energy (maximum change in abundance) is limited below $10\%$ ($0.1$) per substep.

Although Equations~\eqref{e:ODE1} and \eqref{e:ODE2} can be advanced together using an iterative implicit scheme to achieve accuracy and stability, it would incur a significant computational cost. Following the practice introduced by \citet{Anninos97} and adopted in other previous work \citep[e.g.,][]{Rosdahl13, Bryan14, Smith17, Chan21, Katz22}, we instead sequentially update temperature and abundances ($\xHII$ and $\xHH$) explicitly.
We first evolve temperature from $T_{\mu}^n$ to $T_{\mu}^{n+1}$ assuming that $x_s^n$ (and hence $\mu^n$) and other variables remain unchanged. Using the first-order backward difference formula and Taylor-expanding the term $\Lcool(T_{\mu}^{n+1}, \cdots) = \Lcool^n + \partial \Lcool/\partial T_{\mu} \lvert^n ( T_{\mu}^{n+1} - T_{\mu}^{n} )$ on the right hand side of \autoref{e:ODE1}, the temperature is updated to
\begin{equation}
  T_{\mu}^{n+1} 
  = T_{\mu}^n \left( 1 + \dfrac{-\Delta t_{\rm sub}^n/t_{\rm cool}^n}{1 + \frac{\Delta t_{\rm sub}^n}{t_{\rm cool}^n} \frac{T_{\mu}^n}{\Lcool^n}\frac{\partial \Lcool}{\partial T_{\mu}}\big|^n} \right)
\end{equation}
where $t_{\rm cool}^n = -e^n/\mathcal{H}^n$ (see also \citealt{Rosdahl13}) and the derivative is computed numerically using a finite difference approximation. 

To update abundances of \HH\ and H$^+$, we substitute $( 1 - 2\xHH - \xHII )$ for $\xHI$ in Equations~\eqref{e:HH} and \eqref{e:HII}, and regard them as a system of linear ODEs for dependent variables $X = (x_{\rm H_2}, x_{\rm H^+})^\mathsf{T}$ with constant coefficients: $\dot{X} = A X + b$.
After evaluating the rate coefficients using the updated gas temperature $T^{n+1} = \mu^n T_\mu^{n+1}$, we use the backward Euler method to evolve $\xHH$ and $\xHII$ over $\Delta t_{\rm sub}^n$: $X^{n+1} = (1 - A^n \Delta t_{\rm sub}^n)^{-1} ( X^n + b^n \Delta t_{\rm sub}^n)$. Finally, the equilibrium abundances of O$^+$, C$^+$, CO, C, O, and e are computed following the procedure described in \autoref{s:CandO} and \autoref{s:electrons}.

The subcycling of cooling and photochemistry updates relative to the MHD update has been adopted by many previous studies \citep{Anninos97, Whalen04, Baczynski15, Krumholz07, KimJG17}. One important caveat to this approach is that a gas parcel is forced to cool isochorically during $\Delta t_{\rm mhd}$. A large change in internal energy decoupled from hydrodynamics is inappropriate for modeling when the distribution of ``intermediate'' thermal, chemical, and ionization states are of interest, such as production of emission by strongly radiative shocks.
To accurately model the coupling between gas dynamics and thermodynamics/chemistry in those situations, it would be necessary to limit the MHD timestep by the cooling/chemical timescale.

Although the usual practice has been to subcycle the radiation update together with the cooling and photochemistry updates, we do not do so to reduce the computational cost of ray tracing. In \autoref{s:HII-PH}, we demonstrate that performing ray tracing based on $\Delta t_{\rm mhd}$ is sufficient to capture the overall dynamical evolution of expanding \HII\ regions, although the expansion of R-type ionization fronts at an early stage for which $\Delta t_{\rm chem} \ll \Delta t_{\rm mhd}$ cannot be modeled accurately.


\subsection{Radiative Force}\label{s:radforce}

The source terms due to the radiative force can be updated either before or after the update of cooling and chemistry.
We simply add $\bm{f}_{\rm rad} \Delta t_{\rm mhd}$ to the gas momentum density and recalculate the kinetic energy per volume $\rho v^2/2$ with the updated momentum density.

We find that a small number of low-density cells located near a luminous radiation source experience large acceleration by the radiation pressure force. Although this does not affect simulation outcomes in any significant way, it leads to the creation of fast-moving hot gas and reduces the time step. 
To avoid this problem, we apply a simple model for grain destruction and turn off the  radiation pressure force on grains for cells with high temperature and/or high radiation pressure. This is physically motivated by the fact that the low-density cavity created by radiation pressure or a wind/supernova shock is devoid of dust grains because of radial drift relative to gas and/or destruction by sputtering (both thermal and non-thermal). In \autoref{s:dust_dest}, we provide details of our model implementation.

In this work, as we are focused on the effect of radiation on gas thermodynamics and chemistry, we do not directly follow the transfer of 
optical photons, but they can exert a non-negligible radiation pressure force on dust grains. To compensate for this, a boosting factor can be introduced for the radiation pressure force for the PE photons \citep[e.g.,][]{KimJG21}. For a single point source, the sum of radiation pressure force due to PE and optical bands is $f_{\rm PE} + f_{\rm OPT} = f_{\rm PE} \left[ 1 + e^{\tau_{\rm PE} - \tau_{\rm OPT}}(\kappa_{\rm OPT}L_{\rm OPT})/(\kappa_{\rm PE}L_{\rm PE}) \right]$. The ratio $L_{\rm OPT}/L_{\rm PE}$ is in the range $0.5$--$1.0$ for $t_{\rm age} < 20 \Myr$ based on the \SB\ model (see \autoref{f:sb99_Edotpdot}), and $\tau_{\rm OPT}/\tau_{\rm PE} = \kappa_{\rm OPT}/\kappa_{\rm PE} = 2.7$ (see \autoref{t:sb99}) with 
$\tau_{\rm PE}$ from \autoref{e:Erad}.
For the multiple-source case, in principle one can substitute in a time-averaged luminosity ratio and effective optical depth from \autoref{e:tau_eff}, but care should be taken in the use of the boosting factor since the above argument is strictly valid only for a single point source.

\section{Tests}\label{s:tests}

We demonstrate and validate our photochemistry and heating/cooling modules, as implemented in the \Athena\ code, using various test problems. These tests include characterizing gas in thermal equilibrium (we consider unshielded gas, gas with a local shielding approximation, 
and 1D PDR models),  atomic-to-molecular transitions (with and without CR ionization), expansion of a  supernova remnant (focusing on the effects of non-equilibrium cooling and metallicity), and expansion of \HII\ regions. In the tests presented here, we neglect temperature dependence of the \HH\ formation rate on grains in \autoref{e:H2form} and adopt $\kHHform = 3\times 10^{-17}\Zd \second^{-1}$.

\subsection{One-zone Models}\label{s:onezone}

To demonstrate the ability of our model to simulate a wide range of ISM conditions, we 
compute properties of gas in thermal balance for a range of ambient conditions. For a cell of given gas density, incident FUV radiation field, CR ionization rate, gas metallicity, and dust abundance, we 
evolve species abundances (including electron abundance) and temperature over a sufficiently long time ($\sim 10^2 \Myr$) until a steady state is established.  For all of these tests, the EUV radiation energy density $\mathcal{E}_{\rm LyC}$ is set to zero.

We consider both unshielded and shielded cases. In the former case (see \autoref{s:unshielded}), gas is subject to FUV radiation with no shielding and unattenuated CR ionization, pertinent to atomic phases of the ISM. In the latter case, we allow for both FUV radiation and the CR ionization rate to be attenuated by a shielding column.  In \autoref{s:shielded}, we take this column as depending on gas density, and in \autoref{s:plane_par} we fix the density and consider profiles over increasing column.

For notational convenience, we denote the unattenuated (free-space) FUV radiation intensity by $\chiunatt$, unattenuated CR ionization rate by $\xicrunatt$, and scaled abundances by $\Ztot =\Zd = \Zg$ (except in special cases where we fix $\Zg$ and $\Zd$ separately). For the standard solar neighborhood condition, we adopt the optically-thin \citet{Draine78} ISRF with $\chiunatt = 1$, the primary CR ionization rate $\xicrunatt = 2\times 10^{-16} \second^{-1}$, and $\Ztot = 1$. As in \citet{Wolfire95, Bialy19}, we assume that $\xicrunatt$ scales linearly with $\chiunatt$, which is a reasonable assumption for the diffuse ISM in which both FUV radiation and CRs result from recent star formation activity.

\subsubsection{Unshielded Gas}\label{s:unshielded}

\begin{figure*}[t!]
  \epsscale{1.0}\plotone{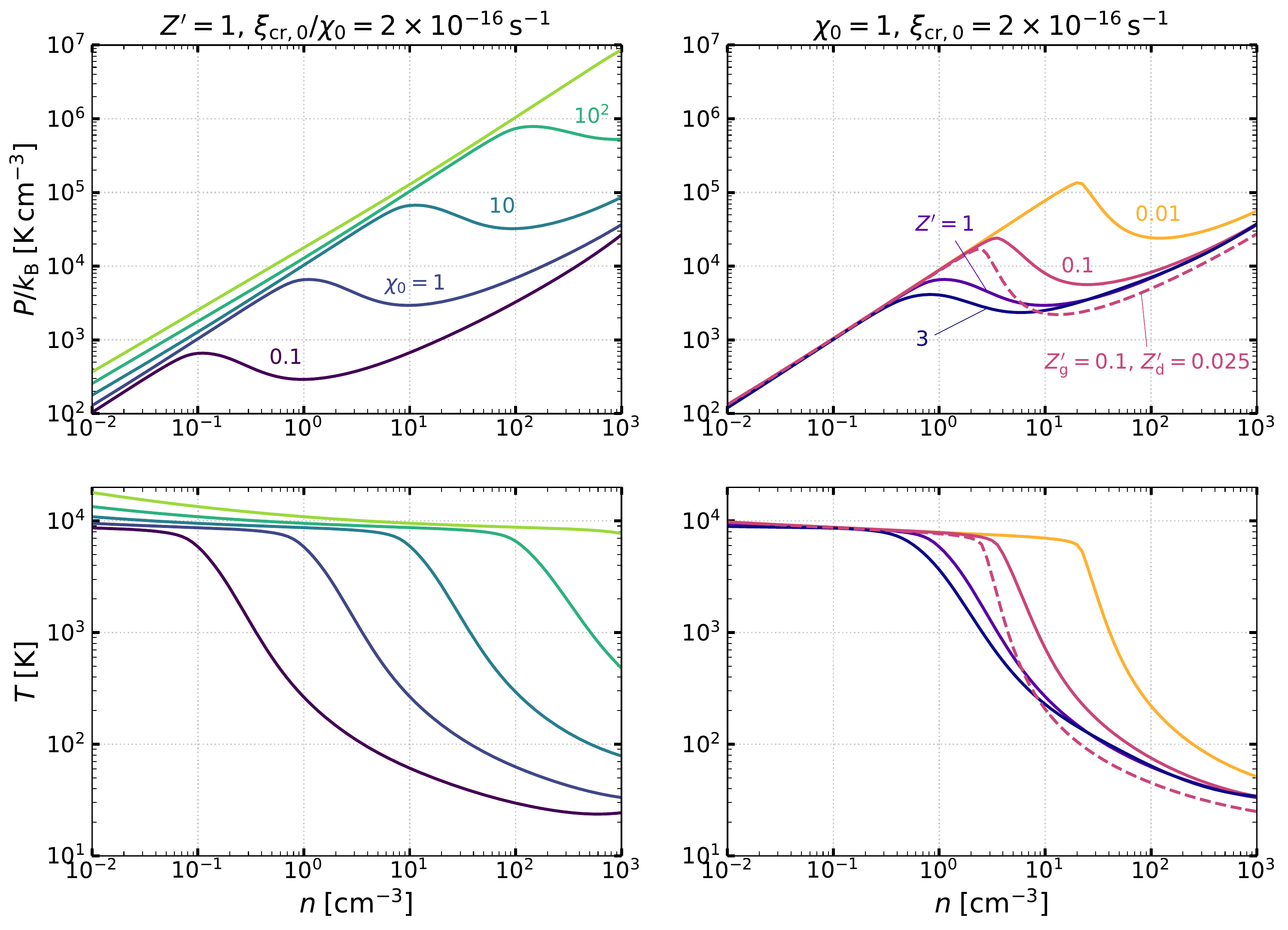}
  \caption{Equilibrium thermal pressure (top) and temperature (bottom) of unshielded gas as a function of hydrogen number density $\nH$. The gas is exposed to FUV radiation that scales with the local ISRF ($\chiunatt$, in units of the \citealt{Draine78} value), with the primary CR ionization rate also scaled as $\xicrunatt = 2\times 10^{-16} \second^{-1} \chiunatt$. The left column shows the case of the standard total elemental abundance ($\Ztot = 1$) with varying $\chiunatt$ and $\xicr$. The right column shows the effects varying $\Ztot$ for $\chiunatt=1$ and $\xicrunatt=2.0\times 10^{-16}\second^{-1}$. Additionally, the dashed lines in the right column show the case with $\Zg=0.1$ and $\Zd=0.025$.
  }\label{f:equil_unshld}
\end{figure*}

\begin{figure*}[t!]
\includegraphics[width=\linewidth]{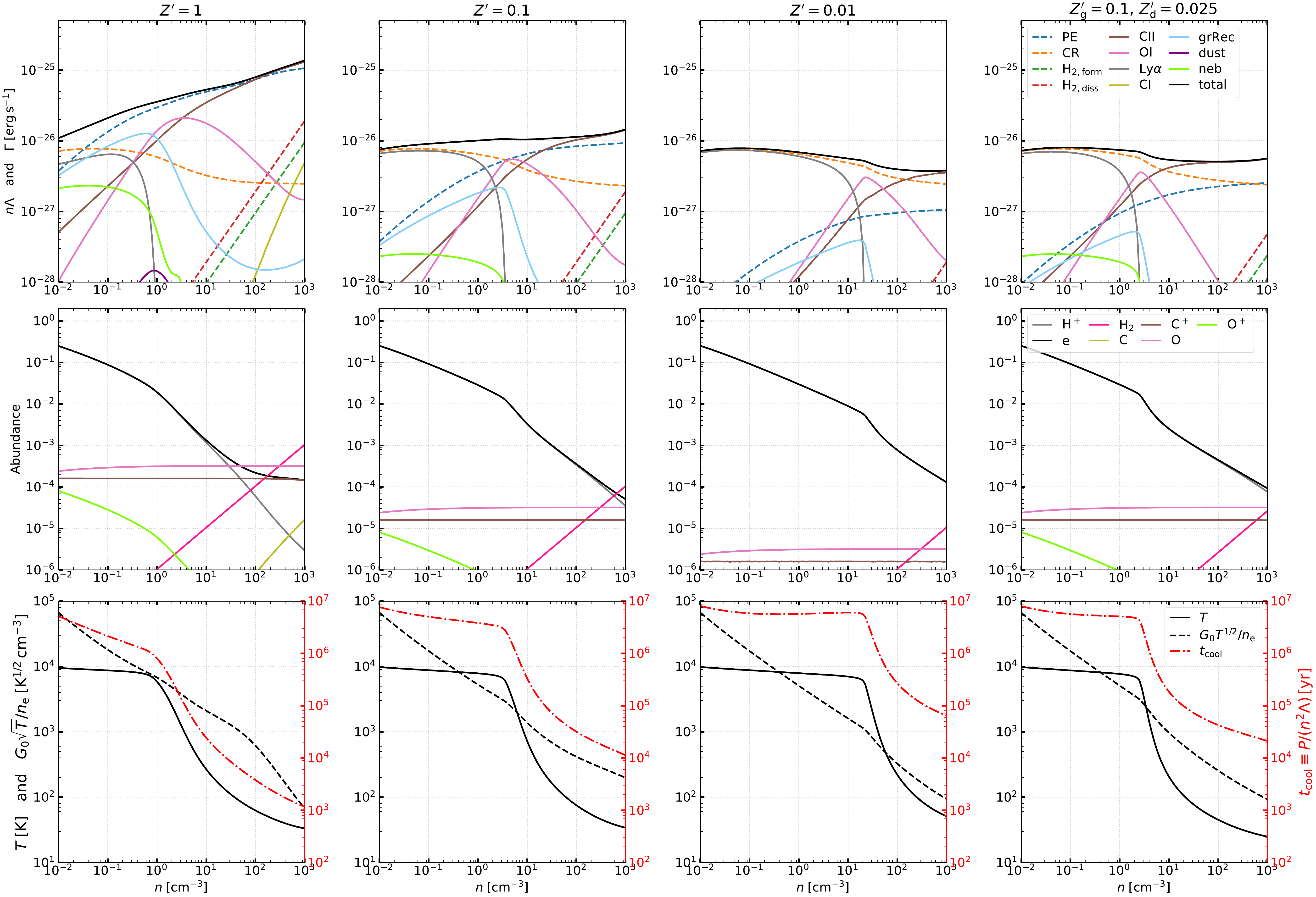}
\caption{Various quantities of gas in thermal/chemical equilibrium exposed to FUV radiation with $\chiunatt=1$ and  CR ionization rate $\xicr = 2\times 10^{-16}\second^{-1}$, considering a range of metal and dust abundance (see also the right column of \autoref{f:equil_unshld}). From left to right, the first three columns show cases with $\Zg=\Zd = 1$, $0.1$, and $0.01$. The fourth column shows the case with $\Zg=0.1$, $\Zd=0.025$.  The top row shows the cooling (solid) and heating (dashed) rates per H nucleon. The total is shown in black, and contributions from individual heating and cooling terms are also shown separately.  
The middle row shows abundances $x_s$ of species that we track, relative to H. The bottom row shows temperature (solid, left axis), grain charging parameter (dashed, left axis), and the characteristic cooling time (dot-dashed, right axis).}\label{f:rate_unshld}
\end{figure*}

\begin{figure*}[t!]
\includegraphics[width=\linewidth]{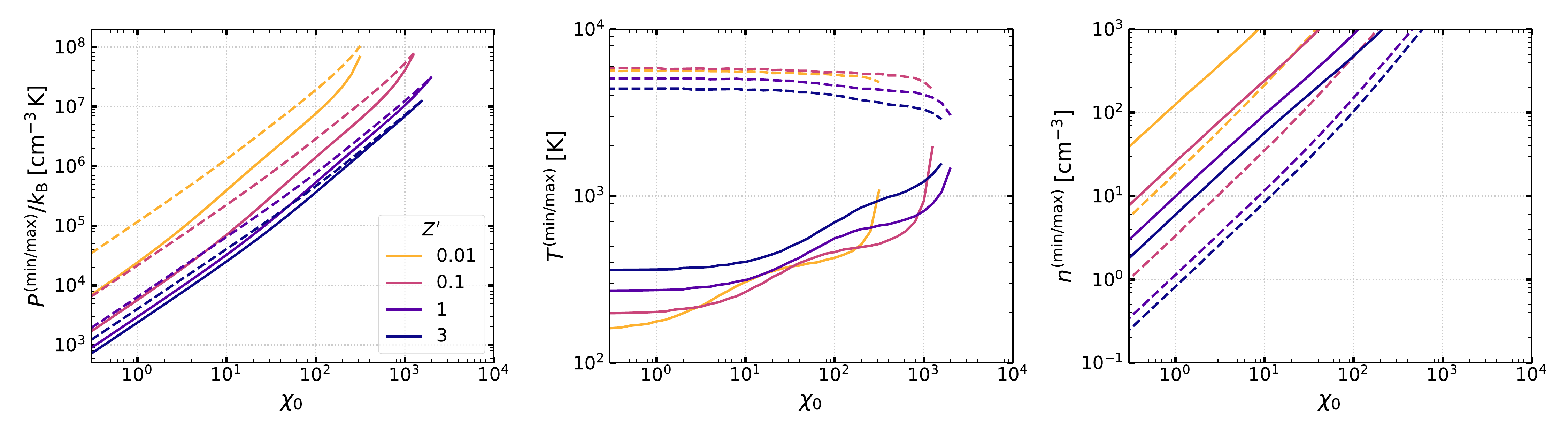}
\caption{{\sl Left}: the minimum/maximum equilibrium thermal pressure $P^{\rm (min/max)}$ for CNM/WNM (solid/dashed) as functions the scaled FUV intensity $\chiunatt$. The CR ionization rate is set to $\xicr = 2\times 10^{-16} \chiunatt \second^{-1}$. For $P^{\rm (min)} < P < P^{\rm (max)}$, two stable equilibrium solutions are possible so that CNM and WNM can coexist in pressure equilibrium. The color of lines indicates the metallicity $\Ztot = \Zg = \Zd$. {\sl Middle and right}: the corresponding temperature and density at $P^{\rm (min)}$ and $P^{\rm (max)}$. Note that $T>T^{\rm (max)}$ ($T<T^{\rm (min)}$) on the warm (cold) branch at $n<n^{\rm (max)}$ ($n>n^{\rm (min)}$). 
}\label{f:minmax}
\end{figure*}


We first explore how equilibrium thermal properties depend on the level of the radiation field $\chiunatt$ (also taking $\xicrunatt\propto \chiunatt$) and on metallicity $\Ztot$.
\autoref{f:equil_unshld} shows the equilibrium pressure (top) and temperature (bottom) as functions of density. The left column shows the solar metallicity case ($\Zg = \Zd = 1$) with varying FUV intensity $\chiunatt=0.1$, $1$, $\cdots$, $10^3$ and CR ionization rate $\xicrunatt = 2\times 10^{-16}\second^{-1} \chiunatt$, while the right column shows the cases with $\Ztot = 0.01$, $0.1$, $1$, $3$ for $\chiunatt=1$ and $\xicrunatt = 2\times 10^{-16}\second^{-1}$. The right column also includes a case with a larger reduction in dust abundance ($\Zd=0.025$) than metallicity ($\Zg=0.1$), as is likely to be applicable in the ISM of high-redshift galaxies (see below).

\autoref{f:rate_unshld} plots the specific heating and cooling rates (top), species abundances (middle), temperature, nominal cooling time $P/(\nH^2\Lambda)$, and grain charging parameter $G_0 T^{1/2}/\nEL$ (bottom). The leftmost column shows the standard case with $\chiunatt=1$, $\xicrunatt = 2\times 10^{-16}\second^{-1}$, and $\Ztot = 1$, while the other columns are for cases with different $\Zg$ and $\Zd$.

The phase diagrams in the $\nH$-$P$ plane show that three steady-state solutions are possible for $P^{\rm (min)} < P < P^{\rm (max)}$, for a wide range of heating rate and abundances.  Here, $P^{\rm (max)}$ is the maximum pressure for which a WNM phase (at low density) exists, while $P^{\rm (min)}$ is the minimum pressure for which a CNM phase (at high density) exists. The warm and cold branches are both thermally stable \citep{Field65}.
The intermediate density (or temperature) range $\nH^{\rm (max)} < \nH < \nH^{\rm (min)}$ where $d{\rm log}\,P / d{\rm log}\,\nH < 0$ 
represents the thermally unstable branch, 
with $\nH^{\rm (max/min)}$ being the maximum/minimum density of the WNM/CNM branch. For $\chiunatt=1$, we find $P^{\rm (min)}/\kB = 2.94 \times 10^3 \pcc\Kel$, $P^{\rm (max)}/\kB = 6.60 \times 10^3 \pcc\Kel$ with $\nH^{\rm (max)} = 1.1\pcc$, $\nH^{\rm (min)} = 10\pcc$ and $T^{\rm (max)} = 5260 \Kel$, $T^{\rm (min)} = 270 \Kel$. The CNM temperature becomes as low as $32\Kel$ at $\nH = 2.0\times 10^3 \pcc$.\footnote{We find that the equilibrium temperature at higher density increases as the heating by \HH\ formation and photodissociation rises rapidly (see also \citealt{Bialy19}). In reality, however, dense gas with $\nH \gtrsim 10^3\pcc$ is likely to be molecular due to shielding unless $\Ztot$ is extremely small (see \autoref{s:shielded}).} These results agree reasonably well with previous equilibrium calculations by \citet{Wolfire03, Bialy19} (see also \autoref{s:comparison}).


For $\chiunatt =1$ and $\Ztot = 1$, \autoref{f:rate_unshld} shows that the heating is dominated by the \PE\ effect (blue dashed), although the CR heating (orange dashed) contributes significantly at the lower end of the WNM branch. The grain charging parameter $\psi = G_0 \sqrt{T}/\nEL$ decreases at increasing density, and due to the negative grain charging, the \PE\ heating efficiency and therefore $\Gamma$ increases with density. 
In the CNM and thermally unstable gas, the main coolants are \ion{C}{2} and \ion{O}{1},
while cooling by electron recombination on PAHs and Ly$\alpha$ become dominant in the WNM.

The dominant ion species at low density is \Hplus, as produced by CR ionization (see \autoref{s:HII}). Therefore, the electron abundance in low-density gas is determined by the balance between CR ionization of \Ho\ and radiative recombination of \Hplus: $x_{\rm e} \approx x_{\rm H^+} \approx \left[ \xicrunatt/(\alpha_{\rm rr,H^+}\nH) \right]^{1/2}$.
$x_{\rm H^+}$ decreases more steeply than $\nH^{-1/2}$ in the unstable and cold branches since $\alpha_{\rm rr,H^+}$ increases with declining temperature and the grain-assisted recombination becomes more important. In high density gas, the dominant ion is C$^+$, produced in unshielded gas by \PI\ by FUV in the LW band (see \autoref{s:carbon}).

With increasing $\chiunatt$ (and $\xicrunatt$), the equilibrium curve shifts diagonally upward and to the right in the $\nH$-$P$ plane (see also \citealt{Wolfire95, Wolfire03, Bialy19}). 
The left and right panels of \autoref{f:minmax} show that 
$P^{\rm (min/max)}$ and $n^{\rm (min/max)}$ scale almost linearly with $\chiunatt$. This scaling arises since the density at a given equilibrium temperature is given by the balance between heating and cooling $n_{\rm eq} = \Gamma /\Lambda$, and the 
\PE\ heating rate increases approximately linearly with $\chiunatt$. The middle panel of \autoref{f:minmax} shows that $T^{\rm (max)}$, corresponding to the temperature of the
highest-pressure warm gas, is nearly 
constant up to very large $\chiunatt$. The temperature of the minimum-pressure cold gas, 
$T^{\rm (min)}$, is flat at low $\chiunatt$ but increases at high $\chiunatt$. Multiphase structure disappears at very high $\chiunatt$, with $T^{\rm (min)}$ increasing and $T^{\rm (max)}$ decreasing.


The right column of \autoref{f:equil_unshld} shows the phase diagrams with varying $\Ztot$. 
The equilibrium pressure and temperature are less sensitive to the variations in $\Ztot$ than in $\chiunatt$, since both the \PE\ heating and metal line cooling
are directly proportional to $\Zd$ and $\Zg$, respectively (see also \citealt{Wolfire95, Bialy19}). Nonetheless, there is a small diagonal shift of the equilibrium curves to higher pressure and density at lower $\Ztot$. 
This is because the CR heating remains relatively constant independent of $\Ztot$ as this heating is via hydrogen/helium ionization and depends only weakly on the chemical state. As a result, a higher equilibrium temperature is needed for cooling to balance heating. These trends are evident in the first three columns in the top row of \autoref{f:rate_unshld} (see also \autoref{f:minmax}).
For the same reason, increasing $\xicrunatt$ at given $\chiunatt$ and $\Ztot$ (not shown) would shift the equilibrium curve to higher pressure and density.

The tests discussed above assume linear scaling between the gas-phase abundance of heavy elements and dust abundance ($\Zd = \Zg$).  However, the degree of metal depletion onto dust grains may be low in metal-poor galaxies with $\Zg \lesssim 0.3$ \citep[e.g.,][]{Remy-Ruyer14, Feldmann15}, such that the dust abundance decreases more rapidly than metallicity. In particular, \citet{Bialy19} adopt a broken power-law relationship between $\Zd$ and $\Zg$, such that $\Zd = 0.025$ at
$\Zg =0.1$. We consider a model with these abundances, with results  shown as dashed lines in the right panels of \autoref{f:equil_unshld} and the rightmost column of \autoref{f:rate_unshld}. The primary effect of lowering the dust abundance relative to the gas metallicity is the reduced \PE\ heating and grain-assisted recombination cooling (compare the second and fourth columns of \autoref{f:equil_unshld}). The net effect is the slightly lower equilibrium pressure and temperature in the unstable and CNM branches.


\subsubsection{Shielded Gas}\label{s:shielded}

\begin{figure*}[t!]
  \epsscale{1.0}\plotone{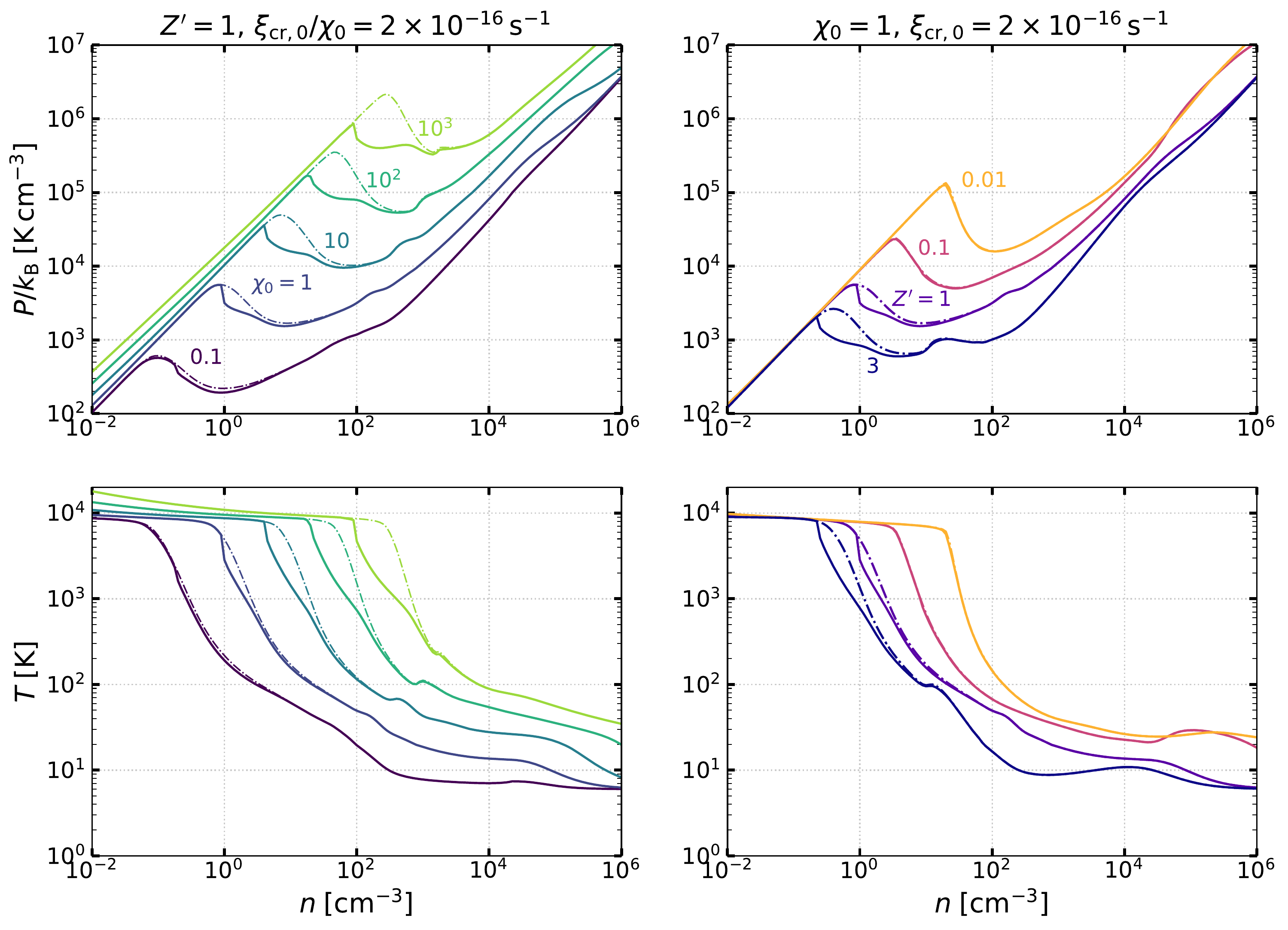}
  \caption{Same as \autoref{f:equil_unshld}, except that we adopt a density-dependent shielding treatment for the FUV intensity and CR ionization rate (see text for details). 
  The equilibrium curves in which \HH\ ro-vibrational cooling is turned off are shown as dot dashed lines.
  }\label{f:equil_shld}
\end{figure*}

\begin{figure*}[t!]
\begin{center}
  \includegraphics[width=1.0\linewidth]{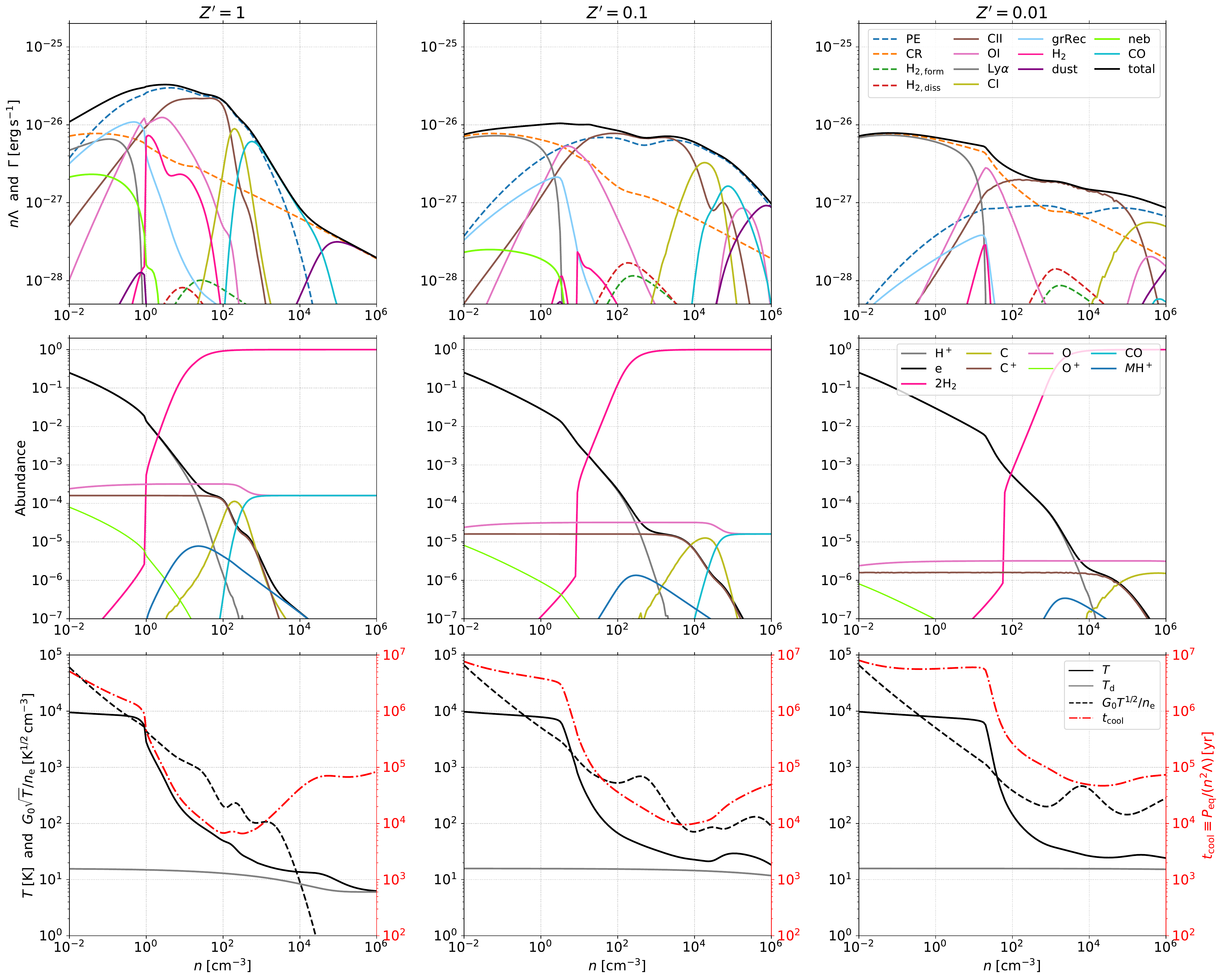}
  \caption{Same as \autoref{f:rate_unshld}, but for shielded gas (with \HH\ rovibrational cooling) as shown in \autoref{f:equil_shld} with $\chiunatt=1$ and $\xicr= 2 \times 10^{-16}\second^{-1}$.}\label{f:rate_shld}
\end{center}
\end{figure*}


Shielding of FUV radiation and CRs in dense gas is key to formation of cold molecular phases of the ISM, as it reduces the heating rate and promotes formation of molecules. The amount of shielding depends on complex distributions of matter and sources in the surroundings, so that while dense gas is often more shielded, there is no general relationship between density and shielding.

Nevertheless, we wish to provide a test that illustrates the general trends in the ISM's dominant heating/cooling agents, chemical abundances, and thermal states as gas tends to become increasingly shielded at high density. For this test, we assume a power-law relation between shielding length and density as $L_{\rm shld} = L_{\rm shld,0} (\nH/\nHinit)^{-a}$ and a shielding column $N_{\rm shld} = \nH L_{\rm shld}$. 
Based on the six-ray approximation, we have found that this functional form describes reasonably well the average FUV radiation field in turbulent, isolated molecular clouds of different radii and metallicities (see also \citealt{KimJG21}).\footnote{This functional form is similar to a shielding length scaling with the Jeans length. This and related approaches (using the density or velocity gradient for a length scale) have been adopted in some simulations to approximately account for the reduction of radiation at high density without the computational cost of radiative transfer \citep[e.g.,][]{Gnedin09, Safranek-Shrader17, Hopkins18, Ploeckinger20}.}
We adopt $\nHinit = 10^2\pcc$, $L_{\rm shld,0} = 5 \pc$, and $a=0.7$. Given the shielding length $L_{\rm shld}$, we calculate the local FUV intensity as $\chi_{\rm PE/LW} = \chiunatt \exp \left( -\sigma_{\rm d,PE/LW}N_{\rm shld} \right)$ and reduce the CR ionization rate according to \autoref{e:cratt} with $N_{\rm eff} = N_{\rm shld}$. For  \HH\ self-shielding, we use the column $N_{\rm shld,H_2} = 0.5 \nHH L_{\rm shld}$ based on the current \HH\ abundance, and similarly for the shielding of C.
We emphasize that the adopted shielding prescription is for demonstrative purposes and is not applicable to general cases.

For CO cooling, we adopt a constant velocity gradient of $|dv/dr| = 1\kms \pc^{-1}$. For the heating of dust grains, we assume that equal heating is provided by FUV and optical photons in the unattenuated limit ($L_{\rm shld} \rightarrow 0$) and that the optical depth in the optical band is smaller than the optical depth in the PE band by a factor 1.87. We also assume that grains are bathed in black body radiation of temperature $6 \Kel$, which serves to set the minimum grain temperature. The adopted blackbody radiation field reflects typical conditions in star-forming regions where there is not just the cosmic background, but also radiation from ambient dusty gas that is heated by embedded stars.
We evolve gas cells over a sufficiently long timescale so that temperature, species abundances, local radiation field, and \HH\ and C shielding factors all converge to steady-state values.
By default, we consider \HH\ ro-vibrational cooling, but we also consider the case in which it is turned off.

Similar to \autoref{f:equil_unshld} and \autoref{f:rate_unshld}, \autoref{f:equil_shld} and \autoref{f:rate_shld} show equilibrium quantities for different values of $\chiunatt$ and $\Ztot$.
As expected, shielding in dense gas reduces equilibrium temperatures and greatly increases  molecular abundances. For the standard case with $\chiunatt=1$ and $\Ztot = 1$, gas becomes molecular ($\xHH > 0.25$) for $\nH \gtrsim 20 \pcc$ and CO-dominated ($\xCO/\xCtot > 0.5$) for $\nH \gtrsim 350 \pcc$. The equilibrium temperature of molecular gas keeps decreasing with increasing $\nH$ and becomes as low as $T_{\rm d} \approx 6\Kel$ at $\nH \sim 10^6 \pcc$, where cooling by dust-gas interaction dominates while the CO cooling becomes subdominant because of thermalized level populations and radiative trapping. The CR heating takes over the \PE\ heating at 
$\nH \gtrsim 10^{3.7} \pcc$ where $N_{\rm shld} = 5 \times 10^{21}\cm^{-2}$. 
The top two rows of \autoref{f:rate_shld} show that as $\Ztot$ drops, the transitions of the dominant C-bearing species (C$^+$/C/CO) and its cooling occur at progressively higher density.

The solid/dot-dashed lines in the left panels of \autoref{f:equil_shld} show the equilibrium curves in which \HH\ rovibrational cooling is turned on/off. They show that, when \HH\ rovibrational cooling is included, even a small fraction (less than 1\%) of molecular hydrogen present in warm/unstable gas lowers the equilibrium temperature and tends to smooth out the two-phase structure, in qualitative agreement with findings by \citet{Bialy19}. However, the \HH\ abundance in reality would not necessarily follow the equilibrium solution presented here (although the general trend may be similar), considering the spatially varying radiation field, shielding, and the long \HH\ formation timescale \citep[e.g.,][]{Gong17, Gong18, Hu21, Chevance22}. Whether the \HH\ abundance in warm/unstable gas is high enough for the ro-vibrational cooling to significantly contribute to the total cooling would depend on the local radiation field and shielding factor, as well as chemical history of a gas parcel.

Finally, the equilibrium curves for gas with different $\Ztot$ show that the temperature of dense gas does not decrease as much at low metallicity, because of weaker dust shielding.

\subsection{1D PDR Models}\label{s:plane_par}

\begin{figure*}[t!]
  \epsscale{1.15}\plotone{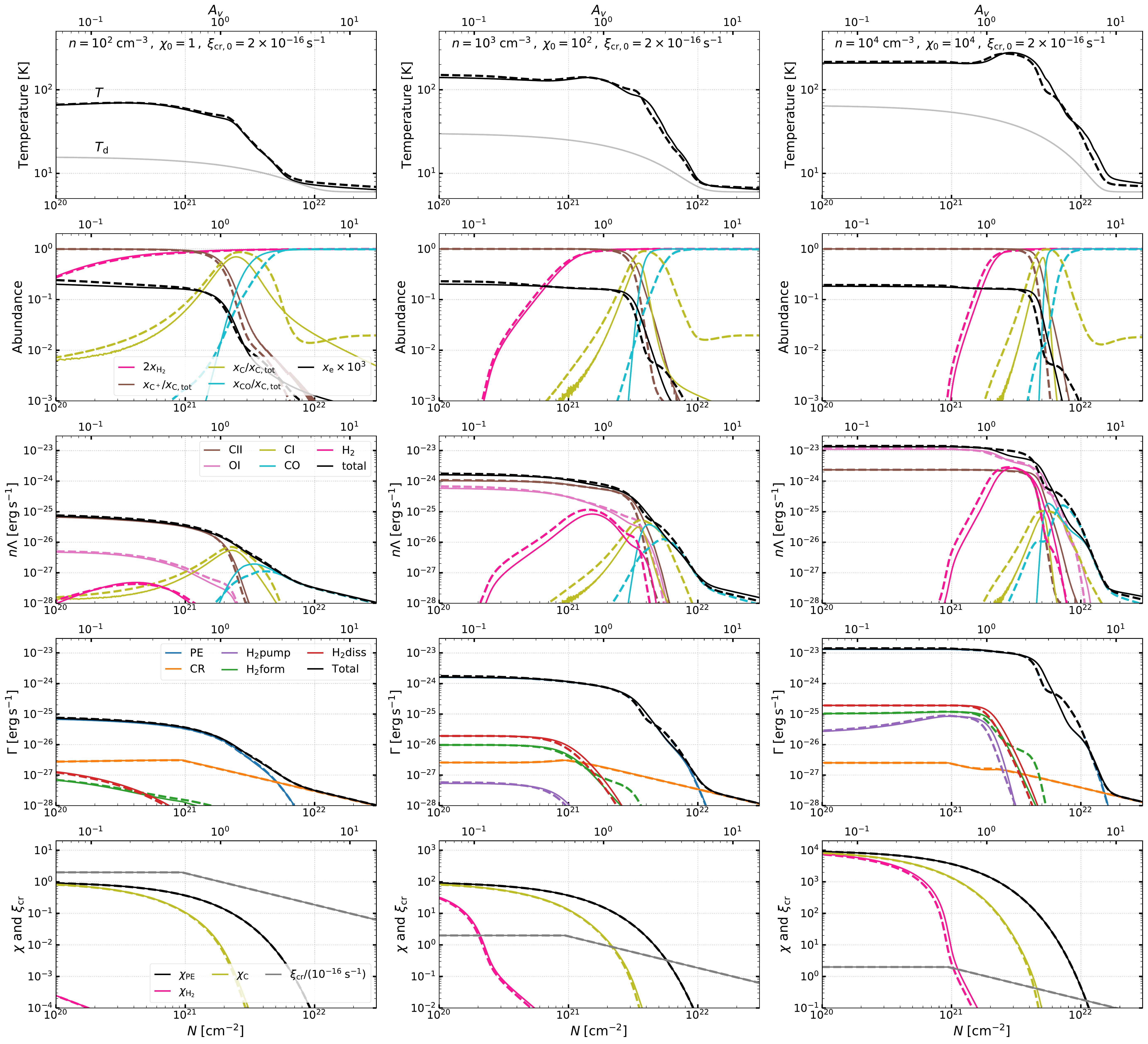}
  \caption{PDR test: a plane-parallel, uniform-density slab of gas in thermal and chemical equilibrium for $\Ztot = 1$. The slab is illuminated by a unidirectional FUV radiation field $\chiunatt$ with the unattenuated CR ionization rate $\xicrunatt = 2\times 10^{-16} \second^{-1}$. From top to bottom, panels show as functions of perpendicular hydrogen column density $\NH$ (or $A_V = \NH/[1.87 \times 10^{21}\cm^{-2}]$; top $x$-axis) the temperature, chemical abundances, cooling rate per H, heating rate per H, and normalized radiation field and CR ionization rate. The left column shows the case with density $\nH=10^2 \pcc$ and low FUV intensity ($\chiunatt = 1$), while the middle and right column shows cases with density $\nH = 10^3,\ 10^4 \pcc$ and high FUV intensity ($\chiunatt = 10^2,\ 10^4$). The results from full chemistry code calculations \citep{Gong17} are shown as dashed lines for comparison (except for dust temperature $T_{\rm d}$ which is set to a fixed value $T_{\rm d}=10\Kel$ in \citet{Gong17}.}
  \label{f:1DPDR}
\end{figure*}

In this test, we construct a standard 1D PDR model: a semi-infinite, uniform-density slab, illuminated from one side by a beam of FUV radiation. We consider three different conditions: $(\nH,\,\chiunatt) = (10^2\pcc,\,1)$, $(10^3\pcc,\,10^2)$, and $(10^4\pcc,\,10^4)$, where $\chiunatt$ denotes the normalized mean intensity at the cloud surface. All models have
$\Ztot=1$ and $\xicrunatt = 2\times 10^{-16}\second^{-1}$. We set up a 1D Cartesian domain consisting of $3\times 10^4$ cells with uniform resolution $10^{-2}$, $10^{-3}$, $10^{-4}) \pc$. The local FUV intensity and \HH\- and \CI-shielding factors are calculated similarly to \citet{Gong17}, assuming that the unidirectional (beamed) flux is incident normal to the cloud surface.
The CR rate is attenuated according to \autoref{e:cratt} using the perpendicular column for $N_{\rm eff}$.
\autoref{f:1DPDR} compares the steady state temperature, chemical abundances, heating and cooling rates, and radiation field strengths from our model (solid lines) to those from the full chemistry code calculations of \citet{Gong17} (dashed lines) as functions of the column density $\NH$ (or $A_V = \NH / 1.87\times 10^{21}\cm^{-2}$). The \citet{Gong17} chemistry includes
  9 additional species ($\mathrm{H_2^+}$, $\mathrm{H_3^+}$, $\mathrm{He}$, $\mathrm{He^+}$,
  $\mathrm{HCO^+}$, $\mathrm{Si}$, $\mathrm{Si^+}$, $\mathrm{CH_x}$, $\mathrm{OH_x}$) 
  and 50 total chemical reactions (including reactions for C$^+$, C, and CO formation and destruction). The chemistry is evolved until it reaches a steady state.

The transition of \HI-to-\HH\ is almost identical between our model and \citet{Gong17}, since we included a similar set of reactions for \HH\ formation and destruction. The decline in the \CII\ abundance toward increasing column is also generally accurate in our model, which includes the most important chemical pathways for \CII, especially at $A_V\lesssim 1$. At higher column, the \CO\ abundance rises, becoming the dominant carbon-bearing species above $A_V \sim 1$--$2$. 
Here, rather than explicitly following chemical reaction networks for molecules including \CO\ and $\mathrm{OH_x}$, the transition to \CO\ is obtained from a fit (see Equations~\eqref{e:CO} and \eqref{e:CO_fit}), which captures the transition region (within 30\%) and the resulting gas cooling.

Although the detailed abundances of \CII\ and \CO\ in the transition region are less accurate, this has little effect on the cooling.  The \CI\ abundance is obtained from the carbon atom conservation, and thus has a large error at high column, but its contribution to cooling there is negligible. 
Our model also reproduces the abundances of electrons, which is important for accurate calculation of the \PE\ heating efficiency.

The steady state temperature and the total heating and cooling rates agree remarkably well between our model and \citet{Gong17}. This is thanks to our careful choice of including all the important heating and cooling processes, as well as the related chemical species and reactions. Compared to the more complete set of chemistry calculations in \citet{Gong17}, our model has a much lower ($\sim 10$ times) computational cost, making it ideal to use for modeling accurate gas thermodynamics in MHD simulations of the ISM.

\subsection{Atomic-to-molecular Transition}\label{s:HIH2}

\begin{figure}[t!]
  \epsscale{1.1}\plotone{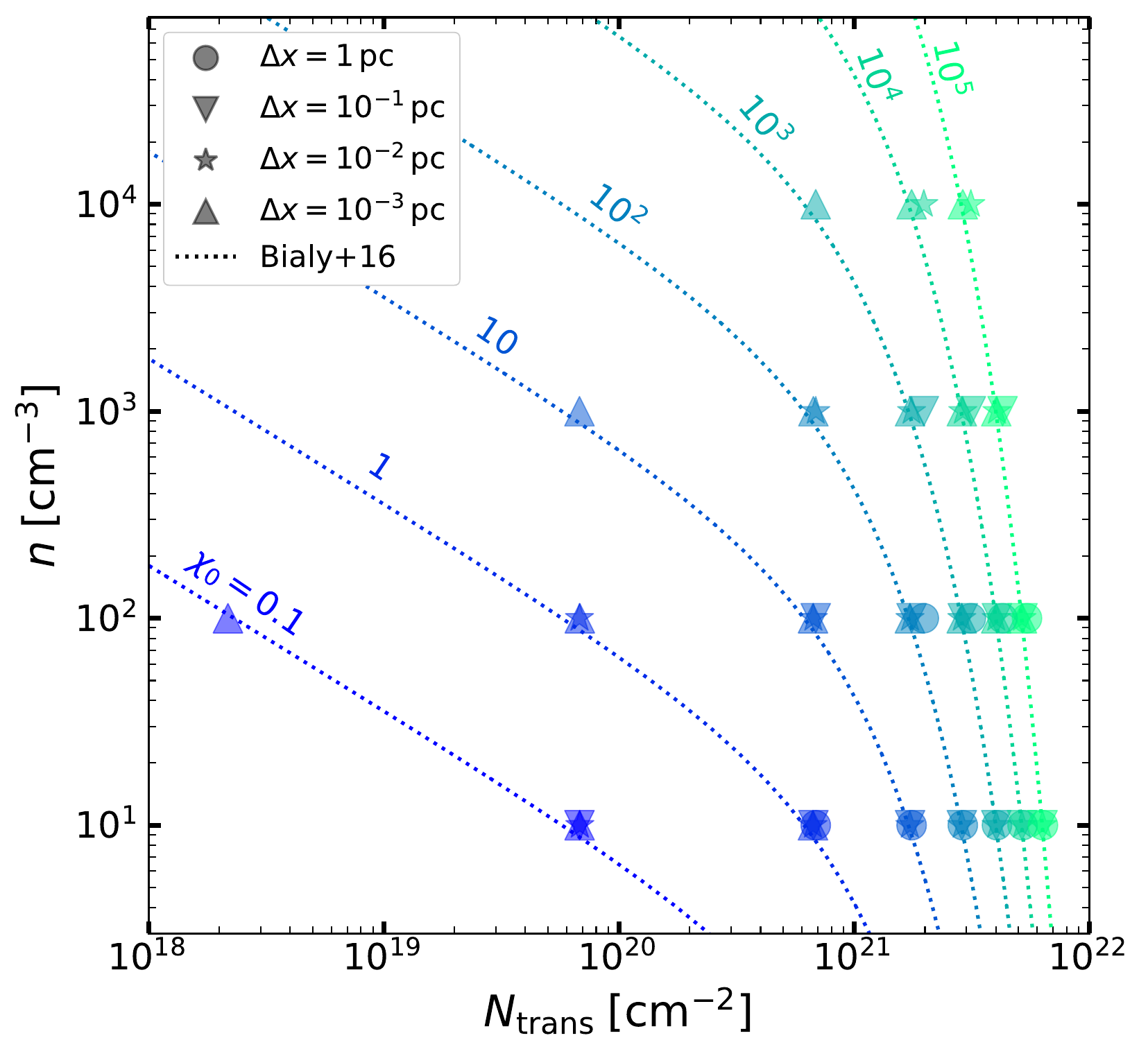}
  \caption{\HI-to-\HH\ transition in a plane-parallel gas slab without CR ionization. The symbols indicate the transition hydrogen column ($N_{\rm 1,trans}$) at which $\xHI=2\xHH$
  for various gas density ($\nH=10$--$10^4 \pcc$) and and FUV incident radiation field ($\chiunatt = 10^{-1}$--$10^{5}$). Results for different spatial resolution ($\Delta x = 10^{-3}$--$1\pc$) are indicated with different symbols, and  
  we only show cases in which the transition column is resolved by more than three cells. Our numerical results agree well with the analytic approximation by \citet{Bialy16} (dotted lines).
  }\label{f:HIH2-noCR}
\end{figure}


Observations of nearby galaxies show that the amount of recent star formation is closely correlated with molecular gas content as observed in CO emission \citep[e.g.,][]{Bigiel08}.
While the chemical state itself 
may not be critical to regulating star formation since temperatures in shielded atomic gas are similar to those in molecular gas \citep{Glover12a, Gong17}, understanding the physics controlling the transition between \HI\ and \HH\ is important in using the available observables to probe key aspects of the physical state, such as the density and radiation field strength.

We consider a 1D PDR model similar to those in Section \ref{s:plane_par}. We first consider the \HI-to-\HH\ transition in the absence of CR ionization. This case has been thoroughly studied by \citet{Sternberg14} and \citet{Bialy16} using full multi-line radiative transfer calculations as well as analytic approximations (see also references therein).
Their analytic formula for the total \HI\ column density produced by beamed FUV fields, and integrated from the cloud surface is
\begin{align}\label{e:N1tot}
  N_{\rm HI,tot} & = \frac{1}{\sigma_{\rm d}}\ln \left[ 1 + \frac{\alpha G}{2} \right] \nonumber \\
                 & = 5.3 \times 10^{20}\tilde{\sigma}^{-1} \ln \left[ 1 + \frac{\alpha G}{2} \right] \cm^{-2}\,,
\end{align}
where we take the dust cross section as $\sigma_{\rm d} = 1.9\times 10^{-21} \tilde{\sigma} \cm^2$.
The parameter 
\begin{align}\label{e:alpha}
\alpha 
 & = 3.8 \times 10^4 \chiunatt \left( \frac{3.0\times 10^{-17}\cm^3\second^{-1}}{\kHHform} \right) \left( \frac{10^2\pcc}{\nH} \right) \,,
\end{align}
is the free-space (unshielded) atomic-to-molecular density ratio\footnote{Note that, as in \autoref{s:plane_par}, we define $\chiunatt$ as the intensity of the FUV radiation field measured at the cloud surface relative to the isotropic Draine field; for a semi-infinite slab  $\chiunatt = 1/2$.}, and
\begin{equation}\label{e:G}
  G = 3.0 \times 10^{-5}\tilde{\sigma}\left( \frac{9.9}{1 + 8.9\tilde{\sigma}} \right)^{0.37} \,,
\end{equation}
is the \HH\ self-shielding factor averaged over the ``\HH-dust'' absorption layer.
The approximate analytic formula for the total hydrogen (\HI\ + \HH)
column density integrated up to the transition point (where $2\xHH = 0.5$) is given by
\begin{equation}\label{e:NH_trans}
  N_{\rm trans} = \frac{\beta}{\sigma_{\rm d}} \ln \left[ 1 + \left( \frac{\alpha G}{2} \right)^{1/\beta} \right] \,,
\end{equation}
for $0.1 \le \tilde{\sigma} \le 10$ with the index of power-law approximation for the shielding function
$\beta \approx 0.7$.
For a given dust opacity $\tilde{\sigma}$, $\alpha G \propto \chiunatt/n$.
In the strong-field limit ($\alpha G \gg 1$), the transition is sharp and most of \HI\ is built up before the transition ($N_{\rm trans} \approx N_{\rm HI,trans} \approx N_{\rm HI,tot} \approx \sigma_{\rm d}^{-1} \ln ( \alpha G )$). In the weak-field limit ($\alpha G \ll 1$), the transition is gradual and most of \HI\ is built up past the transition point.

To test our code's ability to reproduce atomic-to-molecular transition layers in plane-parallel geometry, we run a suite of 1D PDR models, with varying density ($\nH=10$--$10^5\pcc$), incident FUV radiation intensity ($\chi_0 = 0.1$--$10^5$), and spatial resolution ($\Delta x = 10^{-3}$--$1 \pc$). To make comparison with the analytic results easier, we adopt a fixed \HH\ formation rate coefficient $\kHHform = 3\times 10^{-17}\cm^3\second^{-1}$ and no CRs. 
As in \autoref{s:plane_par}, the simulations are run for a sufficiently long time until a steady state is reached.

The results are shown in Figure \ref{f:HIH2-noCR}. Our numerical results match the analytic approximation for $N_{\rm trans}$ by \citet{Bialy16} quite well.
We find that in order for the numerical results to converge, the transition column needs to be resolved by more than three cells, i.e. $3\Delta x < N_{\rm trans}$. Generally, higher spacial resolution (smaller $\Delta x$) is needed for higher density and lower FUV radiation field environments, where $\tilde{\mathcal{D}}$ and $N_{\rm trans}$ are small (Equations \ref{e:alpha} and \ref{e:NH_trans}). 
For the typical solar neighborhood value 
$\chiunatt = 1$ and $\nH=10^2\pcc$, for example, $\Delta x \lesssim 0.1\pc$ is needed to resolve the \HI-to-\HH\ transition.

\begin{figure*}[t!]
\includegraphics[width=\linewidth]{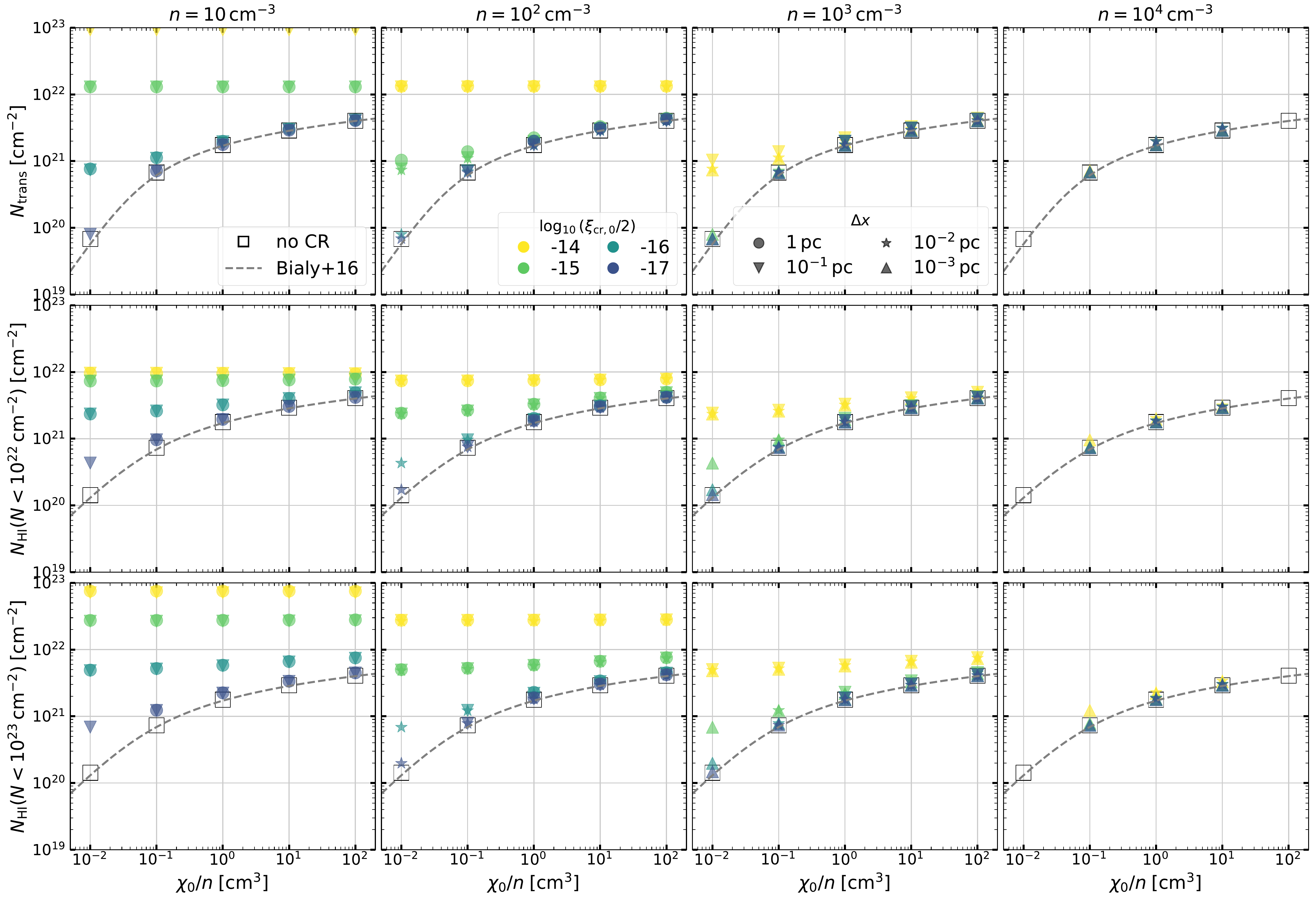}
\caption{\HI-to-\HH\ transition in a plane-parallel slab with CR ionization, as a function of the ratio $\chiunatt/\nH$. The CR ionization rate is attenuated following the prescription given by \autoref{e:cratt}. The top row shows the total hydrogen column $N_{\rm trans}$ at which the transition occurs ($2\xHH = 0.5$), and the middle bottom rows show the \HI\ column integrated up to the point where $\NH = 10^{22}$ and $10^{23} \cm^{-2}$, respectively. Each column shows results from a slab with different gas density. The symbols show the numerical results with (filled) and without (open) CR ionization. The symbol color indicates the strength of the unattenuated CR ionization rate $\xi_{\rm cr,0}$ and shape indicates spatial resolution.
    We only show cases in which $N_{\rm trans}$ is resolved by more than three cells. The gray dashed lines show the analytic approximations for $N_{\rm trans}$ and $N_{\rm HI,tot}$ by \citet{Bialy16} in the absence of CR ionization (Equations~\eqref{e:N1tot} and \eqref{e:NH_trans}). Including CR ionization can significantly increase the shielding required for \HH\ to form.}\label{f:HIH2-CR-Ntrans}
\end{figure*}

In the real ISM, the CR ionization is important in setting the \HH\ abundance \citep[e.g.,][]{Gong18}, so that the idealized results shown above may not hold.
The top row of \autoref{f:HIH2-CR-Ntrans} shows the \HI-to-\HH\ transition column in 1D PDR models now including CR ionization at a range of rates ($\xicrunatt/2 = 10^{-14}$--$10^{-18}\second^{-1}$). Because CRs destroy \HH\ molecules, the transition column density increases relative to the CR-free case.
From \autoref{e:HH} (without \PI\ and collisional dissociation terms), one
obtains a quadratic equation for
the equilibrium \HH\ fraction, which depends only on the ratios $\xicr / (\kHHform\nH)$ and $\zeta_{\rm H_2,pd} / (\kHHform \nH) \propto \chi_{\rm H_2} / (\kHHform \nH)$ (see also \citealt{Sternberg21}). The destruction rates in turn depend just on the column up to that point (providing dust shielding and \HH\ self-shielding).
One can also show from \autoref{e:HH} that for $\chi_{\rm H_2} \ll \xicr / (\nH k_{\rm gr,H_2}) \ll 1$, the equilibrium \HI\ fraction ($1- 2\xHH \ll 1$) is linearly proportional to $\xicr / (\nH k_{\rm gr,H_2})$. This implies that the \HI\ column diverges logarithmically as $\NH \rightarrow \infty$ if the CR ionization rate is attenuated as $\xicr \propto \NH^{-1}$. Of course, the total \HI\ column in reality is limited by the total hydrogen (\HI\ + \HH) column. In the middle and bottom rows of \autoref{f:HIH2-CR-Ntrans}, we show the \HI\ column densities up to the point where the total hydrogen column is $10^{22}$ and $10^{23}\,{\rm cm}^{-2}$, respectively.

In general, the increase in $N_{\rm trans}$ (and \HI\ columns) compared to the no-CR case is larger at higher $\xicrunatt/(\kHHform\nH)$ and at lower $\chiunatt/(\kHHform\nH)$.
Also, there is a larger increase
in the \HI\ columns than the transition column $N_{\rm trans}$.
\autoref{e:HH} and \autoref{eq:H2_cr} imply the ratio of the destruction
  rate to the formation rate $(1.65 \zeta_{\rm cr,H_2} + \zeta_{\rm
    pd,H_2})/(\kHHform \nH) \approx (2.2\xicr + \zeta_{\rm pd,H_2})/(\kHHform\nH) =
  2$ at the transition point. 
  Without any attenuation of CRs, the ratio of the CR ionization rate to the
  formation rate at the transition point would be $2.2 \xicrunatt/(\kHHform \nH) = 0.15 (\xicrunatt/2\times 10^{-16}\second^{-1}) (\nH/10^2\pcc)^{-1}$, which ranges between $1.5\times 10^{-3}$ and $1.5\times 10^2$ for the results shown in \autoref{f:HIH2-CR-Ntrans}. When $\xicrunatt/(\kHHform \nH) \gg 1$ (e.g., the yellow points in the leftmost panel of \autoref{f:HIH2-CR-Ntrans}), the transition column is mainly determined by (the attenuation of) CRs and the gas remains predominantly atomic up to the transition point.
For the typical solar neighborhood environments with $\xicrunatt = 2\times 10^{-16}\second^{-1}$, $\chiunatt \sim 1$, and $n=10^2\pcc$, $N_{\rm trans}$ is similar to the case without CRs. However, $N_{\rm HI}(\NH < 10^{22}\cm^{-2})$ and $N_{\rm HI}(\NH < 10^{23}\cm^{-2})$ are
several times larger than
$N_{\rm HI,tot}$
without CRs from \citet{Bialy16}. This shows that in many cases, including CR ionization is necessary for accurately modeling the \HI-to-\HH\ transition.

\subsection{Expansion of Radiative Supernova Remnants\label{s:SNR}}

\begin{figure*}[t!]
\includegraphics[width=\textwidth]{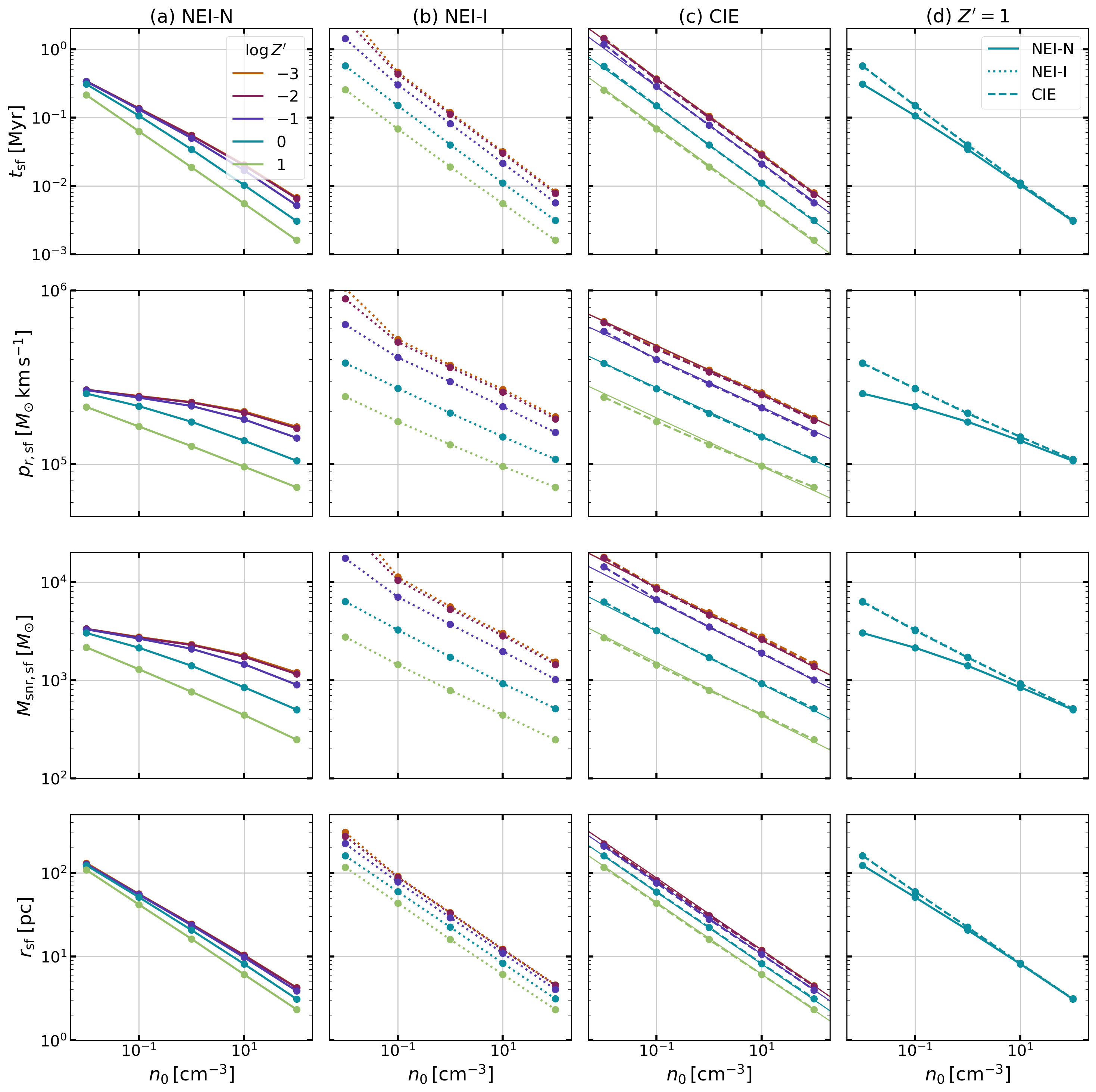}
\caption{SNR properties when cooling becomes important, defined by the time at which the total energy falls below 70\% of the injected energy (the ``shell formation'' time $t_{\rm sf}$).  From top to bottom, we show age, radial momentum, mass, and radius of the  SNRs. In columns from (a) to (c), results are from models using cooling from { (a)} non-equilibrium hydrogen ionization with a neutral background (NEI-N), {(b)} non-equilibrium hydrogen ionization  with an ionized background (NEI-I), and { (c)} collisional-ionization equilibrium (CIE). The thin lines in {\bf (c)} show  fits to the simulation data (\autoref{eq:snfit_tsf}--\autoref{eq:snfit_rsf}). In column { (d)}, we compare all  $\Ztot = 1$ models; note that the NEI-I (dotted) results are nearly identical to CIE (dashed). }\label{f:SN-summary}
\end{figure*}

The idealized evolution of an SNR after ejecta mixing and energy thermalization from interaction with ambient gas is described by a well-known self-similar solution \citep{Sedov59, Taylor50}. The radial momentum of the SNR $p_r = (2E_{\rm kin} M_{\rm snr})^{1/2}\approx 2 E_{\rm kin}/v_{\rm sh}$ grows during the Sedov-Taylor stage \citep{KimCG15} as  the blastwave sweeps up more and more material while conserving total energy, with $E_{\rm kin}=0.28 E_{\rm SN}$ and $v_{\rm sh}$ the leading shock velocity. The Sedov-Taylor stage ends with the onset of cooling in the shocked gas \citep{Cox72}. The SNR becomes radiative (and shell formation occurs) when the forward shock slows to $v_{\rm sh}\sim 200\kms$, so that the temperature of the shocked gas $T\lesssim10^{5-6}\Kel$ (see \autoref{f:transition})
and the cooling time becomes shorter than the expansion time \citep[][]{Draine11book}. 
After shell formation, the temperature of the interior remains high, and the SNR can acquire additional radial momentum during the so-called pressure-driven snowplow stage. However, in contrast to early analytic theory,
numerical studies have found that the momentum boost during the snowplow stage is not very large (a factor of 1.5--2 at most)  
because the interior pressure drops as hot gas condenses into the shell \citep{KimCG15,Walch15b,Iffrig15,Martizzi15}.

Because the shock velocity and post-shock temperature, rather than ambient conditions, determines the onset of cooling, momentum injection by SNe is quite insensitive to  environment \citep[see][for observational characterizations]{Koo20}. Instead, the details of the cooling function determine both the maximum hot gas mass created by SNRs, and the corresponding momentum injection. 
Given the importance of SNe in creating hot gas and injecting momentum in the ISM, 
quantitative evolutionary modeling of radiative SNRs  underpins the theory of ISM phase balance \citep{MO1977}, star formation regulation \citep{Ostriker11,KimCG13}, galactic outflow driving \citep{KOR17, Fielding18, KimCG20}, and galaxy formation and evolution \citep{Kimm14, Hopkins14, Naab17}.

Here, we quantitatively compare radiative stages of SNR for a range of metallicities and ambient densities, measuring the time $t_{\rm sf}$ and radius $r_{\rm sf}$ at the onset of shell formation, and the mass $M_{\rm snr,sf}$ and radial momentum $p_{\rm r,sf}$ that the SNR contains.
There are two major 
methodological extensions 
compared to our previous work \citep{KimCG15} that can affect the radiative SNR evolution. First, the cooling function at high $T$ is now metallicity dependent based on the tabulated CIE cooling function of \citet{Gnat12}. Second, fully 
time-dependent hydrogen species abundance sets cooling by hydrogen over the full temperature range. The effect of metallicity 
has been explored in several previous works, assuming that hydrogen is in CIE \citep{Thornton98, Karpov20, Oku22} or fully time-dependent chemistry \citep{Steinwandel20}. Because cooling by metal ions is generally dominant at $T\sim 10^{5}$--$10^{6}\Kel$, the reduction of metallicity results in delayed shell formation and correspondingly a larger radial momentum and hot gas mass at the shell formation. 
Further dependence on metallicity becomes 
negligible at $\Ztot \lesssim 10^{-2}$ as metal abundances are too low to provide significant cooling \citep{Thornton98, Karpov20}. Implementation of a more complete non-equilibrium ionization network and self-radiation from shocks is described in \citet{Sarkar21a}, and used in \citet{Sarkar21b} for spherically symmetric radiative SNR simulations in an uniform ambient medium with $\nHinit=1\pcc$ and $\Ztot=1$. 
In the future, it will be valuable to extend more sophisticated models of this kind to a range of densities and metallicities, and also to relax the assumption of spherical symmetry.

To investigate the effects of both metallicity and non-equilibrium hydrogen abundance on SNR expansion, 
we conduct a set of 3D HD simulations of SNR expansion using our new chemistry and cooling models. We compare results based on two  
different treatments of cooling by hydrogen at $T>10^4\Kel$: (1) non-equilibrium ionization (NEI) as in our fiducial choice, and (2) collisional-ionization equilibrium (CIE) as in other previous work.
For NEI cooling, the ionization fraction in the ambient medium is also important if the hydrogen recombination/ionization time scale is longer than the cooling or dynamical time scale. Here we consider two extremes for the background gas: fully neutral (NEI-N) and fully ionized (NEI-I). 
As we shall show below, the 
chemical state of the ISM in which the SNR expands is critical in determining the cooling rates.  This underscores the need for further study of NEI effects 
on SNR evolution in a range of environments (e.g., an extension of the work by \citet{Sarkar21b} covering a range of parameters).

We use a simple setup with a uniform background medium, varying the ambient medium density and metallicity over $\nHinit = 0.01, 0.1, 1, 10, 100 \pcc$ and $\Ztot = 10^{-3}, 10^{-2}, 0.1, 1, 10$.
We set the uniform FUV radiation field to $\chiunatt = (\nHinit/\pcc)$ and uniform CR ionization rate to $\xicr = 2\times 10^{-16} \second^{-1} (\nHinit/\pcc)$, such that the background gas sits on the warm branch of the thermal equilibrium curve.

We run a total of 75 models for the NEI-N, NEI-I, and CIE simulation suites. We set the numerical resolution to satisfy $r_{\rm sf}/\Delta x>20$ and inject the SN ejecta with mass $M_{\rm ej}=10\Msun$ and thermal energy $E_{\rm SN}=10^{51}\erg$ within $r_{\rm init}/\Delta x =5$ \citep[our choices of numerical parameters are motivated by the resolution study of][]{KimCG15}. We assume all gas (including SN ejecta) has the same metallicity; i.e., the metal enrichment by SN ejecta is ignored.

In the first three columns, \autoref{f:SN-summary} shows SNR properties for (a) NEI-N (solid), (b) NEI-I (dotted), and (c) CIE (dashed) as a function of the ambient medium density, with colored lines connecting models with the same  metallicity. In the last column (d), we gather the results for the $\Ztot=1$ cases together to ease comparison. From top to bottom, we present age ($t_{\rm sf}$), radial momentum ($p_{r, {\rm  sf}}$), mass ($M_{\rm snr, sf}$), and radius ($r_{\rm sf}$) of SNRs when the total SN energy drops to be smaller than 70\% of the injected energy ; we denote this time by ``sf.''

\autoref{f:SN-summary} shows that the cooling time is 
$t_{\rm sf}<1\Myr$, and that within this time an isolated SNR sweeps up a total mass of order $10^{3-4}\Msun$ within a  radius $1-100 \pc$; $t_{\rm sf}$, $M_{\rm snr,sf}$, and $r_{\rm sf}$ all increase towards lower ambient density  and lower $\Ztot$.
The radial momentum injected by a single SNR is $\sim 10^{5-6}\Msun\kms$, with a weak increase toward lower density and metallicity. The terminal momentum measured at $10\,t_{\rm sf}$ (not shown) is larger than $p_{\rm r,sf}$ by a factor of 1.5-2, similar to the result in \citet{KimCG15}.

For the CIE suite, we obtained fits (thin lines in column (c))  as follows:
\begin{subequations}
\begin{eqnarray}
\label{eq:snfit_tsf}
  t_{\rm sf} & = & 40{\rm\,kyr}  \nonumber\\
  & & \times  (\nHinit/\pcc)^{-0.56}({\Ztot}^2+{\Ztot_0}^2)^{-0.15},\\
\label{eq:snfit_prsf}
  p_{\rm r,sf} & = & 
  2\times  10^5\Msun\kms  \nonumber\\
  & & \times (\nHinit/\pcc)^{-0.14}({\Ztot}^2+{\Ztot_0}^2)^{-0.087},\\
\label{eq:snfit_Msf}
  M_{\rm snr,sf} & = & 1.7\times10^3\Msun  \nonumber\\
  & & \times  (\nHinit/\pcc)^{-0.27}({\Ztot}^2+{\Ztot_0}^2)^{-0.16},\\
\label{eq:snfit_rsf}
  r_{\rm sf} & = & 22\pc  \nonumber\\
  & & \times (\nHinit/\pcc)^{-0.43}({\Ztot}^2+{\Ztot_0}^2)^{-0.061},
\end{eqnarray}
\end{subequations}
where $\Ztot_0=0.04$. 

Our results for the CIE cooling suite are in broad agreement with previous work \citep[e.g.,][]{Thornton98,KimCG15,Karpov20,Steinwandel20,Oku22}, although direct comparison is not straightforward since the SNR characterization and therefore properties that are reported differ between different studies. For example, \citet{KimCG15} used the time at which the hot gas ($T>2\times10^4\Kel$) mass attains its maximum to define the shell formation time, while \citet{Thornton98} reported the age of SNRs at the maximum luminosity (or maximum cooling rate) as a proxy of shell formation. For high metallicity and high density runs where SNRs experience catastrophic cooling, these definitions are in good agreement with each other and also with our present shell-formation time definition. However, in low metallicity and low-density runs, we find that cooling is gradual rather than catastrophic, with multiple peaks in the cooling rate, making some previous definitions ill-defined. Moreover, defining the hot gas mass with a single temperature cut can be problematic since the shell could form at higher temperatures than $2\times10^4\Kel$ and gradually cool. The current definition, based on a fixed fraction of the injected energy, is well-behaved. We also note that extremely low density conditions ($n\sim 0.01\pcc$) would be uncommon for isolated SNe; very  low  density is more often associated with a hot superbubble created by  multiple correlated SNe, in which case the evolution and the momentum injection from each SN is somewhat different \citep{KimCG17,Gentry17,Fielding18}.

At low densities $n<0.1\pcc$ and low metallicities $\Ztot<0.1$, the effect of NEI hydrogen on cooling is significant. For NEI-N, shell formation occurs earlier than that for CIE. The corresponding radial momentum, SNR mass, and SNR size, therefore, are all lower than for CIE. The main reason for this difference is the enhanced Ly$\alpha$ cooling due to higher neutral hydrogen abundance than in CIE at the shock interface. This is not true for NEI-I, in which Ly$\alpha$ cooling is suppressed by the fully ionized background (see \autoref{f:transition}). At high densities, the difference between the NEI-N and CIE models is reduced since 
the post-shock gas reaches collisional ionization/recombination equilibrium on shorter timescales. At high metallicities, the main coolants are metal ions 
and the difference in Ly$\alpha$ cooling becomes less important.

We caution that in these idealized tests (uniform background, purely spherical expansion), the 
under-ionized shocked gas in our NEI-N
models exists mainly for numerical reasons. That is, 
since the shock is captured over a few zones, post-shock gas can contain neutrals due to numerical diffusion from upstream, leading to Ly$\alpha$ cooling even with trace neutral abundance as the rate coefficient increases with temperature (\autoref{f:Lyalpha}).
For ideal spherical expansion, numerical diffusion of this kind can be minimized by using spherical coordinates with a Lagrangian method \citep{Gentry17}. However, in the real ISM, SNR evolution is not a perfectly spherical expansion in a uniform medium. Rather, due to the multiphase and turbulent nature of the ISM, hydrodynamical instabilities may develop and mix gas of different ionization states and temperatures.
In this case, hydrogen that is under (over)-ionized relative to CIE may exist near interfaces, and enhance (reduce) cooling in the ISM at low metallicity. 
CIE cooling, because it implicitly treats ionization as high  when $T\gtrsim 3\times 10^4\Kel$ (see inset in \autoref{f:transition}),  produces weak Ly$\alpha$. CIE cooling  may therefore represent the true momentum injection better than NEI cooling at low density and metallicity if numerical diffusion significantly overestimates the true neutral abundance. In addition, radiation produced by SN shocks will ionize the upstream ambient medium, reducing the effect of non-equilibrium Ly$\alpha$ cooling (by both H and He; see \citealt{Sarkar21b}).
In the future, careful studies at very high resolution that include inhomogeneity in the density and ionization fraction of the ISM are therefore warranted. In practice, however, when SNe are correlated, much of the energy losses occur when shocks propagate into/become radiative in previously-shocked dense gas.  The distinction between NEI-N, NEI-I, and CIE results for clustered SNe would be relatively small even at low $\Ztot$ when $\nHinit \gtrsim 10 \pcc$.

\subsection{Expansion of \texorpdfstring{\HII}\~ Regions}\label{s:HIIreg}

\begin{figure*}[t!]
  \epsscale{1.0}\plotone{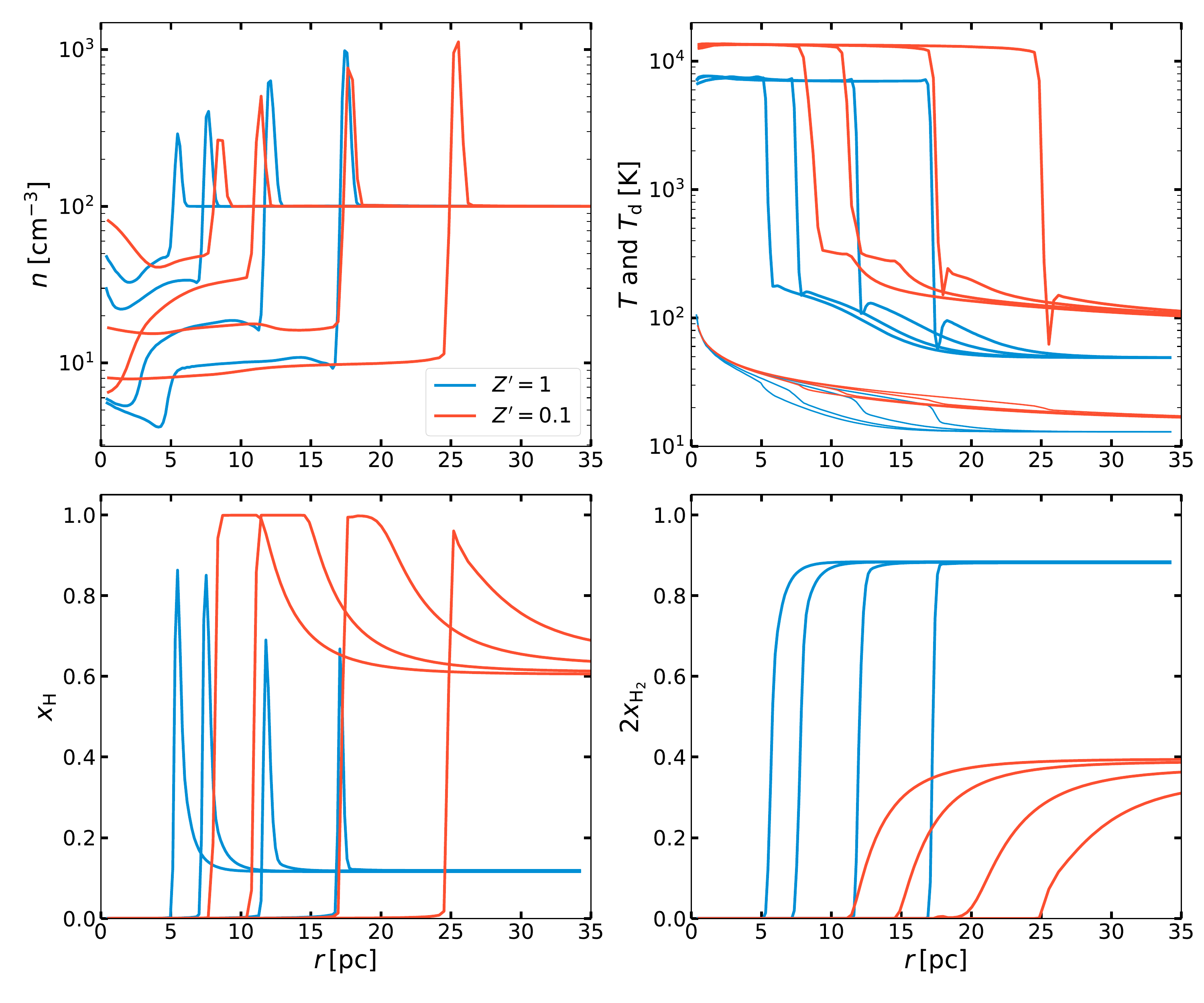}
  \caption{Radial profiles of expanding \HII\ regions driven by thermal pressure of ionized gas. The background medium has density $\nHinit = 10^2 \pcc$ and the metal and dust scaled abundance $\Ztot = 1$ (blue) or $0.1$ (red). We show radial profiles of density, gas and dust (thin lines) temperature, and abundances of \HI\ and \HH\ at times $t = 0.29$, $0.59$, $1.47$, $2.93 \Myr$. The time step size for radiation and hydrodynamics updates is set to 10\% of the global minimum of the chemical and cooling time scales (model A).}\label{f:HII-profile}
\end{figure*}

\begin{figure*}[t!] %
  \epsscale{1.1}\plotone{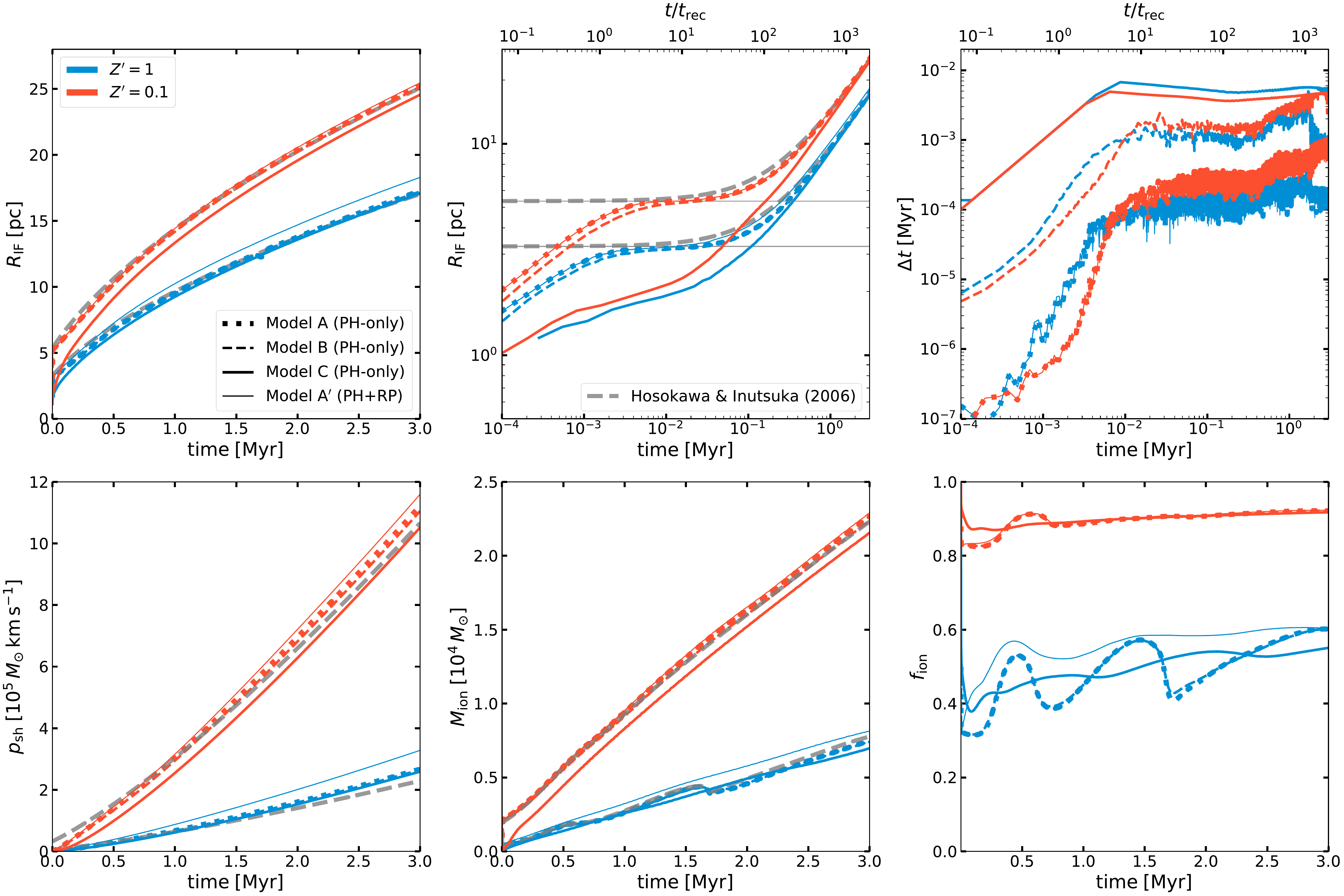}
  \caption{Time evolution of various quantities of an expanding \HH\ region driven by thermal pressure of ionized gas.  Top row shows the IF radius on linear (top left) and log (top middle) axes, and time step size (top right).  Bottom row shows radial momentum (bottom left), ionized gas mass (bottom middle), and the fraction of ionizing photons absorbed by neutral gas (bottom right). Models with $\Ztot=1$ and $0.1$ are shown as blue and red lines, respectively. We compare runs adopting different time step size $\Delta t$ for the hydrodynamics and radiation updates. In Model A/B (dotted/dashed), $\Delta t$ is constrained by 10\%/100\% of the global minimum of the cooling and chemical time scales. In Model C (thick solid), $\Delta t$ is constrained by the usual CFL condition for MHD. We also show the model in which radiation pressure is turned on (Model A$^{\prime}$; thin solid). In the top left and top middle panels, the dashed gray lines show the analytic prediction for the expansion of a D-type IF by \citet{Hosokawa06}. The thin solid grey lines in the top middle panel indicate the initial Str\"{o}mgren radius accounting for the effect of dust absorption \citep{Petrosian72}. The dashed grey lines in the bottom left and bottom middle panels show the scaling relation expected from Equations~\eqref{e:HII_pr} and \eqref{e:HII_Mion} using the \citeauthor{Hosokawa06} solution.
}\label{f:HII-radius}
\end{figure*}

Expansion of a spherical \HII\ region in a uniform medium is a standard test problem that has been widely used for benchmarking radiation hydrodynamics codes \citep[e.g.,][]{Bisbas15}. The implementation of adaptive ray tracing in the \Athena\ code has been tested by \citet{KimJG17} for idealized problems of both dustless and dusty \HII\ regions, adopting the two-temperature isothermal equation of state.
In this work, we present tests of the expansion of an \HII\ region in a dusty medium driven by either thermal pressure or radiation pressure, focusing on the impact of timestep size 
and metallicity $\Ztot$.

\subsubsection{Gas Pressure Driven Expansion}\label{s:HII-PH}

We first perform a test of the expansion of classical \HII\ regions ignoring the radiation pressure force. Consider an ionizing source embedded in a static, uniform medium with hydrogen number density $\nHinit$. The source emits LyC photons at a constant rate $Q_{\rm i}$ at $t=0$. Initially, the ionization front (IF) propagates rapidly as weak R-type without perturbing the gas motion \citep[e.g.,][]{Spitzer78}.
After several recombination timescales, the radius of the IF reaches the initial Str\"{o}mgren radius $R_{\rm St,0} = [3f_{\rm ion} Q_{\rm i}/(4\pi\alpha_{\rm rr,H^+}\nHinit^2)]^{1/3}$, where $f_{\rm ion} = 1 - f_{\rm dust}$ is the fraction of LyC photons absorbed by gas.
Once the IF speed slows down to about twice the sound speed in the ionized gas $c_{\rm s,i}$, the IF transitions to weak D-type and is preceded by a shock front by which the ambient gas is swept up into a thin shell. The time evolution of the IF radius (or the shell radius) is well approximated by
\begin{equation}\label{e:HII_Hosokawa}
    R_{\rm IF} = R_{\rm St,0}\left( 1 + \frac{7}{4}\sqrt{\frac{4}{3}}\frac{t}{t_{\rm St,0}} \right)^{4/7}
\end{equation}
where $t_{\rm St,0} = R_{\rm St,0}/c_{\rm s,i}$ (\citealt{Hosokawa06}; see also \citealt{Spitzer78}).

Assuming that the ionization-recombination balance, 
$f_{\rm ion}Q_{\rm i}=\tfrac{4\pi}{3}\alpha_{\rm rr,H^+}n_{\rm i}^2 R_{\rm IF}^3$ , 
is maintained inside the \HII\ region with uniform ionized gas density $n_{\rm i}$, one can show that the ionized gas mass scales as
\begin{align}\label{e:HII_Mion}
  M_{\rm ion} & = \mu_{\rm H}m_{\rm H} \times \frac{f_{\rm ion} Q_{\rm i}}{\alpha_{\rm rr,H^+}n_{\rm i}} \\
              & \propto (f_{\rm ion}Q_{\rm i}/\alpha_{\rm rr,H^+})^{5/7} c_{\rm s,i}^{6/7} t^{6/7} \nHinit^{-3/7}\,,
\end{align}
where in the proportionality we assumed $t \gg t_{\rm St,0}$. Similarly, the radial momentum of the shell $p_r$ scales as
\begin{align}\label{e:HII_pr}
  p_r & = \dfrac{4\pi}{3} \rho_0 R_{\rm IF}^3 \dfrac{dR_{\rm IF}}{dt} \\
      & \propto (f_{\rm ion}Q_{\rm i}/\alpha_{\rm rr,H^+})^{4/7} c_{\rm s,i}^{16/7} t^{9/7} \nHinit^{-1/7}\,.
\end{align}
The expansion stalls when the pressure of the ionized gas decreases to that of the ambient gas \citep[e.g.,][]{Raga12}.

For our test, we set up a cubic computational domain with a side length $80\pc$ and $256^3$ cells. The background medium has $\nHinit = 10^2\pcc$ and is subject to heating by spatially uniform FUV radiation with $\chiunatt=1$ and CR ionization rate $\xicrunatt = 2\times 10^{-16}\second^{-1}$. The \HH\-dissociation by background radiation is turned off to isolate the dissociation resulting from the central source. A constant-luminosity source with $Q_{\rm i} = 1.24 \times 10^{50}\second^{-1}$ and $L_{\rm FUV} = 1.36\times 10^6 \Lsun$ (corresponding to a $10^3\Msun$, zero-age star cluster according to the \SB\ model) is placed at the domain center.

We consider two different metallicities, $\Ztot = 1$ and $0.1$, for which the gas is initially $88\%$ and $40\%$ molecular (the initial molecular dissociation is by CRs) and has temperature
$46\Kel$ and $33 \Kel$, respectively. For each $\Ztot$, we run three simulations in which the timestep size $\Delta t$ for the hydrodynamics and radiation updates is constrained by 10\%/100\% of the global minimum of the chemical and cooling time scales (Model A/B), or by the usual CFL condition (model C). We also run a reference simulation model A$^{\prime}$, which is the same as model A but the radiation pressure force is additionally turned on.

\autoref{f:HII-profile} shows radial profiles of density, temperature, abundances of \HI\ and \HH\ at four different times for model A. \autoref{f:HII-radius} shows for all models the time evolution of the IF radius (computed as the average distance from the source of the cells with $0.1 \le \xHII \le 0.9$), time step size, radial momentum, ionized gas mass, and gas absorption fraction of ionizing photons.

Our simulations show characteristics that are in good agreement with the classical picture of \HII\ region expansion \citep[e.g.,][]{Spitzer78}, but also exhibit other interesting features: sound waves (launched at the time of shock breakout) traveling back and forth inside the \HII\ region \citep[e.g.,][]{Arthur11, KimJG16}; an \HH\ dissociation front (for $\Ztot=1$) that is trapped between the IF and shock front \citep{Hosokawa06, Krumholz07}; elevated temperature of ionized gas near the center and of neutral gas beyond the IF due to FUV heating by the central source; high dust temperature close to the source.

Compared to the $\Ztot = 1$ case, the \HII\ region formed in low-$\Ztot$ gas has (1) a larger initial size $R_{\rm St,0}$ as the dust absorption fraction is reduced (higher $f_{\rm ion}$), (2) higher ionized gas temperature due to reduced metal cooling (see \autoref{f:photoionized}), which leads to more rapid expansion and higher shell momentum (see below), (3) higher column of atomic layer and more gradual increase (decrease) of the molecular (atomic) fraction, as predicted by models of \HI-to-\HH transitions \citep[e.g.,][]{Sternberg14}, and (4) higher dust temperature away from the source as FUV radiation is attenuated less.



Models A,B,C adopting different time step size criteria yield quite similar results for the overall evolution. Comparison between Model A$^{\prime}$ and other models show the expansion is dominated by thermal pressure of ionized gas, consistent with theoretical expectations for the adopted parameters \citep[e.g.,][]{Krumholz09, KimJG16}; inclusion of radiation pressure only slightly enhances the momentum. In Model A and B, $f_{\rm ion}$ oscillates quasi-periodically because of the density fluctuations produced by sound waves, which in turn affects the total recombination rate of the \HII\ region. The evolution of the D-type IF radius agrees very well with the \citet{Hosokawa06} solution (grey dashed lines) with a typical error of less than $2\%$. Even for Model C, the time-averaged difference is only 3\% and 5\% for $\Ztot = 1$ and $\Ztot = 0.1$, respectively, although the expansion appears ``delayed'' in Model C. As the top middle panel shows, this is because the radius of R-type IF is underestimated (up to $\sim 50\%$) at early times since the IF cannot propagate across more than one cell per radiation update. Nevertheless, the evolution of ionized gas mass and shell radial momentum is consistent with the scaling expected, shown as grey dashed lines (Equations~\eqref{e:HII_Mion} and \eqref{e:HII_pr} with \autoref{e:HII_Hosokawa} for $R_{\rm IF}$).

As shown in the top-right panel of \autoref{f:HII-radius}, simulations using the most restrictive timestep criterion (Model A) would have a cost that is more than 10 times greater than that using the less-restrictive 
Model C criterion. Evidently, the criterion in Model B suffices when it is necessary to follow the early evolution of the R-type front, while the lowest-cost Model C criterion suffices when it is only needed to follow expansion after the D-type front forms.

\subsubsection{Radiation Pressure Driven Expansion}\label{s:HII-RP}

\begin{figure}[t!] %
  \epsscale{1.0}\plotone{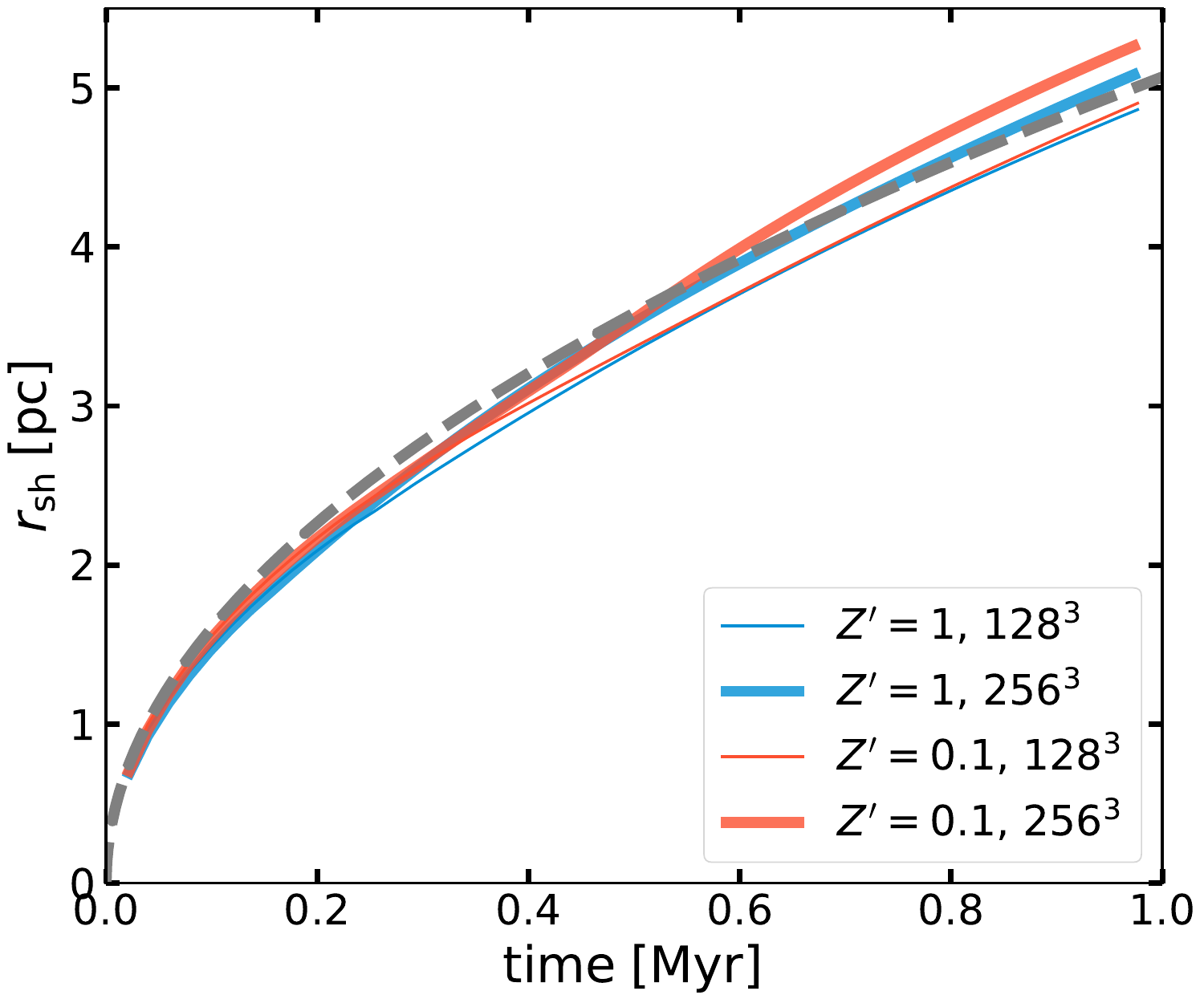}
  \caption{Time evolution of the shell radius for radiation pressure driven bubble expansion in a uniform medium ($\nHinit=10^4 \pcc$) for $\Ztot = 1$ (blue) and $0.1$ (red) and for grid spacings $\Delta x = 9.4 \times 10^{-2}\pc$ (thin) and $4.7\times 10^{-2}\pc$ (thick). The grey dashed line shows the similarity solution for the thin shell expansion (\autoref{e:HII_RP}).}\label{f:HII-RP}
\end{figure}

The direct UV radiation pressure force plays an important role for the dynamics of dense and luminous \HII\ regions \citep[e.g.,][]{Krumholz09, KimJG16}.
Here we perform a test of bubble expansion driven only by the UV radiation pressure force on dust grains. We initialize the simulation with box size $12 \pc$ and two different grid spacings $\Delta x = 12/128 \pc$ and $12/256 \pc$. The background medium has a density of $\nHinit = 10^4 \pcc$ and the central source has a constant UV luminosity of $L_{\rm UV} = L_{\rm LyC} + L_{\rm FUV} = 2.28 \times 10^{7} \Lsun$, characteristic of extreme environments where super star clusters form (this luminosity corresponds to mass $3\times 10^4\Msun$ for a zero-age cluster from \SB). As in Section \ref{s:HII-PH} we consider two values of metallicity $\Ztot = 1$ and $0.1$. We turn off \PI\ by setting H and \HH\ \PI\ cross sections to zero. We do not apply radiation pressure force when the ratio between the local radiation pressure and thermal pressure is greater than 50, as described in \autoref{s:dust_dest}. We also do not consider the radiation force exerted by optical photons (through the use of boosting factor mentioned in \autoref{s:radforce}). The timestep size is solely determined by the CFL condition. For this test, the angular resolution of ray tracing is increased to 20 rays per face for the best comparison to the analytic solution, but we find that the standard 4 rays per face gives quite similar results.

We find that the radiation pressure force creates a cavity filled with low-density ($\nH \sim 1\pcc$), warm ($T \sim 10^{4}\Kel$) gas and a dense, expanding shell, which absorbs most of the UV radiation both for $\Ztot=1$ and $0.1$. 
\autoref{f:HII-RP} shows the time evolution of the shell radius (computed as the mean distance from the source of cells with $\nH > 1.1\nHinit$). Since the shell quickly becomes optically-thick to UV radiation,
we expect the shell radius to follow the similarity solution
\begin{equation}\label{e:HII_RP}
r_{\rm sh} = \left( \frac{3 L_{\rm UV} t^2}{2\pi c \rho_0} \right)^{1/4}\,.
\end{equation}
\citep[e.g.,][]{Krumholz09, KimJG16}. The measured shell radius is in good agreement with the analytic solution (grey dashed line) within 5\%. The discrepancy is acceptable given the finite thickness of the shell, and the breaking of spherical symmetry at late times (when non-linear thin-shell instabilities develop) in high-resolution runs.

The measured gas momentum in the radial direction ($p_r$) increases approximately linearly in time. By comparing radial forces ($F_{\rm thm} = - \int \nabla P \cdot \mathbf{r} \, dV$ with $F_{\rm rad} = L_{\rm UV}/c$), we find that momentum injection by gas pressure (caused by FUV heating of the shell) makes about 3\% contribution to the total injected momentum.
We find that the relative error $(p_r/p_{r,{\rm inj}} - 1)$ is quite small ($\sim 10^{-3}$) at late times, where $p_{r,{\rm inj}} = \int (F_{\rm rad} + F_{\rm thm}) dt$.
The injected momentum by radiation is less than $L_{\rm UV}t/c$ by $1\%$ at early times ($t \lesssim 0.2 \Myr$). This is because the radiation force is not actually applied in zones where dust would be destroyed (see \autoref{s:dust_dest}).


\section{Discussion}\label{s:discussion}

\subsection{Comparison of Photochemistry and Radiation Treatments}\label{s:comparison_methods}


Here we first compare our model with other existing implementations of photochemistry 
and heating/cooling in the ISM simulation literature.
Our model has two key features: (1) following the minimal set of chemical species (and reactions) required for modeling important thermal processes in the neutral ISM---effectively two time-dependent (\HH\ and \Hplus) and two steady-state (C$^+$ and CO) species abundances, with abundances of H, C, O, and e determined from element or charge conservation;
(2) computing the UV radiation field in a few key bands through adaptive ray tracing, accounting for attenuation by hydrogen PI (for LyC), dust, \HH-absorption line self-shielding and cross-shielding, and self-shielding of C. The resulting UV radiation fields are used to treat PI of H, \HH, and C, photodissociation of \HH\ and \CO, 
the PE effect on dust grains, dust heating.  The attenuation of UV is also used to (crudely) estimate the reduction of the low-energy CR rate at high column. Cooling by helium and metals in collisionally ionized hot gas is treated using look-up tables computed from Cloudy. See \autoref{t:chem_cooling} for a summary of included processes.

The SILCC simulations \citep{Walch15, Girichidis16, Gatto17, Girichidis18} model time-dependent hydrogen chemistry based on \citet{Glover07a, Glover07b} and time-dependent CO chemistry based on \citet{Nelson97}, as described in \citep{Glover12b}. 
The hydrogen chemistry is similar to ours, but PI is not included. 
Their CO chemistry includes two processes: formation via CH intermediary (with the rate-limiting step being radiative association of C$^+$ with \HH) and destruction by photodissociation. All atomic carbon (photodissociated CO) is assumed to be singly ionized instantly.\footnote{The \citet{Nelson97} network ignores important chemical processes such as the grain-assisted recombination, and is shown to produce large errors in CO abundances in some cases \citep{Gong17}, although this may not have significant impact on gas dynamics and star formation \citep{Glover12a}.}
Heating and cooling processes modeled (as summarized in Table 1 of \citealt{Glover10} and updated in \citealt{Glover12b}) are similar to ours, although there are minor differences such as the inclusion of fine-structure line cooling by Si$^+$ and Compton cooling. The background (unattenuated) FUV radiation field in the above SILCC simulations is constant spatially and temporally and is set to the \citet{Draine78}'s ISRF ($\chi_0=1$; or scaled appropriately if a pre-defined SN rate is used). The TreeCol method \citep{Clark12, Wunsch18} is used to calculate local (within $50 \pc$) shielding effects (dust-shielding, \HH\ self-shielding, CO self-shielding, and shielding of CO by \HH) for each cell, sampling 48 lines of sight. While this method can obtain the attenuation of isotropic external radiation field with reasonable accuracy, it is not suitable for computing radiation field resulting from discrete sources such as stars.

The SILCC simulations by \citet{Peters17} and \citet{Rathjen21} additionally include the effect of PI, with the former calculating the transfer of ionizing radiation via the adaptive ray tracing method \citep{Baczynski15} and the latter via the TreeRay backward ray tracing method \citep{Wunsch21}. These studies, however, calculated the attenuation of (temporally constant) background FUV for PE heating and \HH- and \CO-dissociation using the TreeCol method. The SILCC simulations adopt a lower value of the CR ionization rate ($3 \times 10^{-17}\second^{-1}$) for the solar neighborhood condition than ours, and did not include the effect of radiation pressure.

The hydrogen chemistry of \citet{Glover07a, Glover07b}, CO chemistry of \citet{Nelson97}, and associated thermal processes \citep{Glover10} (or similar methods)
have been widely adopted in the literature: to name but a few, simulations of isolated dwarf galaxies by \citet{Hu16, Hu17}, dwarf galaxy mergers by \citet{Lahen20} (the GRIFFIN project), Milky Way-like spiral galaxies by \citet{Smith14, Smith20, Jeffreson20, Jeffreson21}, SN-driven ISM by \citet{Steinwandel20}, and kpc-scale stratified-disk simulations by \citet{Kannan20a, Hu21}.

The treatment of UV differs among recent ISM simulations studies, even those with similar chemistry and heating/cooling.
Similar to the SILCC simulations, most work employs the TreeCol method or a similar scheme to model the shielding of FUV radiation as seen by individual cells \citep{Hu16, Hu17, Smith14, Smith20, Lahen20, Steinwandel20}. In \citet{Kannan20a}, chemistry is coupled to moment-based radiation equations with the M1-closure approximation \citep{Kannan19}.
In \citet{Hu16, Smith14, Smith20}, a fixed background FUV ISRF (and CR ionization rate) is adopted and the effect of PI is not included, while the background FUV radiation field and CR ionization rate are scaled with the recent star formation rate in \citet{Hu21}. \citet{Hu17} obtain a spatially and temporally varying (unattenuated) FUV radiation field by summing up contributions from all individual sources assuming an optically thin medium, which is appropriate for the dust-poor systems ($\Zd=0.1$) they considered (see also \citealt{Forbes16}). \citet{Lahen20} adopt a similar approach, but considered sources within $50\pc$ only.
In \citet{Hu17, Hu21, Lahen20}, the effect of (short-range) PI feedback is included using a so-called Str\"{o}mgren-volume approach \citep[see also][]{Hopkins12}, in which cells neighboring an ionizing source are ionized and heated in the order of increasing distance from the source until all ionizing photons are used up. Based on analytic solutions for expanding \HII\ regions \citep{Krumholz09}, \citet{Jeffreson20, Jeffreson21} model effects of PI feedback by direct injection of momentum (and thermal energy) into neighboring cells. Our adaptive ray tracing method covering both ionizing and FUV radiation is a clear improvement upon previous approaches in the literature in terms of self-consistency and accuracy.


In recent years, as numerical resolution has improved, several efforts have been made in the galaxy formation community to
incorporate ISM photochemistry and UV radiation feedback in (M)HD simulations, so as to capture multiphase ISM structure 
\citep[e.g.,][]{Rosdahl13, Katz17, Hopkins18, Nickerson18, Marinacci19, Emerick19, Kannan20b, Smith21, Hopkins22}. 
Most work includes time-dependent primordial chemistry (H and He) and associated heating/cooling processes, often using a dedicated library such as Grackle \citep{Smith17}.
The metal cooling in hot gas is included using tables pre-computed from Cloudy (as a function of density, temperature, and redshift), accounting for the effect of PI by metagalactic UV background (UVB) \citep[e.g.,][]{Vogelsberger13}.
The cooling by metals and molecules in low-temperature ($T \lesssim 10^4 \Kel$) gas is included using Cloudy tables (with varying assumptions about background UV radiation field and shielding columns; e.g., \citealt{Rosdahl15, Hopkins18}) or an approximate fitting function without following relevant chemistry directly \citep[e.g.,][]{Nickerson18, Marinacci19}.

Again, 
different approaches have been taken to treat the effects of UV radiation and CRs. 
For example, in their SMUGGLE model for the ISM and stellar feedback,
\citet{Marinacci19} adopt a spatially and temporally constant FUV radiation field and the \citet{Wolfire03} formula for the PE heating that has explicit dependence on grain charging, but did not include the effect of low-energy CRs.
\citet{Kannan20b} couple the SMUGGLE framework with UV (and IR) radiation moment equations adopting the M1 closure \citep{Kannan19}, thereby calculating time-dependent, spatially-varying LyC and FUV radiation fields using the reduced speed of light approximation for the hyperbolic equations (see also \citealt{Katz17}).
They adopt a constant value of heating/ionization rate by low-energy CRs,
and use a local (Sobolev) approximation for \HH\ self-shielding. Simulations of isolated galaxies by \citet{Emerick18, Emerick19} and \citet{Smith21} calculate the FUV radiation field in the optically-thin limit
and apply dust shielding using a local approximation. Without CR ionization to set the grain charging parameter, they adopt a PE heating efficiency that depends on gas density $\epsilon_{\rm pe} = 0.0148 (\nH/1\pcc)^{0.235}$, which is a power-law fit to the result in \citet{Wolfire03} \citep[see also][]{Smith22}. They also include the effect of UVB with a density-dependent attenuation factor $\fshldH$ from \citet{Rahmati13}. Being computationally efficient and accurate, our photo-chemistry method has the potential to be adopted in galaxy simulations in the future.

\subsection{Comparison of Equilibrium Curves}\label{s:comparison}

\begin{figure*}[t!]
\includegraphics[width=\linewidth]{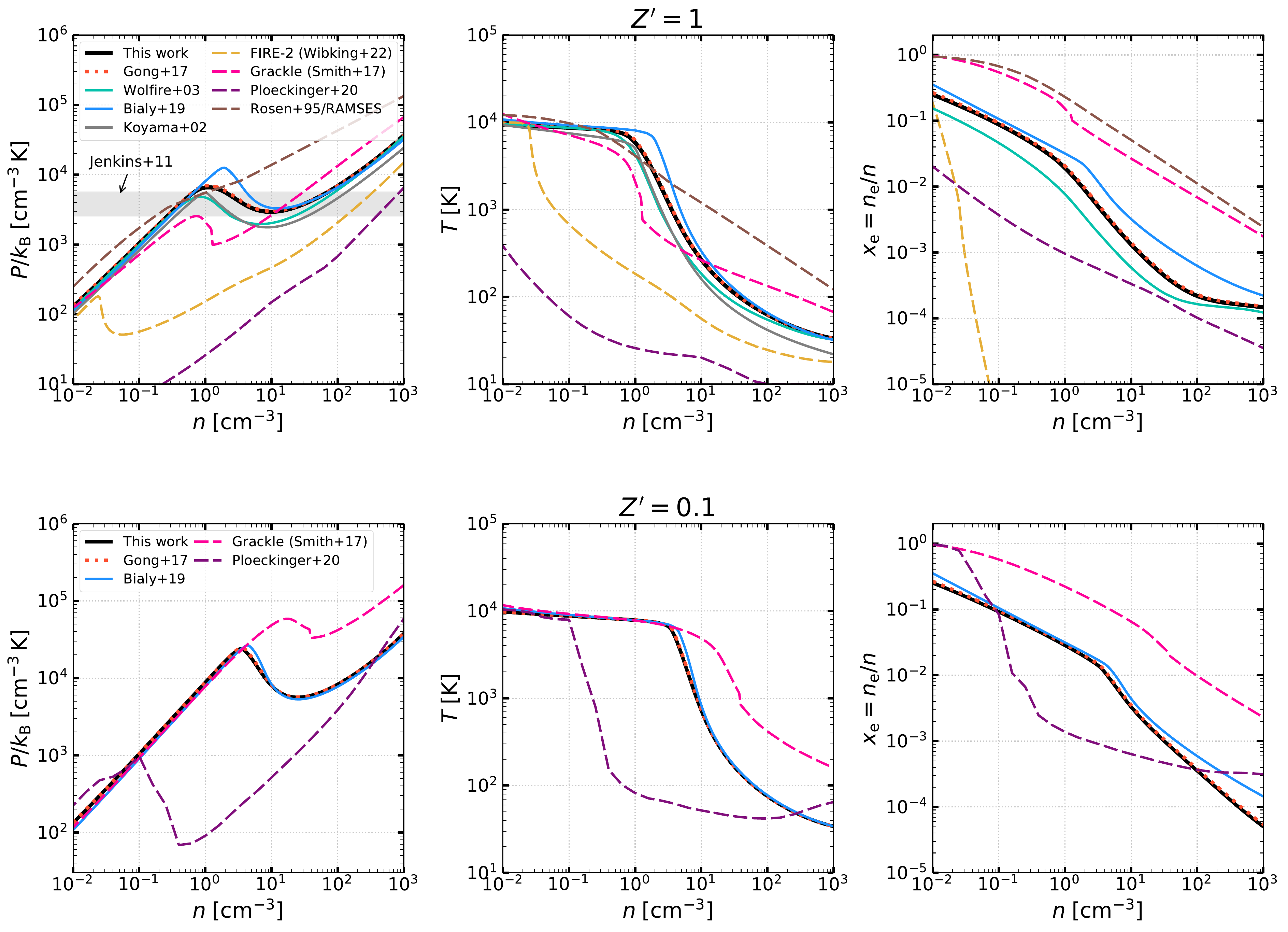}
\caption{Comparison of equilibrium gas properties computed with models originating in the ISM community \citep{Gong17, Wolfire03, Bialy19, Koyama02} and the galaxy formation community \citep{Smith17, Hopkins18, Ploeckinger20, Rosen95}. For the former, the FUV radiation field is representative of solar neighborhood conditions; for the latter, heating and ionization by metagalactic UVB (but not by low-energy CRs) are included (see text for details). From left to right, panels show thermal pressure, temperature, and electron abundance. The top row shows results for solar metallicity  ($\Ztot=1$) and the bottom row for low metallicity ($\Ztot=0.1$). In the top left panel, the grey-shaded region indicates the observed range of thermal pressure in the local diffuse CNM inferred from \ion{C}{1} fine-structure lines \citep{Jenkins11}. Comparison of the cooling functions and specific heating rates for the same models is shown in \autoref{f:heatcool-comp}.}\label{f:equil-comp}
\end{figure*}

\begin{figure*}
  \epsscale{1.0}\plotone{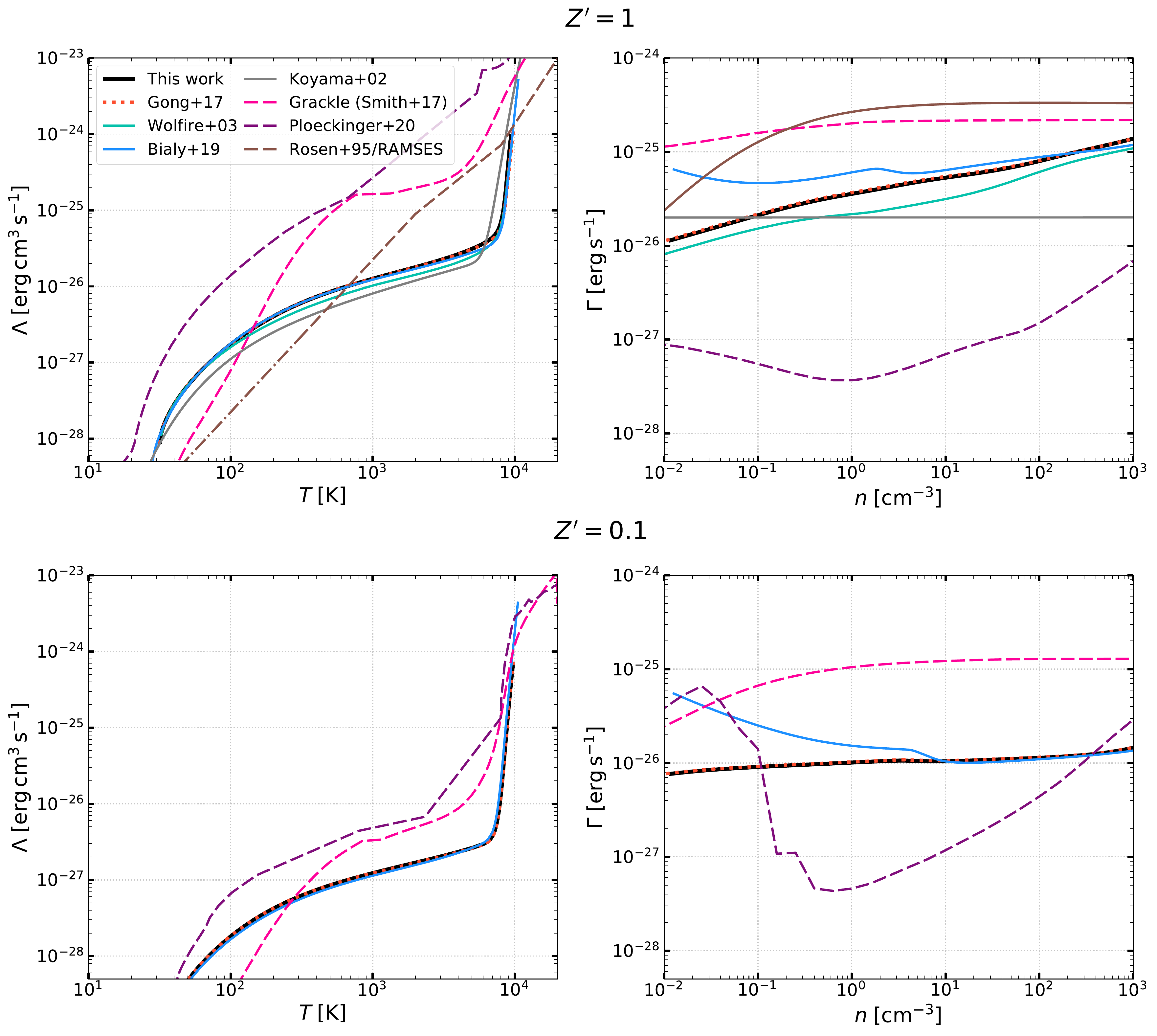}
  \caption{The heating rate per H nucleon ($\Gamma = \nH \Lambda$) as a function of density (left) and the cooling function ($\Lambda$) as a function of temperature (right) for models shown in \autoref{f:equil-comp} with $\Ztot = 1$ (top) and $0.1$ (bottom).}\label{f:heatcool-comp}
\end{figure*}


To provide a sense of similarities and differences in the thermal properties of warm and cold neutral gas across different models in the literature, we compare our results to previous work in Figures \ref{f:equil-comp} and \ref{f:heatcool-comp}.
Our comparison includes four models from the ISM community (solid and dotted lines): \citet{Gong17} (the immediate predecessor of this work), \citet{Wolfire03} (often considered the benchmark Milky Way ISM heating/cooling model), \citet{Bialy19} (a recent study focusing on thermal phases of the neutral ISM across a wide range of metallicities), and \citet{Koyama02} (a simple prescription that has been widely adopted in numerical MHD simulations of the ISM);
as well as four models from the galaxy formation community (dashed lines): the Grackle library for cooling and chemistry as reported by \citealt{Smith17} (version 3.2), the Gizmo cooling module used in FIRE-2 simulations \citep{Hopkins18}, the cooling function based on Cloudy tables provided in \citet{Ploeckinger20}, and the simple \citet{Rosen95} cooling function that has been adopted in many RAMSES simulations \cite[e.g.,][]{Slyz05, Verhamme12, Rosdahl17, Agertz20, Dashyan20, Martin-Alvarez22, Farcy22}.


\citet{Gong17}, \citet{Wolfire03}, and \citet{Bialy19}
calculated the gas properties in
thermal equilibrium including realistic heating and cooling processes.
Although some details differ, such as the adopted CR ionization rate, carbon and oxygen abundances,
and the inclusion or omission of X-rays, the key model ingredients in these
works are largely similar.
In the comparison we make here, both our model and the \citet{Gong17} model adopt a FUV radiation field of $\chiunatt=1$, and a primary CR ionization rate of $\xicr=2 \times 10^{-16} \second^{-1}$ (same as our solar neighborhood model in Figures \ref{f:equil_unshld} and \ref{f:rate_unshld}). For the \citet{Bialy19} model, we adopt $\chiunatt=1$ and a total CR ionization rate of atomic hydrogen $3\times 10^{-16}\second^{-1}$, corresponding to the primary CR ionization rate of $\sim 2 \times 10^{-16} \second^{-1}$.

Focusing on solar metallicity conditions, \citet{Koyama02} proposed a simple analytic cooling function $\Lambda(T)$ (applicable for cold and warm neutral gas with $T \lesssim 10^{4.2}\Kel$). This functional form, when matched to a constant heating rate $\Gamma_0 = 2 \times 10^{-26}\erg \second^{-1}$ (representing \PE\ + CR heating in the solar neighborhood) as $\nH\Lambda(T) = \Gamma_0$, yields an equilibrium curve that matches well to a more detailed, chemistry-based calculation by \citet{Koyama00}.

The Grackle code allows for ionization and heating by the metagalactic UV/X-ray background (UVB, optionally with density-dependent attenuation corrections), but does not include ionization and heating by low-energy CRs. The chemistry and cooling of primordial gas are supported, and the cooling of metal-enriched gas is treated using tables from CLOUDY \citep{Ferland17}.
In our Grackle example, we include primordial chemistry, metal cooling, optically-thin UVB (for \PI\ and heating, from \citet{Haardt12} at redshift $z=0$), and \PE\ heating (from \citet{Wolfire95} with $\chiunatt = 1$).


The FIRE-2 cooling at $\Ztot = 1$ is obtained from the Appendix A in \citet{Wibking22}, and also reproduced from their source code available online\footnote{\href{https://github.com/BenWibking/cooling-curve-comparison}{https://github.com/BenWibking/cooling-curve-comparison}}. The source of ionization is the UVB \citep{Haardt12}, while the source of heating is \PI\ by the UVB plus \PE\ heating \citep[][with $\chiunatt = 1$]{Wolfire03}.

The fiducial model in \citet{Ploeckinger20} ({\tt UVB\_dust1\_CR1\_G1\_shield1}) includes heating from both the UVB \citep{FG20} and a density-dependent ISRF, CR heating, and comprehensive chemical reactions and cooling processes. The strength of the shielded ISRF is computed from the Cloudy code, using the local Jeans length $L_{\rm Jeans} \propto (T/\nH)^{1/2}$ to set the shielding column $N_{\rm Jeans} \propto (\nH T)^{1/2}$ (with some physically motivated limits). The normalization of the unshielded ISRF is set by relating the shielding gas column to a surface density of SFR using the Kennicutt-Schmidt relation \citep{Kennicutt98}, $J \propto \Sigma_{\rm SFR} \propto N_{\rm Jeans}^{1.4}$. The resulting unshielded ISRF strength is similar to the \citet{Draine78} field at $\nH \sim 10^{2.5}\pcc$ with a shielding column of $\sim 10^{21} \cm^{-2}$, and increases roughly linearly with density (Fig 14 in \citealt{Ploeckinger20}). A similar scaling relation is adopted for the CR ionization rate and the dust-to-gas ratio.

The \citet{Rosen95} cooling function is an approximate, piecewise power-law fit to the cooling function in \citet{Dalgarno72} for $T < 10^4\Kel$, and the CIE cooling function in \citet{Raymond76} for $T > 10^4\Kel$ (alternative CIE cooling functions have been adopted in recent RAMSES simulations).
While some simulations included no heating source \citep{Slyz05,Verhamme12},
the usual approach has been to allow for heating and ionization by the redshift-dependent UVB, optionally with an attenuation factor $e^{-\nH/(10^{-2}\pcc)}$ applied. In this comparison, we include heating and ionization by the UVB \citep{FG20} without the attenuation correction. Although the original \citet{Rosen95} cooling function is zero for $T<300\Kel$, we extend it to lower temperature with linear extrapolation in log-space (dot-dashed brown line shown in the left panel of \autoref{f:heatcool-comp}), following the practice adopted in RAMSES galaxy formation simulations (T. Kimm private communication).


We first compare results across different models at $\Ztot=1$ (upper panels in Figures \ref{f:equil-comp} and \ref{f:heatcool-comp}).
By design, our model (black solid) is in excellent agreement with \citet{Gong17} (red dashed).
Our results and those from \citet{Wolfire03} (cyan) are also in good agreement.
The equilibrium pressure and temperature of thermally unstable gas at a given density is slightly lower in \citet{Wolfire03}. The main reason for this difference is that \citet{Wolfire03} adopted a (CR+EUV/X-ray) ionization rate that is lower than the currently accepted value\footnote{Subsequent improved estimates of rate constants and chemical modeling led to the upward revision of the CR ionization rate in the local diffuse ISM \citep[e.g.,][]{Dalgarno06, Indriolo07}.}. They adopted a primary CR ionization rate of $1.8 \times 10^{-17}\second^{-1}$ and a primary EUV/X-ray ionization rate of $1.6 \times 10^{-17}\second^{-1}$ at a column density of $10^{19}\cm^{-2}$. As a result, their electron abundance and \PE\ heating efficiency are about a factor of $\sim 2$ lower than ours. In addition, although less important, their CR+EUV/X-ray heating is about a factor $\sim 3$ lower than our CR heating.

The equilibrium pressure and temperature of thermally stable gas in the \citet{Bialy19} model are similar to ours, but those of thermally unstable gas are somewhat higher. The main reason for the discrepancy is that the hydrogen chemistry in \citet{Bialy19} does not include  grain-assisted recombination of \Hplus\ (although the associated cooling process is included).
  This leads to a slightly higher electron fraction than ours, which in turn leads to lower grain charging parameter and higher PE heating rate.
  Compared to our model, \citet{Gong17}, and \citet{Wolfire03}, the cooling function of \citet{Bialy19} is largely similar, but the heating rate per H nucleon at low density is higher. This is due to an additional heating process associated with the direct interaction between CR protons and free electrons \citep{Goldsmith69}, which becomes important in low density gas with a high electron fraction.
The results from \citet{Koyama02} (grey) are broadly consistent with the other models discussed so far.
We note that the \citet{Koyama02} cooling function has been used in several previous simulations for studying warm/cold ISM
\citep[e.g.,][]{KimCG13, KimCG15b, KimCG17, Vazquez-Semadeni17, Kobayashi20}.

The models discussed so far (solid and dotted curves in Figures \ref{f:equil-comp} and \ref{f:heatcool-comp})
recover diffuse ISM pressure in the two-phase regime consistent with that of the 
CNM inferred from observations of \ion{C}{1} absorption lines
\citep[][grey shaded region in the top left panel of \autoref{f:equil-comp}]{Jenkins11}.

The remaining models originate from the galaxy formation community (Grackle, FIRE-2, \citet{Ploeckinger20}, \citet{Rosen95}; dashed lines). All of them yield equilibrium pressure, temperatures, and ionization that differ significantly from our model and other models from the ISM community (\autoref{f:equil-comp}). This is due to differences both in cooling functions and in treatments of heating and ionization (\autoref{f:heatcool-comp}).

Among the models from the galaxy formation community, the results from Grackle are the closest to ours, but there are still significant departures. The electron fraction set by the photoionizing UVB is about an order of magnitude higher than those set by CR ionization (in the ISM studies), leading to higher grain charging parameter and hence reduced \PE\ heating efficiency. The total heating is however dominated by \PI\ heating from the UVB and is considerably higher. Despite this, the equilibrium pressure at $\nH < 10\pcc$ ($T > 300 \Kel$) is lower because of more efficient cooling.


Similarly, the FIRE-2 model also includes heating and ionization by the UVB and \PE\ heating from grains.
However, as pointed out by \citet{Wibking22}, since the equilibrium electron abundance is based on UVB with an attenuation factor that exponentially decreases with density above $\nH \sim 10^{-2} \pcc$ (Appendix B in \citealt{Hopkins18}), the electron abundance drops very sharply for $\nH \gtrsim 10^{-2}\pcc$. As a result, both the \PE\ and \PI\ heating efficiencies are significantly reduced and the equilibrium pressure is nearly two orders of magnitude lower than that in our model at $\nH=1 \pcc$.



The thermal equilibrium curves of the standard model in \citet{Ploeckinger20} fall at pressure and temperature far below ours (by nearly three orders of magnitude at $\nH=1 \pcc$). Because their adopted ISRF (and CR ionization rate) decreases at lower density, the discrepancy is largest at low densities. Their heating rate is an order of magnitude below ours, and their cooling function exceeds ours by two orders of magnitude at WNM temperature.


The \citet{Rosen95} result also shows large departures from ours. The specific heating rate $\Gamma$ that is often paired with the \citet{Rosen95} cooling in RAMSES simulations is quite similar to that adopted in Grackle (although without \PE\ heating), but the equilibrium pressure is higher than the Grackle result because cooling is less efficient. Strikingly, the equilibrium curve does not have a thermally unstable branch with $dP_{\rm eq}/d\nH < 0$ because the cooling function increases steeply with temperature.\footnote{If the cooling function can be locally approximated by a power-law function $\Lambda = \Lambda_0 T^{\delta}$ and the specific heating rate $\Gamma$ is relatively constant, the condition for thermal instability is $$d\ln P_{\rm eq}/d\ln\nH \approx 1 + d\ln T_{\rm eq}/d\ln\nH \approx 1 - \delta^{-1}<0\,;$$ thus if $\delta > 1$ there is no thermally unstable regime (see \citealt{Balbus86} for a more general criterion).}

\begin{figure*}[t!]
\includegraphics[width=0.8\linewidth]{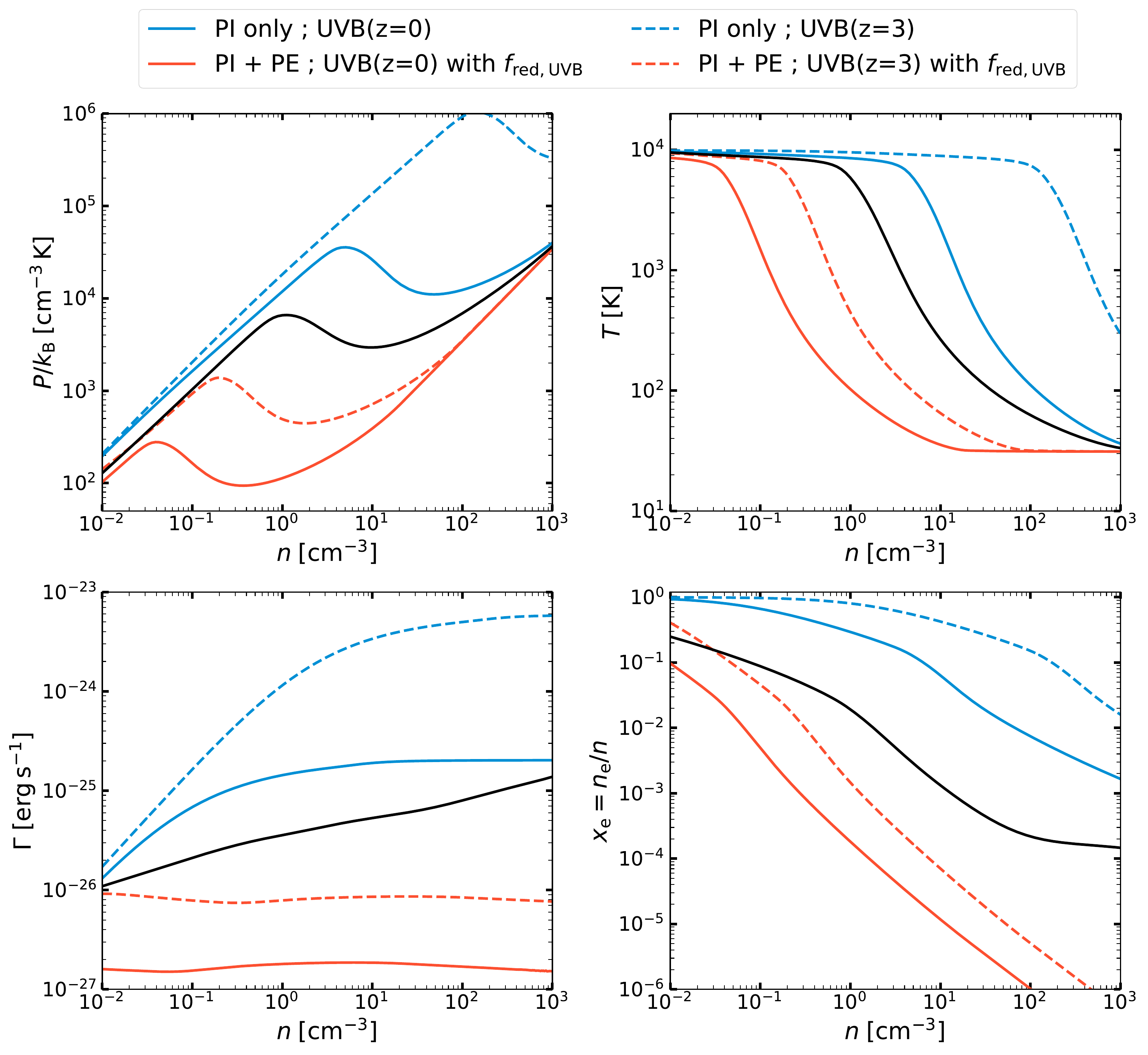}
\caption{Colored lines show equilibrium phase diagrams (top) and heating/ionization curves (bottom) in the presence of UVB of \citet{FG20} at two redshifts $z=0$ (solid) and $3$ (dashed). The equilibrium electron abundance is obtained assuming ionizing UVB with (red) and without (blue) the attenuation prescription of \citet{Rahmati13}. For the latter, we additionally include grain \PE\ heating ($\Gamma_{\rm pe}$) with $\chi=1$ and $\Zd=1$. The \PE\ heating efficiency is computed using the \citet{Weingartner01b} formula, which depends on the grain charging parameter $\propto \chi T^{1/2}/\nEL$. For comparison, the results of our cooling module for the solar neighborhood condition (without UVB) are shown as black solid lines. All curves shown here assume the same cooling function $\Lambda(T)$ obtained from our model for the solar neighborhood condition (black curve shown in the upper left panel of \autoref{f:heatcool-comp}).}\label{f:equil-UVB}
\end{figure*}

In the $\Ztot=0.1$ case (bottom panels in Figures \ref{f:equil-comp} and \ref{f:heatcool-comp}), results from our model and \citet{Gong17} again are almost identical. Also, results from the \citet{Bialy19} model are in better agreement than in the $\Ztot=1$ case as the grain-assisted recombination does not play a significant role. As we noted in \autoref{s:unshielded}, at lower metallicities, cooling at $T \lesssim 8000\Kel$ is reduced due to lower abundances of the main coolants (metals and dust), but heating is not reduced as much because CR heating is insensitive to metallicity and becomes more dominant.
This results in a higher equilibrium pressure at a given density. A similar trend is seen in the Grackle result, but the increase in equilibrium pressure is more dramatic as the main heating source (UVB) remains unchanged and the cooling function is reduced by more than an order of magnitude. Similar to the $\Ztot=1$ case, the \citet{Ploeckinger20} equilibrium curve lies well below ours.

The discrepancy between pressure equilibrium curves from the ISM community and the galaxy formation community is perhaps not surprising, given that the latter has been focused on different scientific issues (such as recovering cosmic star formation and metallicity history, galaxy-halo correlations, etc).
However, as numerical techniques and resolution have improved, galaxy simulations are beginning to resolve the neutral ISM. For cosmological zoom and other galaxy simulations that do attempt to follow the neutral gas dynamics and thermodynamics, we caution that some previous treatments of ionization and heating/cooling may no longer be adequate.

\subsection{Equilibrium Curves with Metagalactic UVB}\label{s:comparison_UVB}

The comparison we made in \autoref{s:comparison} shows that models for cooling and heating adopted in the galaxy formation community lead to equilibrium pressure solutions that significantly differ from those in the solar neighborhood at the same density. A common practice widely adopted in galaxy formation simulations -- but not in ISM simulations -- is to include heating and ionization attributed to the metagalactic UVB; the models discussed above follow this practice. In the following exercise, we directly investigate how the ISM equilibrium state is affected by the inclusion of UVB with or without a density-dependent attenuation factor $\fshldH < 1$ representing a reduction of the UVB to a lower level under denser intragalactic conditions (see Eq. (A1) in \citealt{Rahmati13}).

We consider cases with (1) unattenuated UVB heating and no \PE\ heating, i.e. $\Gamma = \Gamma_{\rm pi}$, and (2) heating from the UVB with the density-dependent reduction of \citet{Rahmati13} turned on, plus \PE\ heating, i.e. $\Gamma = \Gamma_{\rm pi + pe}$ with a factor $\fshldH$ applied to the UVB.

Our simplified procedure for obtaining the equilibrium state is as follows. For a given density $\nH$, we use a root finding algorithm (Brent's method in {\tt scipy}) to find temperature and electron fraction that satisfy the thermal and ionization balance equations simultaneously. The condition for thermal equilibrium is
\begin{equation}\label{e:therm_bal}
  \Gamma = \nH \Lambda(T) \,,
\end{equation}
while the ionization balance requires %
\begin{equation}\label{e:ion_bal}
  \xHI \zeta_{\rm pi,H} 
   + \xHI ( 1 - \xHI )\zeta_{\rm ci,H}
  = \alpha_{\rm rr,H^+}\nH ( 1 - \xHI )^2 \,,
\end{equation}
which is a quadratic equation for $\xHI$ \citep[e.g.,][]{Altay13}. Note that we ignore free electrons produced by metal ions and assume that the molecular fraction is zero, so that $\xEL = \xHII = 1 - \xHI$. We also ignore heating and free electrons due to PI of helium, but they are expected to make only small differences.

For all cases considered here
we adopt the cooling function $\Lambda(T)$ obtained for equilibrium gas in the solar neighborhood condition with $\Ztot = 1$ 
(black solid line in the left panel of \autoref{f:heatcool-comp}). For heating sources, we consider hydrogen \PI\ heating ($\Gamma_{\rm pi}= \xHI q_{\rm pi,H}\zeta_{\rm pi}$) by the UVB, and (Case 2 only) grain \PE\ heating ($\Gamma_{\rm pe}$) at the \citet{Weingartner01b} rate with $\chi = 1$ and $\Zd = 1$, where $\nEL = (1- \xHI)\nH$ is used in the grain charging parameter $\psi = G_0 T^{1/2}/\nEL$ that sets the \PE\ efficiency.

We consider the intergalactic UVB model of \citet{FG20}. At two redshifts $z = 0$ and $3$, Table D1 in \citet{FG20} gives the hydrogen PI rate of
  $\zeta_{\rm pi,H} = 3.62 \times 10^{-14} \second^{-1}$
  and $9.15 \times 10^{-13} \second^{-1}$; the PI heating rate of $\Gamma_{\rm pi}/\xHI = q_{\rm pi,H}\zeta_{\rm pi,H} = 3.22\times 10^{-25} \erg \second^{-1}$ and $6.03\times 10^{-24} \erg \second^{-1}$. We ignore the FUV part of the UVB which is generally small compared to the ISRF; $J_{\rm FUV,UVB}$ at redshift 0 (3) is about a factor 220 (3.4) times smaller than $J_{\rm FUV}^{\rm Draine}$
  (see also the bottom panel of \autoref{f:sb99_spec} in \autoref{s:ISRF}). Following typical practices, a reduction in the ionization and heating rates associated with the surrounding galactic ISM can be optionally included via a density-dependent shielding factor
  $\fshldH < 1$, using the form proposed by \citet{Rahmati13} (their Table A1 and Eq. A1); we apply shielding in Case 2.

Our results are shown in \autoref{f:equil-UVB}. Even though all models adopt an identical cooling function (black curve in the top left panel of \autoref{f:heatcool-comp}), results are strikingly different. 
A two-phase equilibrium is possible for all cases,
but for the unshielded UVB \PI\ heating at $z=0$ (blue solid curves), the range of two-phase thermal pressure and density (see top-left panel) is several times higher than that of the local solar neighborhood. 
With this ``high'' equilibrium curve, all diffuse gas in the solar neighborhood that has thermal pressure falling in the observed range ($\log_{10} P/\kB = 3.58 \pm 0.175\,{\rm dex}$; \citealt{Jenkins11}) would be in the warm phase.
The temperature (top-right panel) on the low-density, warm branch, which is controlled by Ly$\alpha$ cooling, is essentially the same as in our fiducial model.
Without reduction of the UVB, the ionization rate is quite high and the equilibrium ionization (bottom-right panel) on the low-density, warm branch is an order of magnitude above that in our fiducial model.
The higher $\xEL$ also raises the pressure on the warm branch (up to a factor of 2).
We note (not shown) that if unattenuated UVB is assumed, adding the effect of \PE\ heating (with $\chiunatt = 1$) only mildly increases thermal pressure (about a factor 2).

If galactic shielding of the UVB is included following \citet{Rahmati13} (red solid curves), the \PI\ heating rate becomes very low: $\Gamma_{\rm pi} \sim 10^{-28}(\nH/n_{\rm gal\textnormal{-}shld})^{-1} \erg \second^{-1}$ where $n_{\rm gal\textnormal{-}shld} \sim 10^{-3}\pcc$ is the density above which $\fshldH$ drops significantly below 1.
The equilibrium ionization fraction is also extremely low (bottom-right panel).
Even though \PE\ heating is included with FUV radiation the same as the solar neighborhood ($\chiunatt = 1$), the two-phase medium regime shifts to very low pressure and density.
For the maximum/minimum pressure of the WNM/CNM branch, we find $P^{\rm (min)}/\kB = 94 \pcc\Kel$, $P^{\rm (max)}/\kB = 278 \pcc\Kel$ with $\nH^{\rm (max)} = 0.04\pcc$, $\nH^{\rm (min)} = 0.36\pcc$ and $T^{\rm (max)} = 6350 \Kel$, $T^{\rm (min)} = 238 \Kel$.\footnote{In a somewhat different context, \citet{Sternberg02} calculated thermal and ionization equilibrium of \HI\ clouds exposed to metagalactic UVB.
They considered low-metallicity ($\Ztot=0.1$) gas, did not include PE heating from interstellar FUV, and assumed constant (density-independent) \HI\ shielding columns ranging between $\sim 10^{18}$--$10^{22}\cm^{-2}$. Although direct comparison is difficult, their phase diagram and $P^{\rm (max/min)}$ at shielding column $\sim 5\times 10^{19}\cm^{-2}$ (Figures 5 and 6; for which the primary H ionization rate $\sim 10^{-17}$--$10^{-18}\second^{-1}$ is comparable to our value at $\nH \sim 0.1 \pcc$) are broadly consistent with our result.}
At the observed solar neighborhood pressure, all gas would be in the cold phase. As pointed out by \citet{Wibking22}, the reason for this low \PE\ heating rate is the very low electron abundance (two orders of magnitude below our fiducial model at $\nH = 1 \pcc$)
This (unphysically) low $\xEL$ increases the grain charging parameter, which in turn dramatically reduces the efficiency of \PE\ heating (see \autoref{s:heating_pe}) compared to our fiducial model (bottom-left panel).

Finally, we note that ionization and heating by unattenuated UVB at high redshift ($z \sim 2$--$3$) is much higher, which results in even higher two-phase pressure and density.
When the density-dependent reduction factor is included, however, the heating and ionization rates are still low enough that the maximum warm-phase pressure is an order of magnitude lower than in our fiducial model even at $z \sim 2$--$3$.

This simple exercise demonstrates that not only the cooling function (held fixed here) but also the heating rate and ionization fraction need to be treated carefully to produce realistic ISM phase structure. In particular, if one adopts a \PE\ heating rate that sensitively depends on grain charging, a realistic electron abundance must be calculated in order for that heating rate to be meaningful. The two-phase equilibrium pressure can be severely underestimated if the source of ionization (low energy CRs and/or X-rays) is not modeled realistically. For example, \citet{Hu17} adopted the PE heating prescription of \citet{Wolfire03}, but turned off ionization by low-energy CRs; this led to an extremely low two-phase equilibrium pressure of $\sim 10^2\cm^{-3}\Kel$ at $\chi=1$ and $\Ztot=0.1$.

Of course, the \PE\ heating rate is proportional to the FUV intensity, so a realistic estimate of this quantity (which depends on recent local star formation) is also needed in order to obtain realistic phase structure. Too high (low) a value of $\chiunatt$ could lead to an all-warm (all-cold) ISM (see \autoref{f:equil_unshld}).

For our model and most other contemporary ISM models, low-energy CRs provide the ionization in most of the ISM's mass; soft X-rays would also be able to produce an ionization level high enough to neutralize positive grain charging and the resulting drop in \PE\ efficiency \citep{Wolfire95, Wolfire03}.
The heating and ionization by metagalactic UVB are negligible in the main portions of star-forming galaxy disks\footnote{Using the \citet{Haardt96} UVB model, \citet{Wolfire03} estimated that EUV/X-ray from the UVB provides only $\sim 1\%$ of the total ionization rate at the solar circle.}, although they play a role in quenching star formation in outer regions of galaxy disks \citep{Schaye04} and in determining multiphase structure of gas well above the disk midplane \citep{Sternberg02}.


\subsection{Effect of Photoionization in hot gas cooling}

It should be noted that our adopted CIE cooling in hot gas neglects the effect of \PI, which can change the ionization balance and the cooling rate at a given temperature \citep[][]{Wiersma09, Gnedin12, Oppenheimer13, Kannan14}.
However, a significant change is expected only for a small volume of the ISM exposed to very strong and hard radiation fields (e.g., near hot stars). In addition, because the cooling time can be short compared to the recombination time in low-density plasma, the ``recombination lag'' can cause gas to be overionized relative to CIE \citep[e.g.,][]{Gnat07}. This effect can partially cancel out the reduction of cooling rate by \PI.

\section{Summary}\label{s:summary}

The gas in the ISM spans a wide range of density, temperature, and ionization states. Realistic modeling of the multi-phase structure of the ISM requires self-consistent calculations of gas (M)HD coupled with thermodynamics, the latter of which depends on chemistry and UV radiation. 

In this paper, we present a comprehensive, accurate, and computationally efficient model for ISM photochemistry and heating/cooling processes. This new model has been implemented in the MHD code \Athena, and incorporated in the 3D MHD galactic disk simulation framework, TIGRESS (papers upcoming), and has been used in MHD simulations of individual turbulent, star-forming GMCs with radiation feedback \citep{KimJG21}.

We summarize our model approach as follows.
\begin{enumerate}
    \item Our model includes key microphysical processes for all the most important ISM phases (\autoref{s:overview}): molecular gas, atomic gas (CNM and WNM), \HII\ regions, warm ionized medium (WIM or DIG), and the hot ionized medium (HIM). We also follow interactions between the gas and dust grains, UV radiation, and CRs. Our models are applicable over the full range of densities and temperatures found in the multiphase ISM.
   
    \item We have identified an essential set of chemical species and reactions (\autoref{s:chem} and \autoref{t:chem_cooling}), as needed to follow time-dependent thermodynamics realistically in (M)HD simulations. We trace the main ionized, atomic, and molecular forms of hydrogen, carbon, and oxygen, as well as the electron abundance. The hydrogen chemistry (\HH, \Hplus, and \Ho) is calculated in a time-dependent fashion. In atomic gas, the \CII\ abundance is calculated assuming equilibrium chemistry. In molecular gas, the abundances of \CO\ and molecular ions are obtained from fits. The \OII\ abundance follows that of \Hplus. Finally, the abundances of \CI, \OI, and electrons are obtained from conservation laws.
    
    \item We implement the major heating and cooling processes in the ISM (\autoref{s:heating_cooling} and \autoref{t:chem_cooling}), allowing for metallicity and dust abundance to be free parameters. Heating includes PI, photodissociation, the \PE\ effect on small grains, CR ionization, UV pumping, and \HH\ formation and dissociation heating.  Cooling includes molecular, atomic, and ionic lines, free-free and recombination emission, and collisions with dust grains.
    The cooling by hydrogen (\HH, \Hplus, and \Ho) is calculated self-consistently with chemistry across the full temperature range. The cooling by metals is calculated from the chemical abundances in atomic and molecular gas, and from fits to nebular emission lines in photoionized gas. At $T > 2\times 10^4~\Kel$, the cooling by He and metals is transitioned to tabulated CIE cooling functions.
    
    \item We follow key thermal and chemical effects on the gas due to UV radiation and CRs (\autoref{t:chem_cooling}). For the radiation itself, we follow three different bands (EUV, PE, LW) separately using the adaptive ray tracing method for radiative transfer of photon packets that originate from stellar clusters, including extinction by dust and shielding by chemical species (\autoref{s:radcr}). The diffuse background FUV radiation is calculated from the simple six-ray method. The CRs are attenuated using the effective column density obtained from the adaptive ray tracing or six-ray method.
    
    \item Our method is implemented in the \Athena\ code, and we use a first-order operator split method to update the energy, chemistry, and radiation source terms (\autoref{s:updates}). We first calculate the radiation field after the MHD updates, and then use sub-cycling to evolve the gas temperature and chemical species abundances.
\end{enumerate}

We have run a series of tests to validate our methods and to obtain insights into the effects of photochemistry on realistic problems. We summarize the main findings from these tests as follows.
\begin{enumerate}[label=(\alph*)]
    \item Equilibrium curves of thermal pressure and temperature of the neutral ISM depend sensitively on the FUV radiation, CR ionization rate, metallicity, and shielding (\autoref{s:onezone}). Accurately modeling the heating and cooling rates accounting for these environmental variations is crucial in determining the equilibrium pressure, density, and temperature of the WNM and CNM.
    
    \item We have successfully reproduced the chemical and thermal structure of 1D PDRs calculated from a full chemistry code with high accuracy (\autoref{s:plane_par}). Compared to the full chemistry code, our method has a much lower computational cost. 
    
    \item For the \HI-to-\HH\ transition (\autoref{s:HIH2}), our model agrees very well with the analytic theory by \citet{Bialy16} in the case without CRs, as long as the transitional layer is numerically resolved. We note, however, that CR dissociation is important under realistic ISM conditions; if CRs are not included the \HI\ column is underestimated, 
    especially in low density or low UV radiation environments.
    
    \item In the SNR expansion test (\autoref{s:SNR}), we 
    varied the ambient medium density and metallicity over $\nHinit = 0.01, 0.1, 1, 10, 100 \pcc$ and $\Ztot = 10^{-3}, 10^{-2}, 0.1, 1, 10$, and computed the mass, radius, and momentum at the time of shell formation. We also provide fits for these quantities as functions of background density and metallicity for the CIE cooling case. When we use non-equilibrium ionization and cooling, the shell evolution and momentum injection can be modified compared to calculations adopting CIE cooling, especially in low density and low metallicity environments, and depending on upstream ionization. 
   
    \item We perform tests of expanding \HII\ regions, where the expansion is driven by either gas or radiation pressure (\autoref{s:HIIreg}). In both cases, our numerical results agree with the known analytic solutions. We confirm that radiation only needs to be updated based on the MHD CFL timestep criterion during dynamical expansion stages. In the gas pressure driven case, the expansion rate and momentum injection are higher at low metallicity due to the higher ionized gas temperature.
\end{enumerate}

In addition, we compare our model to other models of heating and cooling that have been adopted in the ISM and galaxy formation literature (\autoref{s:discussion}). We find that:
\begin{itemize}
\item We can
  reproduce the equilibrium pressure and temperature curves for the neutral ISM when compared to a full chemistry model, and the pressure is consistent with observational constraints. However, this is not the case for other heating/cooling models used in the galaxy formation literature (Figures \ref{f:equil-comp} and \ref{f:heatcool-comp}). Discrepancies by more than an order of magnitude are common, and are more extreme for the low metallicity case. As galaxy simulations reach higher resolution 
    and achieve conditions under which the gas should become neutral, some previous approaches will not be adequate to produce realistic thermal properties of the ISM.
    
    \item In both ISM and galaxy (M)HD simulations, the main motivation for following cooling in detail is that thermodynamics affects dynamics. However, the heating function strongly affects thermal transition points, and ionization strongly affects heating. Even with our cooling function, adopting common approaches to heating and ionization based on metagalactic UVB will lead to pressures discrepant with observations (Figure \ref{f:equil-UVB}). For star-forming galaxies, heating and ionization of the ISM are internally generated by stellar radiation and CRs, and must be calculated self-consistently in order to obtain  physically meaningful ISM pressure and temperature.
\end{itemize}

Finally, we provide some thoughts on promising future directions to further improve physical ingredients and modeling methods presented in this work. 
First, our chemical network that follows a limited number of metal ion species can be extended to follow non-equilibrium states of multiple metal ions and resulting cooling, as in e.g.~\citet{Sarkar21a, Sarkar21b, Katz22}; at the same time, the number of radiation energy bins can be increased to improve the accuracy of PI calculation.
Second, our model assumes that grains to be perfectly coupled to gas with a fixed dust-to-gas ratio.
In the real ISM, dust moving through gas experiences a drag force that depends on grain size; the abundance and size distribution of dust grains also evolve over time due to multiple physical processes such as accretion of gas-phase metals, sputtering, shattering, coagulation, and injection by evolved stars and SNe. These physical processes can be modeled directly by utilizing ``live'' dust particles \citep{McKinnon18}, as well as the self-consistent coupling with radiation field \citep{McKinnon21}.

\acknowledgements

We are grateful to the referee for a careful reading of the manuscript and critical comments for its improvement. We thank Shmuel Bialy and Mark Wolfire for sharing their data, and Bruce Draine, Taysun Kimm, and Aaron Smith for helpful discussions.
J.-G.K. acknowledges support from the Lyman Spitzer, Jr. Postdoctoral Fellowship at Princeton University and from the EACOA Fellowship awarded by the East Asia Core Observatories Association. M.G. acknowledges support from Paola Caselli and the Max Planck Institute for Extraterrestrial Physics. J.-G.K. and M.G. acknowledge Paris-Saclay University's Institut Pascal program ``The Self-Organized Star Formation Process'' and the Interstellar Institute for hosting discussions that nourished the development of the ideas behind this work. This work was supported in part by NASA ATP grant No. NNX17AG26G, NSF grant AST-1713949, and by grant No. 510940 from the Simons Foundation to E.~C. Ostriker. 

\appendix

\begin{figure*}
  \epsscale{1.0}\plotone{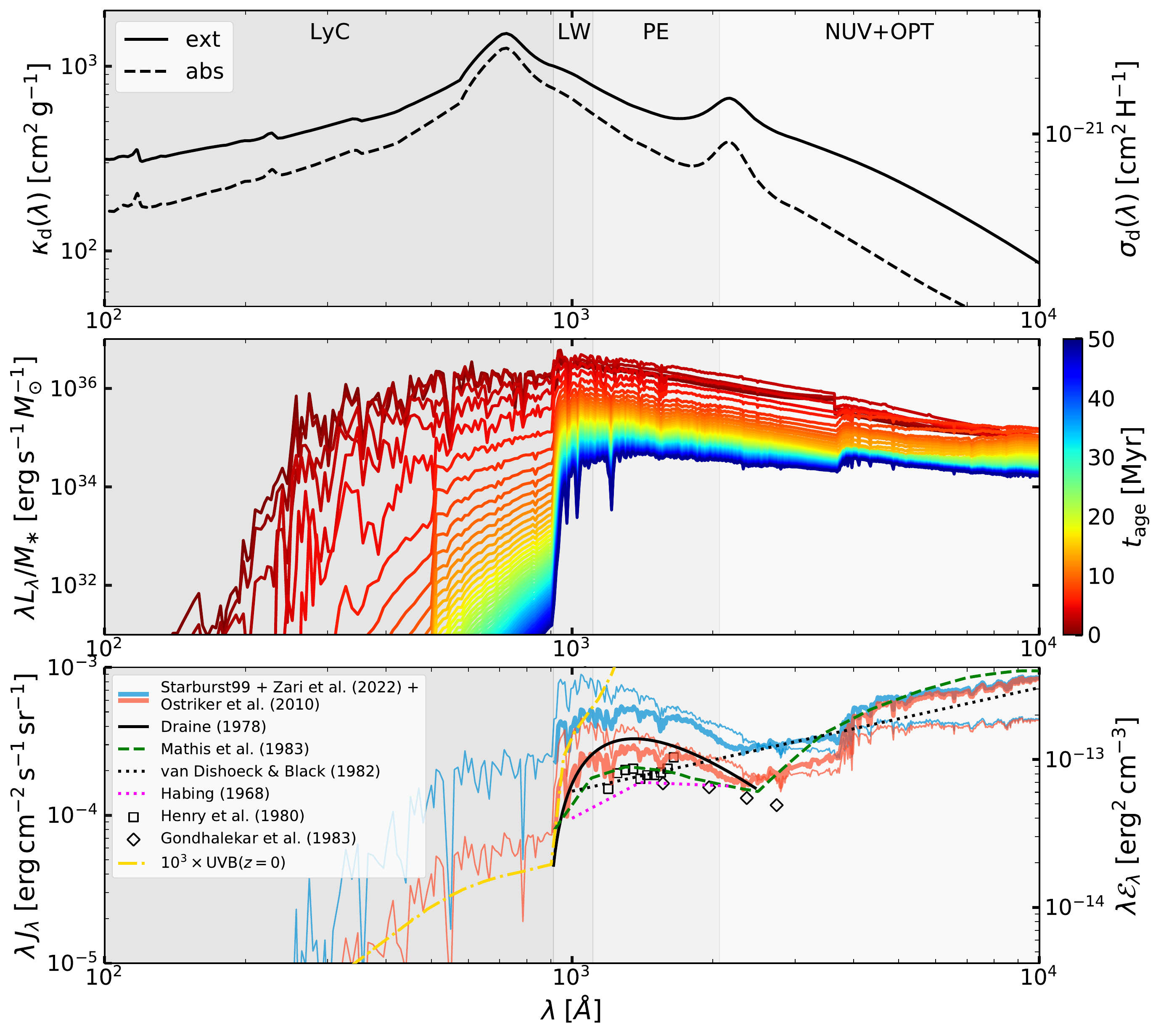}
  \caption{(Top) Dust extinction (solid) and absorption (dashed) cross sections at UV and optical wavelengths for the \citet{Weingartner01a} Milky Way grain size distribution with ratio of visual extinction to reddening $R_V = 3.1$ and C abundance parameter $b_{\rm C}=55.8\,{\rm ppm}$. (Middle) SED in the UV and optical bands for a coeval population of stars sampling the Kroupa IMF, with ages $1$, $2$, $\cdots$, $50 \Myr$. (Bottom) Various estimates of the local ISRF. The thick blue and red lines show the estimate at the disk midplane using (1) the light-to-mass ratio shown in the middle panel, (2) history of SFR at the solar circle based on \citet{Zari22}, and (3) an attenuation factor that depends on dust optical depth of the gas disk \citep{Ostriker10}; the thin blue and red lines show the corresponding estimate without the attenuation factor (see text for details). Other lines are the estimate of the local ISRF by \citet{Draine78} (black solid), \citet{Mathis83} (green dashed), \citet{van-Dishoeck82} (black dotted), and \citet{Habing68} (magenta dotted). For comparison, we also show the UVB model of \citet{FG20} at zero redshift, scaled by $10^3$ times (yellow dot-dashed). Estimates based on observations are shown as open squares \citep{Henry80} and diamonds \citep{Gondhalekar80}.}\label{f:sb99_spec}
\end{figure*}

\section{Radiation Output from Starburst99}\label{s:sb99}

\begin{figure*}
  \epsscale{1.0}\plotone{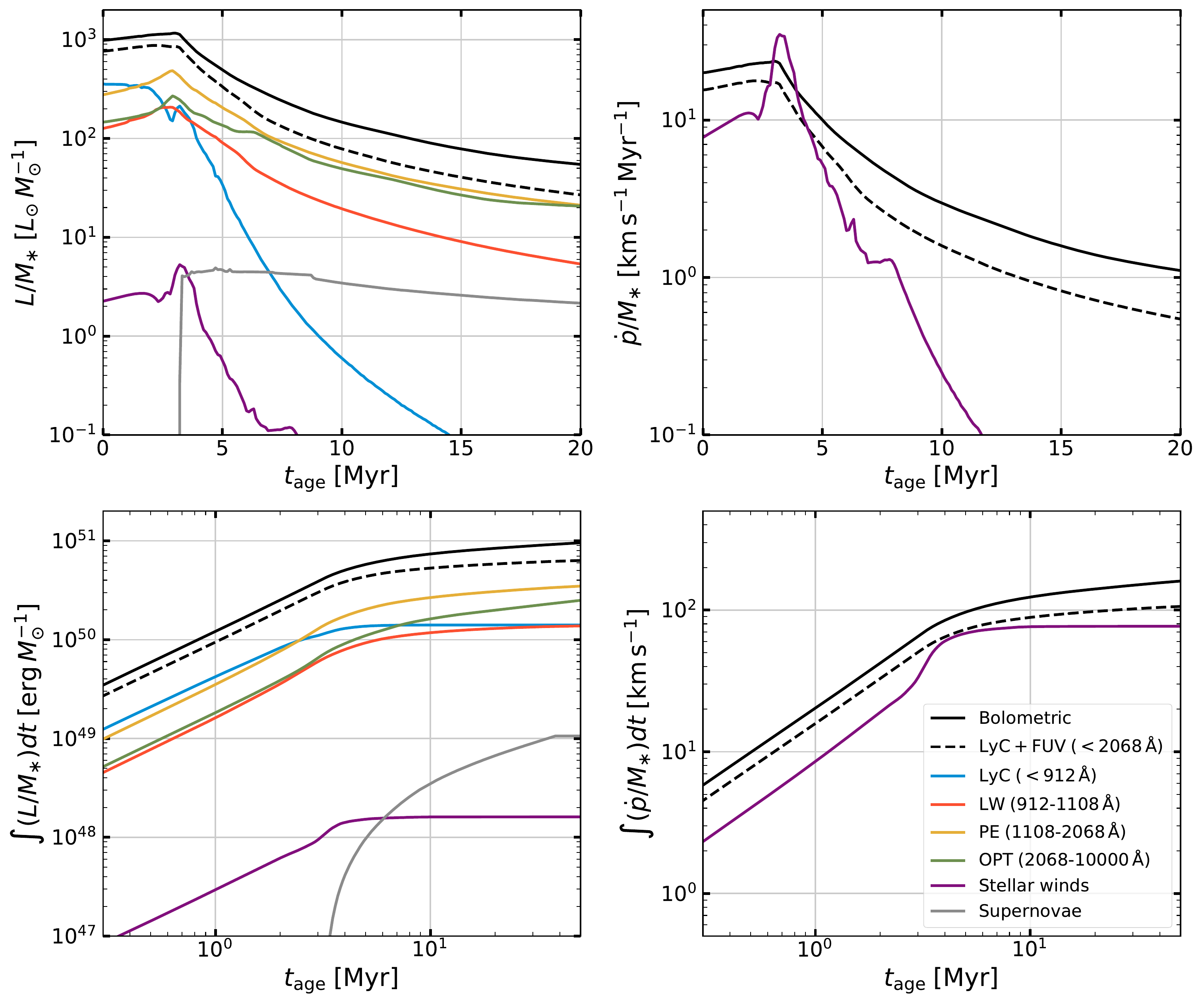}
  \caption{The top left panel shows the luminosity per unit mass of a coeval stellar population in the UV and optical bands as a function of age, based on \SB. The purple and grey lines show the energy output rate per unit mass from stellar winds ($L_{\rm w}/M_{\ast}$) and supernovae ($L_{\rm sn}/M_{\ast}$), respectively. The corresponding cumulative energy output per unit mass is shown in the bottom left panel. The right panels show the instantaneous and cumulative momentum output per unit mass.}\label{f:sb99_Edotpdot}
\end{figure*}

\begin{figure}
  \epsscale{1.0}\plotone{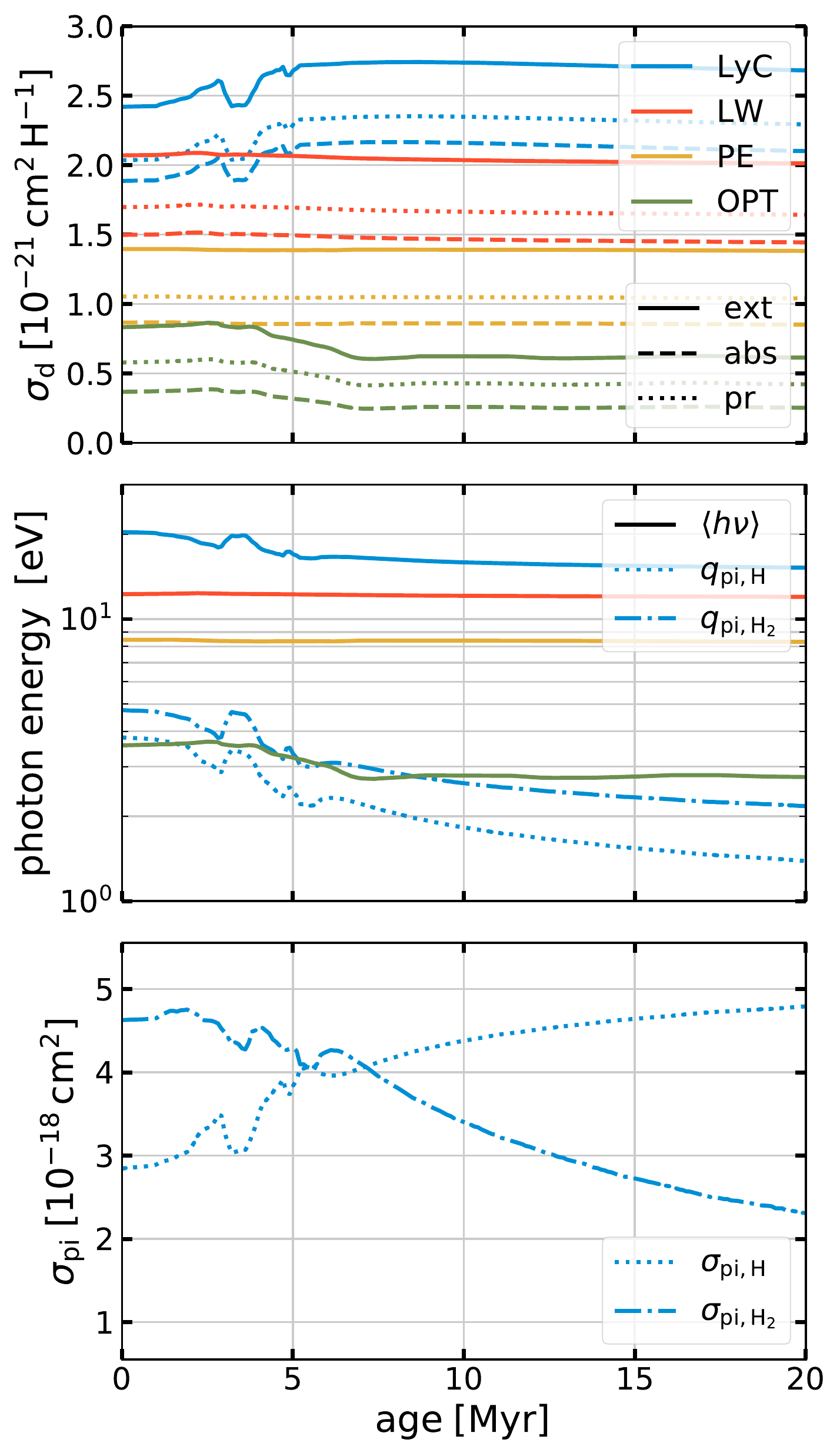}
  \caption{(Top) Absorption, extinction, and radiation pressure cross sections per H nucleon, averaged over the SED of a star cluster (middle panel of \autoref{f:sb99_spec}), as a function of the cluster age. (Middle) Mean energy of photons (solid) and average energy of photoelectrons for \PI\ of H and \HH. (Bottom) Average \PI\ cross sections for H and \HH.}\label{f:sb99_hnu_sigma}
\end{figure}

We use \SB\ \citep{Leitherer99, Leitherer14} to calculate the radiation output from a coeval population of stars sampling the Kroupa IMF for cluster age up to $t_{\rm max} = 50 \Myr$. 
For the results shown here, we adopt the Geneva evolutionary tracks for non-rotating and solar metallicity stars.
The middle panel in \autoref{f:sb99_spec} shows the specific luminosity per unit wavelength ($\Psi_{\lambda} = L_{\lambda}/M_{\ast}$)
at $t_{\rm age} = 1$, $2$, $\dots$, $50 \Myr$, multiplied by wavelength. The left column of \autoref{f:sb99_Edotpdot} 
shows the specific luminosity (top) and cumulative specific energy output (bottom) in our different frequency bins (plus OPT) while the right column shows the corresponding specific momentum outputs. Most of radiation energy is emitted in  UV wavelengths. The UV (bolometric) specific luminosity is $760 \Lsun\Msun^{-1}$ ($986 \Lsun\Msun^{-1}$) at $t=0$, reaches a maximum $870 \Lsun\Msun^{-1}$ ($1166 \Lsun\Msun^{-1}$) at $t_{\rm age} \approx 3\Myr$ and decreases afterwards.
In the top three rows of \autoref{t:sb99}, we present for our different wavebands the average time at which a photon is emitted after the the birth of a cluster ($\langle t \rangle = \int_0^{t_{\rm max}} \Psi t dt / \int_0^{t_{\rm max}} \Psi dt$), and the times at which 50\% and 90\% of the total cumulative radiation energy is emitted (e.g., $\int_0^{t_{\rm cumul,50\%}} L dt = 0.5 \int_0^{t_{\rm max}} L dt$) for $t_{\rm max}=100\Myr$. The fourth row  presents the total radiation energy emitted per stellar mass over $t_{\rm max}=100\Myr$ period.

\autoref{f:sb99_Edotpdot} also shows the energy and momentum output from stellar winds and supernovae for comparison. 
For comparison to the integrated radiation outputs in \autoref{t:sb99},  the specific energy output for stellar winds and SNe are $1.6 \times 10^{48}\erg\Msun^{-1}$ and $1.1 \times 10^{49}\erg\Msun^{-1}$, respectively.
The UV luminosity  ($L_{\rm UV}$) dominates over the wind luminosity ($L_{\rm w}$) by more than two orders of magnitude, but the wind momentum input rate
($\dot{p}_{\rm w} = 2L_{\rm w}/V_{\rm w}$) is comparable to that of radiation ($\dot{p}_{\rm UV} = L_{\rm UV}/c$) because $(c/V_{\rm w})( L_{\rm w}/L_{\rm UV})$ is order unity for the wind velocity $V_{\rm w} \sim (1$--$3) \times 10^3\kms$.
In \autoref{f:sb99_Edotpdot} we do not show the direct momentum from supernovae because the actual momentum deposition in the ISM depends on work done by supernova shocks during adiabatic expansion stages, and therefore depends on supernova energy rather than momentum \citep[e.g][]{KimCG15, KOR17}.



\section{Dust Opacity and SED-averaged Cross Sections}\label{s:sigma_ave}

We adopt the \citet{Weingartner01a} grain optical properties for size distribution A (that minimizes the amount of C and Si in grains) and $b_{\rm C}=4 \times 10^{-5}$ (C abundance in very small grains in log-normal distribution) for the ratio of visual extinction to reddening $R_V \equiv A_V / E(B-V) = 3.1$, appropriate for diffuse regions of the local ISM. The top panel of \autoref{f:sb99_spec} shows the absorption and scattering cross sections per gas mass $\kappa_{\rm d}$ (left) and per hydrogen nuclei $\sigma_{\rm d} = \muH m_{\rm H} \kappa_{\rm d}$ (right) in the UV and optical wavelength range for $\Ztot = 1$.

We calculate dust cross sections for absorption $\sigma_{\rm d,abs}$, extinction $\sigma_{\rm d,ext} = \sigma_{\rm d,abs} + \sigma_{\rm d,sca}$, and radiation pressure ($\sigma_{\rm d,pr} = \sigma_{\rm d,abs} + (1 - \langle \cos \theta \rangle) \sigma_{\rm d,sca}$) averaged over the (optically-thin) \SB\ SED as a function of $t_{\rm age}$. Each of these averages is defined as
\begin{equation}
  \sigma_{\rm d} = \frac{\int \sigma_{\rm d}(\nu) L_{\nu} d\nu}{\int L_{\nu} d\nu}.
\end{equation}
We also calculate the photon-rate weighted mean photon energy and mean cross sections and photoelectron energy for H- and \HH-photoionization: 
\begin{equation}
  \langle h\nu \rangle = \frac{\int L_{\nu} d\nu}{\int L_{\nu}/(h\nu) d\nu}
\end{equation}
\begin{equation}
  \sigma_{\rm pi} = \frac{\int  \sigma_{\rm pi}(\nu) L_{\nu}/(h\nu) d\nu}{\int L_{\nu}/(h\nu) d\nu}
\end{equation}
\begin{equation}
  q_{\rm pi} = \frac{\int  \sigma_{\rm pi}(\nu) L_{\nu}/(h\nu) \sigma_{\rm d}(\nu) (h\nu - h\nu_0) d\nu}{\int \sigma_{\rm pi}(\nu) L_{\nu}/(h\nu) d\nu}
\end{equation}
These are shown as a function of cluster age in \autoref{f:sb99_hnu_sigma}. \autoref{t:sb99} presents luminosity- or photon-rate weighted cross sections and photon energies. These temporal averages involve an additional integral over time in the numerator and denominator of each expression.

We note that since we weight cross sections in given radiation bands under the assumption that the radiation field derives from a young stellar population, some adjustments would be needed for applications to elliptical galaxies or bulge regions where the UV radiation field is dominated by an older stellar population.

\section{Theoretical Estimate of the Local ISRF}\label{s:ISRF}

The SED of a coeval stellar population as a function of age (as shown in the middle panel of \autoref{f:sb99_spec}) can be used to estimate the intensity of the local ISRF at FUV and optical wavelengths. We first calculate the luminosity per area of all stars formed until time $t$ as
\begin{equation}
    \Sigma_{\lambda}(t) = \int_{-\infty}^{t} \Psi_{\lambda}(t - t^{\prime}) \Sigma_{\rm SFR}(t^{\prime}) dt^{\prime}\,,
\end{equation}
where $\Sigma_{\rm SFR}(t)$ is the SFR per unit area, and we adopt the same \SB\ parameters as in \autoref{s:sb99}.
The midplane intensity of a dusty slab of uniform density and emissivity is given by
\begin{equation}\label{e:isrf_att}
    J_{\lambda,{\rm mid}} = \frac{\Sigma_{\lambda}}{4\pi \tau_{\rm d,\lambda}} \left( 1 - E_2(\tau_{\rm d,\lambda}/2) \right)\,,
\end{equation}
where $\tau_{\rm d,\lambda} = \kappa_{\rm d,abs,\lambda}\Sigma_{\rm gas}$ is the dust optical depth and $E_2$ is the second exponential integral \citep{Ostriker10, Ostriker2022}. We take $\Sigma_{\rm gas} = 10 \Sunit$ for the gas surface density.

Using a sample of hot and luminous stars drawn from \citet{Zari21}, \citet{Zari22} mapped the stellar age distribution and star formation history of the extended solar neighborhood. They found that there is a significant increase in the SFR in the very recent past ($5\Myr < t - t^{\prime} < 10\Myr$) (see their Figs 5 and 6).
Since \citet{Zari22} 
did not consider stars younger than $5\Myr$ and hence the star formation history in the past $5\Myr$ in their analysis, here we consider two simple star formation histories to bracket a range of plausible values: (1) $\Sigma_{\rm SFR,\textnormal{-3}} = 4$ for $t - t^{\prime} < 10\Myr$ and $2$ otherwise; and (2) $\Sigma_{\rm SFR,\textnormal{-3}} = 1$ for $t - t^{\prime} < 5\Myr$; $4$ for $5\Myr < t - t^{\prime} <10\Myr$, and $2$ otherwise, where $\Sigma_{\rm SFR,\textnormal{-3}} = \Sigma_{\rm SFR}/(10^{-3}\Msun\yr^{-1}\kpc^{-2})$.

The resulting local ISRF including attenuation (\autoref{e:isrf_att}) based on star formation histories (1) and (2) is respectively shown as blue and red thick lines in the bottom panel of \autoref{f:sb99_spec}. The thin blue and red lines indicate the corresponding naive estimate $J_{\lambda,{\rm mid}}=\Sigma_{\lambda}/(4\pi)$. Other (theoretical and observational) estimates of the local ISRF are shown as lines and symbols. History (1) with elevated star formation for all of the past $10\Myr$ (thick blue) lies slightly above the \citet{Draine78} estimate (solid black) in the FUV wavelengths. History (2) with a lower level of star formation in the past $5\Myr$ has a factor $\sim 2$ lower intensity and agrees better with the \citet{Draine78} and \citet{Mathis83} models and observational estimates \citep{Henry80, Gondhalekar80}. We also examined the case with a constant SFR $\Sigma_{\rm SFR,\textnormal{-}3} = 2.4$, an estimate of the local SFR averaged over the past $\sim 3\Gyr$ by \citet{Fuchs_2009}. The result is intermediate between the above two cases (not shown).
Integrated over the FUV (LW+PE) band, the luminosity per unit area for the three cases considered is $\Sigma_{\rm FUV} = 15.6$, $8.1 \Lsun \pc^{-2}$, and $10.9\Lsun \pc^{-2}$, while the midplane intensity is $J_{\rm FUV} = 3.6\times 10^{-4}$, $1.9\times 10^{-4}$, and $2.5\times 10^{-4}\erg \second^{-1} \cm^{-2}\,{\rm sr}^{-1}$.
For steady star formation, the ratio between luminosity and SFR is $1.1 \times 10^3$, $1.1 \times 10^3$, and $3.2 \times 10^3 \Lsun/(\Msun \Myr^{-1})$ for LyC, LW, and PE bands, respectively.

Finally, the yellow dot-dashed line 
shows the UVB model of \citet{FG20} at zero redshift scaled by a factor $10^3$. Even though the UVB is orders of magnitude lower than the ISRF strength in UV wavelengths, its \PI\ heating rate can be higher than the grain PE heating rate by the ISRF when no attenuation factor is applied (see \autoref{s:comparison} and \autoref{s:comparison_UVB} for details).

\begin{deluxetable}{lcccccc} 
\tabletypesize{\scriptsize}
\tablecaption{Time-scales \REV{and specific energy} of radiation from Starburst99 simulation and time-averaged cross sections and energies of photons and photoelectrons}
\tablewidth{0pt}
\tablehead{
\colhead{} &
\colhead{LyC} &
\colhead{LW} &
\colhead{PE} &
\colhead{FUV} &
\colhead{OPT} &
\colhead{Bol}
}
\startdata
\tableline
\multicolumn{7}{c}{Timescales (Myr)} \\
\tableline
(1) $\langle t \rangle$ & 1.9 & 6.6 & 11.9 & 10.4 & 20.7 & 12.9 \\
(2) $t_{\rm cumul,50\%}$ & 1.7 & 3.4 & 4.3 & 4.0 & 7.8 & 4.1 \\
(3) $t_{\rm cumul,90\%}$ & 3.7 & 14.9 & 34.3 & 28.6 & 63.4 & 40.8 \\
\tableline
\multicolumn{7}{c}{Specific energy ($10^{50} \erg \Msun^{-1}$)} \\
\tableline
(4) $\smallint \Psi dt$ & 1.4 & 1.4 & 3.7 & 5.1 & 2.9 & 10.4 \\ 
\tableline
\multicolumn{7}{c}{Dust cross sections ($\sigma_{\rm d}/10^{-21}\cm^{2}$)} \\
\tableline
(5) $\langle \sigma_{\rm d,abs} \rangle$ & 1.94 & 1.50 & 0.86 & 1.00 & 0.32 & - \\
(6) $\langle \sigma_{\rm d,ext} \rangle$ & 2.48 & 2.07 & 1.39 & 1.54 & 0.75 & - \\
(7) $\langle \sigma_{\rm d,pr} \rangle$ & 2.10 & 1.69 & 1.05 & 1.19 & 0.52 & - \\
\tableline
\multicolumn{7}{c}{Photoionization cross sections ($\sigma_{\rm pi}/10^{-18}\cm^{2}$)} \\
\tableline
(8) $\langle \sigma_{\rm pi,H} \rangle$ & 3.1 & - & - & - & - & - \\
(9) $\langle \sigma_{\rm pi,H_2} \rangle$ & 4.6 & - & - & - & - & - \\
\tableline
\multicolumn{7}{c}{Energy (${\rm eV}$)} \\
\tableline
(10) $\langle h\nu \rangle$ & 19.4 & 12.2 & 8.4 & 9.3 & 3.2 & - \\
(11) $\langle q_{\rm pi,H} \rangle$ & 3.4 & - & - & - & - & - \\
(12) $\langle q_{\rm pi,H_2} \rangle$ & 4.4 & - & - & - & - & - \\
\enddata
\tablecomments{Columns show photon energy bands (see \autoref{s:rt}). Each row shows: (1) luminosity-weighted average time a photon is emitted; (2)--(3) time needed to emit 50\% and 90\% of the total energy emitted over $t_{\rm max} = 100\Myr$ period; \REV{(4) total energy emitted per mass over $t_{\rm max} = 100\Myr$ period;} (5)--(7) Luminosity-weighted average dust absorption, extinction, and radiation pressure cross sections; (8)--(9) photon rate-weighted average photoionization cross sections for H and H$_2$; (10) photon rate-weighted average photon energy; (11)--(12) photon-weighted mean energy of photoelectrons produced by photoionization of H and H$_2$ (LyC only).}\label{t:sb99}
\end{deluxetable}

\section{Implementation of Grain Destruction}\label{s:dust_dest}

Dust grains are believed to be destroyed by sputtering, which can be either thermal or nonthermal \citep[e.g.,][]{Hu19}.  To model effects of thermal sputtering, 
we simply assume that grains are destroyed instantly by thermal sputtering if $T > 10^6 \Kel$. 

For non-thermal sputtering, we assume that grains are destroyed if the terminal drift velocity $v_{\rm d}$ is greater than $10^2 \kms$. The terminal velocity for a grain 
is found by balancing the radiation pressure force with the gas drag force (both per unit area):
\begin{equation}\label{e:vd}
    \langle Q_{\rm pr} \rangle \frac{|\bm{F}|}{c} = 
2 \nH \kB T G(s) \,,
\end{equation}
where $\langle Q_{\rm pr} \rangle$ is the radiation pressure efficiency factor, and the drag function $G(s)$ accounts for both collisional and Coulomb drag (e.g., Eq. 24 in \citealt{Draine11}) with $s = v_d/\sqrt{2\kB T/\mH}$.
Taking $\langle Q_{\rm pr} \rangle \approx 1$, $P \approx 2\nH \kB T$, $P_{\rm rad}=|\bm{F}|/c$, the drift velocity can be found by solving $P_{\rm rad}/P \approx G(s)$. For representative values of grain charge and Coulomb logarithm adopted by \citet{Draine11}, $G(s)$ attains local maximum of $\sim 42$ when $s = 0.9$, declines with $s$ due to Coulomb focusing, and rises again and behaves as $G(s) \approx s^2$ for $s \gtrsim 5$ as the collisional drag dominates. The approximate solution of \autoref{e:vd} relevant for non-thermal sputtering is $s \approx \sqrt{P_{\rm rad}/P} \gtrsim \sqrt{50}$. 
\footnote{For $20 < G(s) < 42$, there are two other solutions at lower value of $s$, but one is unstable and the other is too low for non-thermal sputtering to occur \citep{Draine11}.}
Before applying the momentum source term update, we check whether either the condition on temperature or $s$ is  met, and if so we omit the force due to radiation on dust grains.

\bibliographystyle{aasjournal}
\bibliography{references.bib}{}

\end{document}